\documentclass[a4paper,11pt]{article}
\usepackage{jcappub} 
\usepackage{lineno}
\usepackage{graphicx}	
\usepackage{amsmath}	
\usepackage{subfigure}
\usepackage{ifthen}
\usepackage{verbatim}
\usepackage{multicol}
\usepackage{pdflscape}
\usepackage{tabularx}
\usepackage{booktabs} 
\usepackage{rotating} 
\usepackage{adjustbox} 
\usepackage{lipsum} 
\usepackage{float} 
\usepackage{enumitem} 
\usepackage{makecell} 
\usepackage{url}
\usepackage{xcolor}
\usepackage{tikz}
\usepackage{cancel}

\newcommand{\wignerThree}[6]{
  \begin{pmatrix}
    #1 & #2 & #3 \\
    #4 & #5 & #6
  \end{pmatrix}
}

\definecolor{lime}{HTML}{A6CE39}
\DeclareRobustCommand{\orcidicon}{
\begin{tikzpicture}
\draw[lime, fill=lime] (0,0)
circle[radius=0.16]
node[white]{{\fontfamily{qag}\selectfont \tiny \.{I}D}};
\end{tikzpicture}
\hspace{-2mm}
}
\foreach \x in {A, ..., Z}{%
\expandafter\xdef\csname orcid\x\endcsname{\noexpand\href{https://orcid.org/\csname orcidauthor\x\endcsname}{\noexpand\orcidicon}}
}

\arxivnumber{2507.19897} 
\title{From South to North: Leveraging Ground-Based LATs for Full-Sky CMB Delensing and Constraints on $r$}

\author{
Wen-Zheng Chen\hspace{-1.5mm}\orcidA{}$^{a,b}$,  
Yang Liu\hspace{-1.5mm}\orcidB{}$^{a,*}$,  
Yi-Ming Wang\hspace{-1.5mm}\orcidC{}$^{a,b}$,  
Hong Li\hspace{-1.5mm}\orcidD{}$^{a,b,*}$  
}

\begingroup
 
\footnotetext{\mbox{$^*$ Corresponding author.}} 
\endgroup

\affiliation{$^{a}$Key Laboratory of Particle Astrophysics, Institute of High Energy Physics, Chinese Academy of Sciences, Beijing 100049, People's Republic of China}
\affiliation{$^{b}$University of Chinese Academy of Sciences, Beijing 100049, People's Republic of China}

\emailAdd{liuy92@ihep.ac.cn}
\emailAdd{hongli@ihep.ac.cn}

\abstract{
Delensing—the process of mitigating the lensing-induced B-mode contamination in cosmic microwave background (CMB) observations—will be a pivotal challenge for next-generation CMB experiments seeking to detect primordial gravitational waves (PGWs) through B-mode polarization.
This process requires an accurate lensing tracer, which can be obtained either through internal reconstruction from high-resolution CMB observations or from external large-scale structure (LSS) surveys.  Ground-based large-aperture telescopes (LATs) are crucial for internal reconstruction, yet existing and planned facilities are confined to the southern hemisphere, limiting effective delensing to that region.
In this work, we assess the impact of introducing a northern hemisphere LAT, assumed to be situated near AliCPT (hence termed Ali-like LAT, or \texttt{LATN}), on delensing performance and PGW detection, using simulations. Our baseline setup includes a space-based small-aperture mission (LiteBIRD-like, \texttt{SAT}) and a southern LAT (SO-like, \texttt{LATS}). External LSS tracers, which have been shown to play an important role in delensing before the availability of ultra-sensitive polarization data, are also considered.
We find that southern-hemisphere internal delensing reduces the uncertainty in $r$ by $\sim$17\% compared to the no-delensing case. Adding \texttt{LATN} enables full-sky internal delensing, achieving a further $\sim$18\% reduction—comparable to that from including LSS tracers ($\sim$13\%). Once \texttt{LATN} is included, the marginal benefit of LSS tracers drops to $\sim$10\%. These results highlight the significant role of \texttt{LATN} in advancing delensing capabilities and improving PGW constraints.}


\begin{document}
\maketitle
\flushbottom

\section{Introduction}
The Cosmic Microwave Background (CMB) radiation, a faint glow left over from the early universe, serves as a crucial window into the cosmos' infancy. As one of the most important sources of information about the early universe, the CMB anisotropies provide a detailed snapshot of cosmic conditions just a few hundred thousand years after the Big Bang, offering critical insights into cosmology. 
In particular, the study of CMB B-mode polarization patterns has emerged as a powerful tool in the search for primordial gravitational waves (PGWs) \cite{kamionkowski1997probe,seljak1997signature,kamionkowski2016quest,komatsu2022new}, which are predicted to be a direct consequence of inflation in the early universe. Detecting these modes would open a unique observational window into the physics of the very early universe, enabling us to probe fundamental questions about the origin of cosmic structures, the dynamics of inflation, and the fundamental nature of gravity.

The BICEP/Keck Array collaborations have recently placed constraints on PGWs \cite{ade2021improved,collaboration2021bicep}, parameterized by the tensor-to-scalar ratio $r$ (evaluated at a pivot scale of $0.05 \,\text{Mpc}^{-1}$), obtaining an upper bound of $r < 0.036$ at $2\sigma$ confidence for a fixed cosmology. Additionally, Ref.~\cite{tristram2022improved} combines BICEP/Keck Array data with $Planck$ PR4 and baryon acoustic oscillations (BAO), yielding a tighter constraint of $r < 0.032$ at $2\sigma$ by simultaneously fitting cosmological parameters.
Several ongoing and upcoming CMB experiments, including the BICEP Array \cite{hui2018bicep}, Simons Array \cite{suzuki2016polarbear}, Simons Observatory (SO) \cite{lee2019simons}, AliCPT \cite{ghosh2022performance}, \emph{LiteBIRD} \cite{litebird2023probing}, and CMB-S4 \cite{abazajian2022cmb}, aim to detect primordial B-mode polarization over the next decade, further advancing our ability to probe the early universe.


The precise measurement of the tensor-to-scalar ratio $r$ through cosmic microwave background (CMB) B-mode polarization is hindered by various sources of contamination. One of the major challenges arises from gravitational lensing by large-scale structure (LSS), which introduces a significant and unavoidable source of confusion. This lensing converts primordial E-modes into secondary B-modes, thereby mimicking the signal from primordial gravitational waves and obscuring the detection of a true $r$. When $r \lesssim 0.03$\cite{litebird2023probing}, lensing-induced B-modes dominate the signal, necessitating their removal through delensing to recover the primordial component. Next-generation CMB experiments aim to suppress this lensing contamination via high-resolution lensing reconstruction and cross-correlation with external LSS tracers, thus tightening constraints on $r$ and enabling more robust tests of inflationary cosmology \cite{namikawa2022simons,hertig2024simons,namikawa2024litebird,abazajian2022cmb}.
Polarized Galactic emission—mainly from thermal dust and synchrotron radiation by relativistic electrons—is another major contaminant in CMB observations \cite{krachmalnicoff2016characterization}. To mitigate this, component separation methods such as \texttt{Commander}\cite{eriksen2008joint,adam2016planck}, \texttt{NILC}\cite{basak2012needlet}, \texttt{SEVEM}\cite{fernandez2012multiresolution}, and \texttt{SMICA}\cite{cardoso2008component} have been developed to isolate foregrounds based on their distinct spectral signatures.


The removal of lensing-induced contamination from CMB anisotropy maps has been a major focus in recent literature~\cite{diego2020comparison, Diego-Palazuelos:2020lme, Carron:2017vfg, baleato2021limitations}. Accurate delensing requires prior knowledge of the lensing potential $\phi$, which can be obtained through various methods in practice. One approach involves internal reconstruction using quadratic estimators based on CMB two-point statistics~\cite{Okamoto:2003zw}, while Bayesian iterative methods~\cite{carron2017maximum,belkner2024cmb} provide a statistically optimal alternative. 
Additionally, external mass tracers, such as the Cosmic Infrared Background and high-redshift galaxy surveys, act as effective proxies for lensing and have shown significant improvements, particularly in the absence of future polarization surveys with deeper sensitivities~\cite{smith2012delensing, sherwin2015delensing, simard2015prospects}.
For the internal lensing reconstruction, high-resolution (high-$\ell$) observation with low instrumental noise is necessary to reduce reconstruction uncertainty. Ground-based telescopes are exceptionally well-suited for this task, as their large apertures—readily achievable in ground-based installations—enable precise measurements of small-scale CMB fluctuations. 
Although atmospheric noise typically degrades large-scale sensitivity, its impact on lensing reconstruction is limited if properly accounted for. Furthermore, foreground contamination across multiple frequency channels (27–280,GHz) can be effectively mitigated using the Needlet Internal Linear Combination (NILC) method.

Satellite-based small-aperture missions such as LiteBIRD enable full-sky measurements of large-scale CMB B-mode polarization, offering significant advantages over ground-based SATs due to their wide sky coverage and broad frequency range. The extensive sky coverage helps reduce sample variance in measuring the tensor-to-scalar ratio $r$, while the richer frequency information enhances component separation between the CMB and foregrounds. However, to fully realize the potential of such missions in constraining $r$, delensing is essential.
Ref.\cite{namikawa2024litebird} reports a 20\% reduction in the uncertainty of $r$ using a multi-tracer delensing approach, despite residual foreground contamination. This study employs mock internal lensing reconstruction based on CMB-S4 LAT data (primarily from the Southern Hemisphere) and incorporates external LSS tracers, such as the Cosmic Infrared Background (CIB), Euclid\cite{laureijs2011Euclid}, and LSST~\cite{abell2009lsst}, to enhance the mass distribution estimate and demonstrate the effectiveness of multi-tracer delensing for future full-sky surveys.

For future ultra-low-noise observations~\cite{belkner2024cmb}, internal lensing reconstruction is expected to provide more accurate lensing estimates.
In this context, achieving a full-sky internal reconstruction of the lensing potential using ground-based large-aperture telescopes (LATs) is crucial for enabling full-sky delensing and tightening constraints on the tensor-to-scalar ratio $r$. 
In this work, to emphasize the importance of full-sky internal delensing, we focus on a near-future achievable configuration, using a southern LAT (SO-like, denoted as \texttt{LATS}) as the baseline, and assess the additional contribution of a northern LAT, assumed to be situated near AliCPT\cite{Li:2017drr} (hence termed Ali-like LAT, or \texttt{LATN}), to full-sky primordial gravitational wave (PGW) detection through delensing. 
Our simulations combine polarization data from a hypothetical satellite mission (\texttt{SAT})—featuring ultra-low instrumental noise, no atmospheric contamination, and wide frequency coverage (40–402 GHz across 15 channels)—with data from both LATs. Foreground removal is performed using the NILC method. We demonstrate that, even with a near-future LAT configurations, the inclusion of \texttt{LATN}, enabling full-sky internal delensing, yields a substantial improvement in constraints on $r$—comparable to that achieved through the use of external large-scale structure tracers.


This paper is organized as follows. In Section \ref{sec: delensing_theory}, we introduce the delensing methodology along with the associated debiasing formalism. Section \ref{sec: pre_processing} outlines the complete analysis pipeline, including map simulation, foreground cleaning, map co-addition, lensing reconstruction, CMB B-mode delensing, and parameter estimation. In Section \ref{sec:results}, we present the performance of each step in the pipeline, culminating in the final parameter constraints. We conclude the paper in Section \ref{sec: conclusion}. Further discussion and clarification of the delensing methods are provided in Appendix~\ref{sec: delens_bias_ana}. Additional technical details relevant to the likelihood analysis are presented in Appendix~\ref{sec: likelihood_app}. The contribution from the LT at locations without LAT observations is discussed in Appendix~\ref{sec: joint_delensing_app}. An alternative choice of lensing estimators for mitigating extragalactic foregrounds is presented in Appendix~\ref{sec: bh_app}.

\section{Large-scale B-mode Delensing}\label{sec: delensing_theory}
As cosmic microwave background (CMB) photons travel from the last-scattering surface to Earth, they encounter large-scale structures (LSS), which induce a cumulative gravitational lensing effect. This effect displaces the primordial CMB photons, causing them to reach us from directions that deviate from their original line of sight (see \cite{lewis2006weak} for a comprehensive review). Mathematically, this process can be described as a remapping of the CMB temperature and polarization anisotropies:
\begin{equation}\label{EQ:lensedT}
    \begin{aligned}
        \tilde{\Theta}(\hat{\mathbf{n}}) &= \Theta(\hat{\mathbf{n}} + \mathbf{d}(\hat{\mathbf{n}})) 
        = \Theta(\hat{\mathbf{n}}) + \nabla_i\Theta(\hat{\mathbf{n}})\mathbf{d}^i(\hat{\mathbf{n}}) + \mathcal{O}(\phi^2),
    \end{aligned}
\end{equation}
\begin{equation}\label{EQ:lensedQU}
    \begin{aligned}
        \relax[\tilde Q\pm i\tilde U](\hat{\mathbf{n}}) &= [Q\pm iU](\hat{\mathbf{n}} + \mathbf{d}(\hat{\mathbf{n}})) 
            = [Q\pm iU](\hat{\mathbf{n}}) + \nabla_i[Q\pm iU](\hat{\mathbf{n}})\mathbf{d}^i(\hat{\mathbf{n}}) + \mathcal{O}(\phi^2),
    \end{aligned}
\end{equation}
where tildes indicate lensed quantities and \( \mathbf{d} \) is the deflection angle, given by the gradient of the lensing potential \( \mathbf{d}(\hat{\mathbf{n}}) = \nabla \phi(\hat{\mathbf{n}}) \) (here we ignore the curl modes) \cite{namikawa2014lensing}, where the lensing potential is the projection of gravitational potential along line-of-sight \cite{Okamoto:2003zw}:
\[
\phi(\hat{\mathbf{n}}) = -2 \int d\eta \frac{\chi(\eta - \eta_s)}{\chi(\eta) \chi(\eta_s)} \Psi(\chi \hat{\mathbf{n}}, \eta),
\]
where $\eta$ is the conformal time, $\eta_s$ is the epoch of last scattering and $\chi$ is the angular diameter distance in comoving coordinates.

One interesting aspect for polarization is that in harmonic space, the remapping will convert part of the E-modes into B-modes, known as lensing B-modes. At linear order in $\phi$, these are given by (neglecting the tensor B-modes) \cite{namikawa2022simons}:
\begin{equation}\label{EQ:lensed_B}
	\begin{aligned}
		{B}_{\ell m}^\text{lens} &= i \sum_{\ell' m'}\sum_{LM} \wignerThree{\ell}{\ell'}{L}{m}{m'}{M} p^{-}F^{(2)}_{\ell L \ell'} E^*_{\ell'm'}\kappa^*_{LM},
	\end{aligned}
\end{equation}
where $p^{+} (p^{-})$ is unity if $\ell+L+\ell'$ is even (odd) and zero otherwise. Lensing induced mode coupling for spin-$s$ fields can be expressed as:
\begin{equation}\label{EQ:mode_coupling}
	\begin{aligned}
	F^{(s)}_{lLl'} = \frac{2}{L(L+1)} \left[ l'(l' + 1) + L(L + 1) - l(l + 1) \right] \\
    \times \sqrt{\frac{(2l + 1)(2l' + 1)(2L + 1)}{16\pi}} \wignerThree{\ell}{\ell'}{0}{-s}{s'}{0}.
	\end{aligned}
\end{equation}

As instrumental sensitivity improves, lensing-induced B-modes have become an increasingly significant source of contamination in the observation of primordial gravitational waves (PGWs).  
Considerable attention has been devoted to mitigating lensing B-modes from observations, and various methods have been developed to achieve this goal (see, e.g., \cite{diego2020comparison,Diego-Palazuelos:2020lme,Carron:2017vfg,baleato2021limitations}). This process, known as CMB B-mode delensing, aims to reduce the lensing contribution to enhance the detectability of primordial signals. Here, we briefly describe two delensing methods used in this work.

\subsection{Delensing Method Description}\label{sec: method}

\paragraph{\textbf{Gradient-Order Template Method}} 
A straightforward approach to mitigating lensing B-modes is to construct a lensing B-mode template at gradient order, which serves as a good approximation for lensing B-modes on large scales \cite{challinor2005lensed,baleato2021limitations}. It is convenient to work in real space, drawing inspiration from Eq.~(\ref{EQ:lensedQU}):  
\begin{equation}
    \nabla_i[Q\pm iU](\hat{\mathbf{n}}) \mathbf{d}^i(\hat{\mathbf{n}}).
\end{equation}
In practice, the gradient of a field can be efficiently computed using the ladder operators \cite{Okamoto:2003zw}:
\begin{equation}\label{eq:ladder}
	\begin{gathered}
		D_i [{_s}f(\hat{\mathbf{n}})] = -\frac{1}{\sqrt{2}} \left\{ \sharp {_s}f(\hat{\mathbf{n}}) \bar m + \flat {_s}f(\hat{\mathbf{n}}) m \right\},
	\end{gathered}
\end{equation}
where the covariant derivative $\nabla_i$ operating on the spin-$s$ field piece of a tensor is equivalent to a gradient operation $D_i$ on its spin-$s$ weighted representation, $\sharp$ and $\flat$ denote the ladder operators, and the complex-conjugated vectors $\bar{m}$ and $m$ serve as the basis.  

A crucial point to note is that the $Q$ and $U$ fields used to construct the template must exclude the B-modes. Otherwise, the lensing operation would be applied again, introducing redundant terms. The resulting template can then be subtracted from the observed map (map-level delensing) or treated as a pseudo-channel in the likelihood analysis when constraining parameters (cross-spectral method delensing), as discussed in Section~\ref{sec: params}.

\paragraph{\textbf{Inverse-lensing Method}} Another method is more intuitive, by reversing the lensing effect through remapping the observed photons back to their original positions, we can achieve a more accurate and optimal approach for delensing.
The inverse deflection angle is well-defined because the points are remapped onto themselves after being deflected back and forth, as discussed in \cite{Diego-Palazuelos:2020lme}:
\begin{equation}\label{hat_n}
	\hat{\mathbf{n}} + \mathbf{d^{inv}}(\hat{\mathbf{n}}) + \mathbf{d}(\hat{\mathbf{n}} + \mathbf{d^{inv}}(\hat{\mathbf{n}})) = \hat{\mathbf{n}},
\end{equation}
where the primary CMB fields can be recovered from the lensed field by remapping the latter using the inverse deflection angle $\mathbf{d^{inv}}(\hat{\mathbf{n}})$:
\begin{equation}
	\tilde{X}(\hat{\mathbf{n}}) = X(\hat{\mathbf{n}} + \mathbf{d}(\hat{\mathbf{n}})) \Leftrightarrow X(\hat{\mathbf{n}}) = \tilde{X}(\hat{\mathbf{n}} + \mathbf{d^{inv}}(\hat{\mathbf{n}})).
\end{equation}
In practice, at low-order approximation, the inverse deflection angle is approximately the negative of the deflection angle. A more accurate estimate can be obtained by solving Eq.~(\ref{hat_n}) using a Newton-Raphson scheme, where the inverse deflection angle $\mathbf{d^{\text{inv}}}(\hat{\mathbf{n}})$ is iteratively computed, as described in \cite{Carron:2017vfg}. The lensing B-mode template is then constructed by subtracting the inverse-lensed B-mode map from the observed data.

\subsection{Bias Analysis}\label{sec: bias}
Although the delensing procedure mitigates part of the lensing effect in observations, it also introduces biases in the measurement of $r$. These biases arise from inherent correlations between the lensed CMB, noise and foreground residuals, and the lensing potential. Such correlations manifest across different statistical orders, including the two-point correlation (angular power spectrum), three-point correlation (bispectrum), and four-point correlation (trispectrum).  
Below we do not explicitly derive each analytical term. Instead, we present the analysis in a way that facilitates understanding their origins and estimating their magnitudes.

The NILC cleaned E-modes, B-modes and the estimated lensing potential with Wiener filters (for a detailed description, see Section \ref{sec: delensing}), are as follows:
\begin{equation}
	\begin{aligned}
		&E = \mathcal{W}^E(E^\text{lens}+E^\text{res,NILC}), \\
		&B = \mathcal{W}^E(B^\text{lens}+B^\text{res,NILC}), \\
		&\hat \phi = \mathcal{W}^{\phi}(\phi + \phi^\text{noise}),
    \end{aligned}
\end{equation}
where $\mathcal{W}^E$ and $\mathcal{W}^{\phi}$ are the Wiener filters applied to CMB polarization observation and reconstructed lensing potential, respectively.

\subsubsection{Bias Analysis of the Gradient-Order Template Method}

Starting from the expression above, the lensing B-mode template can be written as:
\begin{equation}\label{EQ:template}
	\begin{aligned}
    		B^\text{temp} &= \mathcal{B}^{(1)}[\mathcal{W}^E(E^\text{lens}+E^\text{res,NILC}) \ast \mathcal{W}^{\phi}(\phi + \phi^\text{noise})] \\
		  &= \begin{aligned}[t] &\mathcal{B}^{(1)}[\mathcal{W}^E E^\text{lens} \ast \mathcal{W}^{\phi}\phi] 
		  	+ \left\{ \mathcal{B}^{(1)}[\mathcal{W}^E E^\text{res,NILC} \ast \mathcal{W}^{\phi}(\phi + \phi^\text{noise})] 
		  	+ \mathcal{B}^{(1)}[\mathcal{W}^E E^\text{lens} \ast \mathcal{W}^{\phi}\phi^\text{noise}] \right\}
		  \end{aligned} \\
		  &= B^\text{temp}_S + B^\text{temp}_N,
	\end{aligned}
\end{equation}
where $\mathcal{B}^{(1)}[E\ast\phi]$ represents the operation of constructing the gradient-order template using E-mode field and $\phi$. Here, the first term represents the signal component of the template, while the noise arises from two distinct sources: (i) foreground residuals after NILC processing, which contaminate the lensing template through the reconstruction pipeline; and (ii) the spurious lensing effect on the true $E$-modes induced by the lensing reconstruction noise.

\subsubsection{Bias Analysis of the Inverse-Lensing Method}

Similarly, we start from the remapped B-modes:
\begin{equation}\label{EQ:remap}
	\begin{aligned}
    		B^\text{del} &= \mathcal{B} [\mathcal{W}^E (B^\text{lens} + B^\text{res,NILC}) \star \mathbf{d^{inv}} ] \\
				& \approx  \mathcal{B} [\mathcal{W}^E (B^\text{lens} + B^\text{res,NILC}) \star (\mathbf{d^{inv}_\phi} + \mathbf{d^{inv}_\text{noise}}) ] \\
				& \approx \left\{ \mathcal{B} [\mathcal{W}^E B^\text{lens} \star \mathbf{d^{inv}_\phi} ] \right\} 
				+ \left\{ [ \mathcal{B} [\mathcal{W}^E B^\text{lens} \star \mathbf{d^{inv}_\text{noise}} ] - \mathcal{W}^E B^\text{lens}] + \mathcal{B} [\mathcal{W}^E  B^\text{res,NILC} \star \mathbf{d^{inv}}  ]  + B^\text{de,hi} \right\} \\
				&= B^\text{del}_S + B^\text{del}_{N}, 
	\end{aligned}
\end{equation}
where $\mathcal{B} [B\star \mathbf{d^{inv}}]$ represents the inverse
remapping operation of lensed B mode with $\mathbf{d^{inv}}$ to obtain
the delensed B-modes. Notice that we have applied a linear approximation when moving from the second row to the third, and all higher-order terms neglected in this step are absorbed into $B^{\mathrm{de,hi}}$. This should be distinguished from the higher-order lensing contributions contained in the term $\mathcal{B}[\, B \star \mathbf{d}^{\mathrm{inv}} ]$.
The lensing B-mode template:
\begin{equation}\label{EQ:remap_template}
\begin{aligned}
B^\text{temp} &= \mathcal{W}^E B^{obs} - B^\text{del} \\
&= \mathcal{W}^E[B^\text{lens} + B^\text{res,NILC}] - [B^\text{del}_S + B^\text{del}_{N}] \\
&\approx [\mathcal{W}^E B^\text{lens} - B^\text{del}_S] + [\mathcal{W}^E B^\text{res,NILC} - B^\text{del}_N] \\
&= [\mathcal{W}^E B^\text{lens} - B^\text{del}_S] 
    + \left\{
      \begin{aligned}
        &\mathcal{W}^E B^\text{res,NILC}
          -\mathcal{B}(\mathcal{W}^E B^\text{res,NILC}\star\mathbf{d^{inv}}) \\
        &\quad + [\mathcal{W}^E B^\text{lens}
          -\mathcal{B}(\mathcal{W}^E B^\text{lens}\star\mathbf{d^{inv}_\text{noise} })] -B^\text{de,hi}
      \end{aligned}
      \right\} \\
&= B^\text{temp}_S + B_N^\text{temp},
\end{aligned}
\end{equation}
where $B_N^{\mathrm{temp}}$ corresponds to $B^{\mathrm{temp}}_N$ in Eq.~\ref{EQ:template}. 
For both the Gradient-Order Template Method and the Inverse-Lensing Method, we find that the noise terms arise from similar physical origins, although the latter includes additional higher-order contributions through $B^{\mathrm{de,hi}}$ beyond the gradient approximation. 
A comparison of these terms for the two methods is shown in Figure~\ref{fig:delens_terms_map}.
Overall, the two approaches exhibit consistent behavior \textbf{on large angular scales}.

\subsubsection{Bias Analysis Formula on power spectral level}
With the unified notation for the terms in Eq.~\ref{EQ:template} and 
Eq.~\ref{EQ:remap_template}, we proceed to compute the power spectra of the LT 
and the observed maps. These spectra form the data vector used in the likelihood 
analysis to obtain constraints on $r$.
The auto-power spectrum of the lensing template, given by \( B^\text{temp} = B^\text{temp}_S + B_N^\text{temp} \), is:
\begin{equation}\label{EQ:auto_temp}
	\begin{aligned}
        C^\text{temp} &= \langle B^\text{temp}_S B^\text{temp}_S \rangle + \langle B_N^\text{temp} B_N^\text{temp} \rangle + 2\langle B^\text{temp}_S B_N^\text{temp} \rangle.
	\end{aligned}
\end{equation}
For the cross-power spectrum with the observed B-modes, \( B^\text{obs} = B^\text{lens} + B^\text{res,NILC} \), we have:
\begin{equation}\label{EQ:cross_temp}
	\begin{aligned}
        C^{\text{cross}} &= \langle B^\text{lens} B_S^\text{temp} \rangle + \langle B^\text{lens} B_N^\text{temp} \rangle + \langle B^\text{res,NILC} B^\text{temp} \rangle.
	\end{aligned}
\end{equation}
We present a detailed discussion of the comparison and origin of these terms in Appendix~\ref{sec: delens_bias_ana}. 
The cross-bias terms discussed above are expected to be consistent with zero when either large-scale polarization modes are removed from the internal lensing reconstruction or an external lensing proxy is used. 
In contrast, $\langle B_N^{\mathrm{temp}} B_N^{\mathrm{temp}} \rangle$ is found to be of the same order as the CMB lensing $B$-mode signal. 
Hence, this auto-power term represents the dominant bias that must be carefully accounted for to avoid introducing significant bias in the likelihood analysis.

\section{Step-by-Step Delensing Process: Techniques and Implementation}\label{sec: pre_processing}

In this section, we present the complete delensing pipeline implemented in our simulation-based analysis. The workflow consists of the following key stages:
We begin with an overview of the simulated datasets used throughout this work (Section \ref{sec: sim}), followed by a description of the component separation procedure implemented via NILC (Section \ref{sec: nilc}). We then detail the implementation of lensing reconstruction—both internal and external (Section \ref{sec: rec}). These reconstructions are then used to construct large-scale CMB lensing B-mode templates (Section \ref{sec: delensing}). Finally, we perform a likelihood analysis using Markov Chain Monte Carlo (MCMC)
methods to derive constraints on cosmological parameters (Section \ref{sec: likelihood}).
A schematic representation of the full workflow is provided in Fig.~\ref{fig:pipeline}, highlighting the key steps and their interdependencies.

\subsection{Data Simulation}\label{sec: sim}
We employ simulated data to evaluate the feasibility of the delensing pipeline. Specifically, we simulate future CMB observations from various experiments, including a large-aperture, ground-based $6$-m telescope in the Northern Hemisphere (\texttt{LATN}), a large-aperture, ground-based $6$-m telescope in the Southern Hemisphere (\texttt{LATS}), and a small-aperture $1$-m satellite telescope (\texttt{SAT}). We then assess the effectiveness of different delensing methods in reducing the uncertainty in $r$ as shown in Section \ref{sec: params}.

\begin{figure}
\centering
\includegraphics[width=1\textwidth]{./fig/pipeline_revise.pdf}
\caption{The flowchart of the entire analysis pipeline is presented for clarity. The workflow is broadly divided into five main components: component separation using NILC, map combination, lensing reconstruction, construction of the lensing $B$-mode template, and parameter constraint estimation. While the aim is not to detail every operational step, the flowchart highlights the logical structure and data dependencies throughout the process. For reference, the resulting reduction of uncertainties in $r$ are also summarized alongside the corresponding configurations. Note that the arrow from $\kappa$ to the lensing template (LT) construction indicates a computational pathway only, and does not imply that the lensing B-modes are physically induced by the convergence field.}
\label{fig:pipeline}
\end{figure}

\begin{figure}
\centering
\includegraphics[width=1\textwidth]{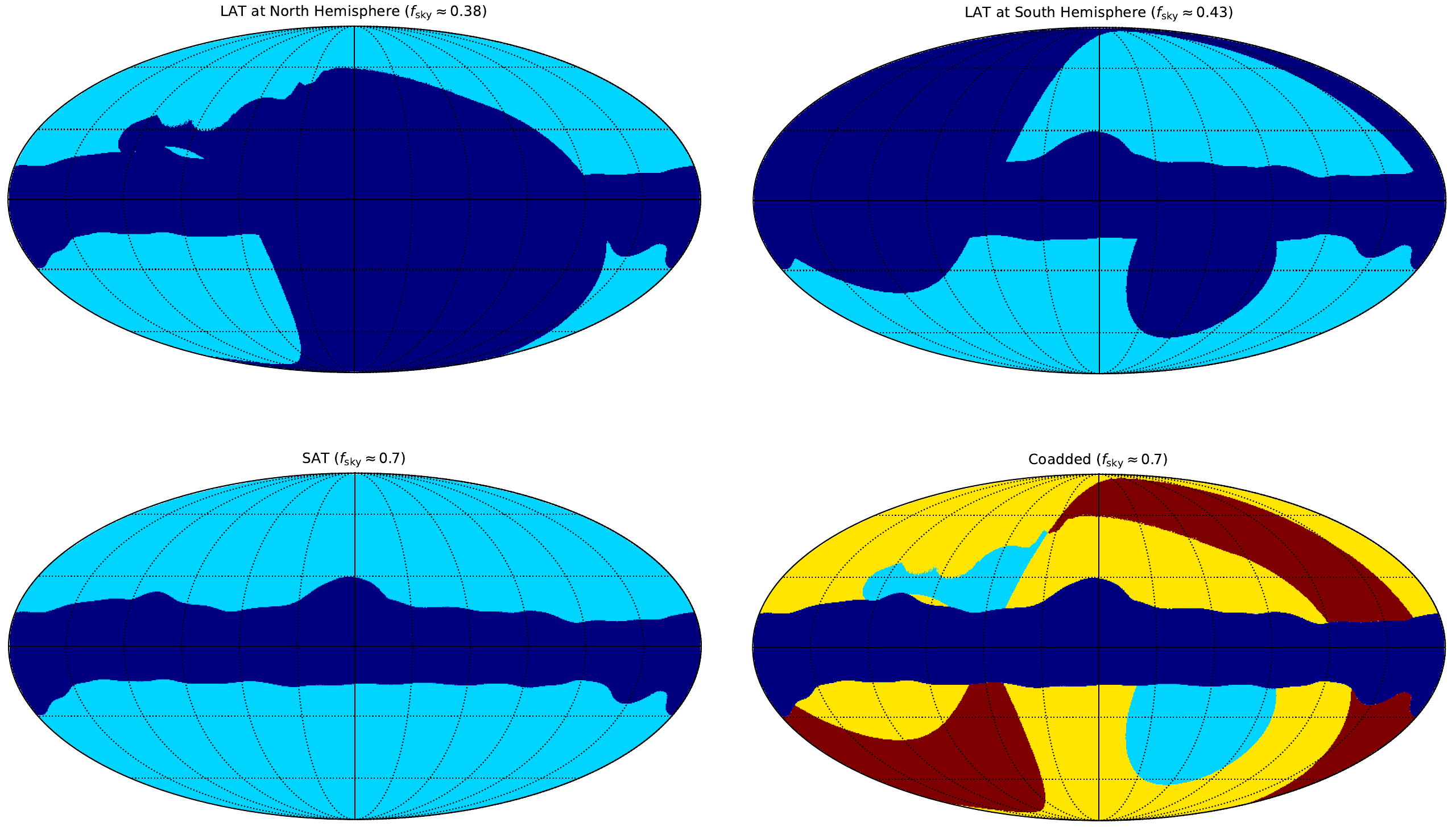}
\caption{Survey windows for each observation in Celestial coordinates used in the analysis. For the coadded case, we combine two ground-based LATs with the satellite SAT. The yellow region represents the overlap between SAT and one of the LATs, while the brown region indicates the common region covered by all three observations.}
\label{fig:mask_plot}
\end{figure}

The survey windows for these observations are shown in Fig.~\ref{fig:mask_plot}. The two ground-based LATs have sky coverages of approximately 38\% and 43\% after masking the Galactic disk, respectively, with an overlap of about 20\% (depicted in brown in the bottom-right panel of Fig.~\ref{fig:mask_plot}). The satellite \texttt{SAT} covers 70\% of the sky after masking the Galactic disk, with an overlap of approximately 61\% with the two LATs, represented by the combined brown and yellow regions in the bottom-right panel of Fig.~\ref{fig:mask_plot}

We include both Galactic and extragalactic foregrounds in our simulations. For the Galactic foregrounds, we use \texttt{PySM3} \cite{zonca2021python,thorne2017python}, which generates full-sky simulations of Galactic emissions in both intensity and polarization based on realistic observational data (e.g., Planck and WMAP). 
The modeled diffuse Galactic emissions include thermal dust (\texttt{d9} model), synchrotron (\texttt{s4} model), spinning dust (\texttt{a1} model), and free-free emission (\texttt{f1} model). 
Since these foreground templates are derived from actual observations, only a single realization is available for each component.

For extragalactic foregrounds—including tSZ, kSZ, the CIB, and radio point
sources—these components are intrinsically correlated with the lensing
potential because they trace the same underlying matter distribution. In
addition, they exhibit strong non-Gaussianity on small angular scales, primarily due to the discrete and clustered nature of the underlying astrophysical sources, with additional contributions from the non-linear evolution of large-scale structure. Both the correlation and the non-Gaussianity introduce biases and increase the variance in CMB lensing
reconstruction, and numerous works have proposed strategies to mitigate these
effects \cite{osborne2014extragalactic,sailer2023foreground,schaan2019foreground,
sailer2020lower,maccrann2024atacama,lizancos2025halo}. These
foreground-induced systematics propagate into the reconstructed lensing map
and, consequently, into the delensing procedure, ultimately leading to
increased bias and uncertainty in the inferred constraint on $r$
\cite{baleato2022impact}. 
Several methods have been proposed to mitigate extragalactic contamination in CMB lensing reconstruction. The standard internal linear combination (ILC) method can leave point-like residuals, such as from point sources or tSZ, but known foregrounds can be deprojected at the cost of increased map variance \cite{kusiak2023enhancing}. Modified lensing estimators, including bias-hardened  estimators \cite{namikawa2014bias,osborne2014extragalactic,sailer2020lower}, shear estimators \cite{qu2023cmb,schaan2019foreground} and symmetric multi-frequency–cleaned estimators~\cite{darwish2023optimizing,madhavacheril2018mitigating,darwish2021atacama}, further reduce foreground biases, albeit with some loss in signal-to-noise. Since techniques for mitigating biases from extragalactic foregrounds in lensing reconstruction are already well established, we do not address this issue in detail in the present work. An alternative foreground-mitigation approach is briefly discussed in Appendix~\ref{sec: bh_app}.

To fully capture the influence of extragalactic foregrounds, multiple realizations are required in the analysis pipeline. Realistic correlated mock data are typically generated using N-body simulations, which model dark matter halos through the gravitational interactions of a large number of particles. These halos form the scaffolding for galaxies and other cosmic structures. Major projects such as the Millennium Run \cite{springel2005simulations}, Bolshoi \cite{klypin2011dark}, and the Outer Rim Simulation \cite{heitmann2019outer} have simulated the evolution of large-scale structures and have been widely used to generate realistic extragalactic foregrounds, including the Sunyaev-Zeldovich (SZ) effect, the Cosmic Infrared Background (CIB), and galaxy distributions \cite{sehgal2010simulations,stein2020websky,omori2024agora}. 

Although N-body simulations can capture realistic correlations and non-Gaussian features, they are computationally expensive, and therefore only a limited number of microwave-sky realizations are currently available (e.g., \cite{sehgal2010simulations,stein2020websky,omori2024agora}). Recently, \cite{han2021deep} presented 500 high-resolution, full-sky millimeter-wave deep-learning (DL) simulations (``mmDL'' simulations\footnote{\url{https://portal.nersc.gov/project/cmb/data/generic/mmDL/}}). These simulations are based on the ``S10'' simulation \cite{sehgal2010simulations} and employ multiple neural-network architectures, including Generative Adversarial Networks (GANs \cite{goodfellow2020generative}), to generate new realizations that preserve the statistical properties of the original “S10” data.

In this work, we use the extragalactic foreground realizations provided by "mmDL" simulation. The simulation are provided at 30 GHz, 90 GHz, 148 GHz, 219 GHz, 277 GHz and 350GHz, we therefore interpolate the maps to the frequencies we need. Note that these foregrounds are only used for mocking LAT temperature observation since they are largely unpolarized under current sensitivity, except for the polarized point sources which have been proved to be sensitive to the PGW detection \cite{remazeilles2018exploring,tucci2005limits,tucci2012impact}. Besides, as described in \cite{han2021deep}, the flux of point sources at 148GHz has been truncated to 7 mJy, above which can be safely removed by point-source removal procedure under current and future sensitivity. Since the influence of polarized point sources is expected to be subdominant for the LAT configuration considered in this work \cite{sailer2023foreground}, we leave a more detailed assessment to future work.

The input unlensed CMB maps are Gaussian realizations generated from a specific power spectrum obtained using the Boltzmann code \texttt{CAMB} \cite{lewis2011camb}, based on the Planck 2018 best-fit cosmological parameters \cite{aghanim2020planck} with a tensor-to-scalar ratio $r = 0$. The lensing potential maps are converted from the convergence maps given by "mmDL" simulation. We then utilize the \texttt{Lenspyx} \cite{carron2020lenspyx,reinecke2023improved} package, which implements an algorithm to distort the primordial signal based on the lensing deflection ($d_{LM}=\sqrt{L (L+1)} \phi_{LM}$) generated above.

\begin{figure}[htbp]
    \centering
    \subfigure[LAT temperature]{
        \includegraphics[width=0.47\linewidth]{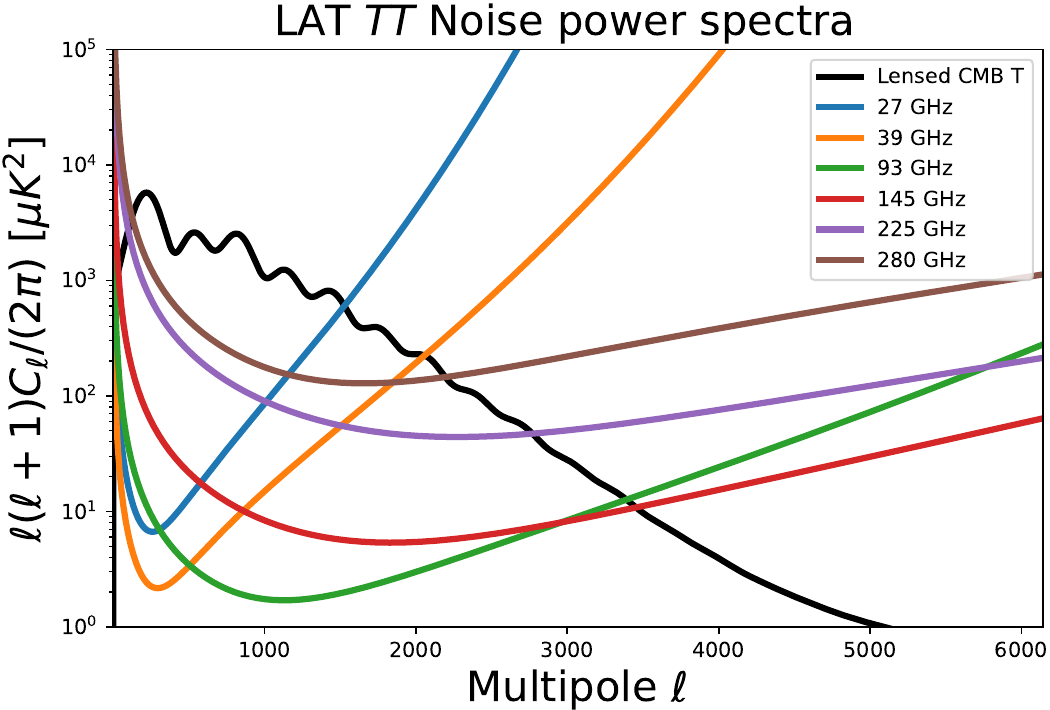}
    }
    \hspace{0.01\linewidth} 
    \subfigure[LAT polarization]{
        \includegraphics[width=0.47\linewidth]{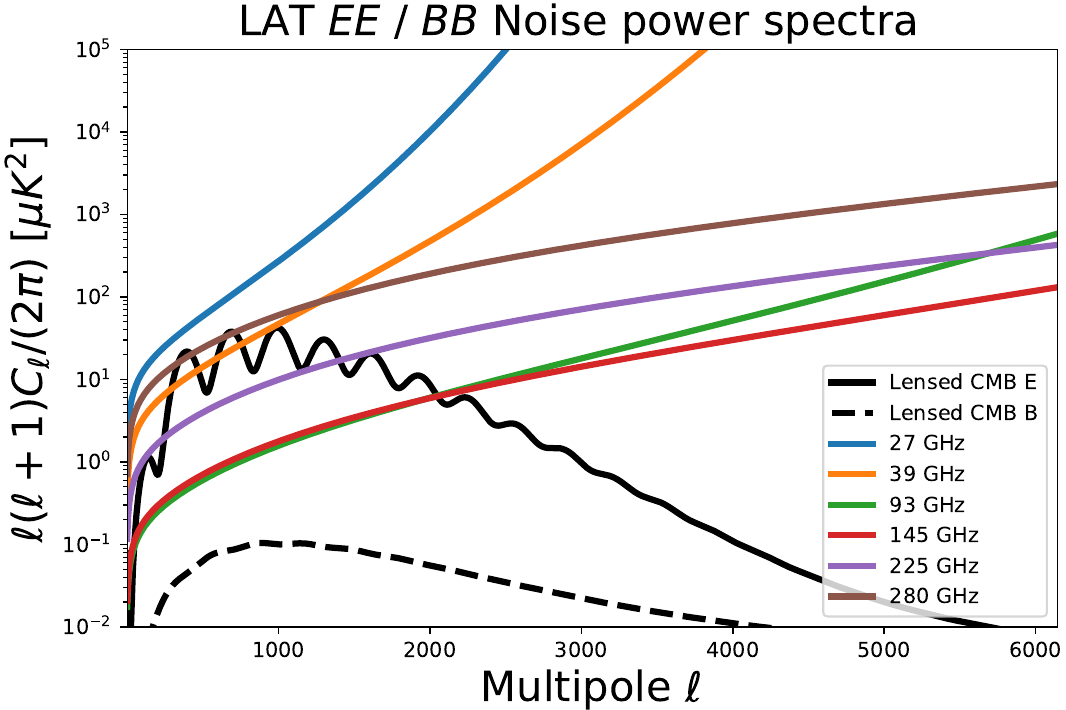}
    }
    \caption{The total noise power spectra (beam-corrected) for the ground-based LATs.}
    \label{fig:noise_lat}
\end{figure}

For the noise simulation of ground-based LATs, we assume that both telescopes share the same experimental configuration and observational conditions, differing only in their survey windows. We adopt the noise model described in \cite{ade2019simons}, which accounts for both detector white noise and a $1/f$ component. The total noise power spectrum is given by:
\begin{equation}\label{eq:comb}
    N_{\ell} = N_\text{red} \left(\frac{\ell}{\ell_\text{knee}}\right)^{\alpha_\text{knee}} + N_\text{white},
\end{equation}
where $N_\text{white}$ represents the white noise component, while $N_\text{red}$, $\ell_\text{knee}$, and $\alpha_\text{knee}$ characterize the contribution from $1/f$ noise. The total noise maps for LATs are generated as Gaussian realizations from the total noise power spectrum (with $\ell_\text{max} = 6143$), using the parameters listed in Table~\ref{tab:exp_params_lat}. The beam-corrected noise power spectra of the LATs are shown in Fig.~\ref{fig:noise_lat}.

\begin{table} 
    \caption{The experimental parameters of the LAT experiments. $\sigma_\text{white}$ give white noise levels for temperature, with polarization noise $\sqrt{2}$ higher as both $Q$ and $U$ Stokes parameters. Both the LATN and LATS configurations adopt the "goal" specifications define for SO surveys~\cite{ade2019simons}.}
    \label{tab:exp_params_lat}
    \renewcommand{\arraystretch}{1.0} 
    \makebox[\textwidth]{ 
    \begin{tabularx}{1.1\textwidth}{l *{1}{>{\centering\arraybackslash}X} *{1}{>{\centering\arraybackslash}X} *{1}{>{\centering\arraybackslash}X} *{3}{>{\centering\arraybackslash}X} *{3}{>{\centering\arraybackslash}X}} 
        \hline
        \quad & \multicolumn{1}{c}{} & \multicolumn{1}{c}{}  & \multicolumn{3}{c}{LAT Temperature} & \multicolumn{3}{c}{LAT Polarization} \\
        \cmidrule(lr){4-6} \cmidrule(lr){7-9}
        Freq. [GHz] & FWHM [arcmin]  & $\sigma_\text{white}$ [$\mu$K-arcmin]  & $N_\text{red} \cdot t$ [$\mu$K$^{2}$-s] & $\ell_\text{knee}$ &$\alpha_\text{knee}$ & $N_\text{red}$ [$\mu$K$^{2}$] & $\ell_\text{knee}$ &$\alpha_\text{knee}$ \\
        \hline
        27  & 7.4   & 52   & 100  & 1000  & -3.5 & $N_\text{white,pol.}$ & 700  & -1.4\\
        39  & 5.1   & 27   & 39   & 1000  & -3.5 & $N_\text{white,pol.}$ & 700  & -1.4\\
        93  & 2.2   & 5.8  & 230  & 1000  & -3.5 & $N_\text{white,pol.}$ & 700  & -1.4\\
        145 & 1.4   & 6.3  & 1500 & 1000  & -3.5 & $N_\text{white,pol.}$ & 700  & -1.4\\
        225 & 1.0   & 15   & 17000& 1000  & -3.5 & $N_\text{white,pol.}$ & 700  & -1.4\\
        280 & 0.9   & 37   & 31000& 1000  & -3.5 & $N_\text{white,pol.}$ & 700  & -1.4\\
        \hline
    \end{tabularx}
    }
\end{table}

\begin{table}[htbp]
\centering
\caption{Parameters of the SAT experiment. For the simplicity of simulation, the values shown are approximate estimates based on the science goals of LiteBIRD; for an updated and precise description of the instrumentation, see \cite{hazumi2020litebird,litebird2023probing}.}
\label{tab:exp_params_sat}
\begin{tabular}{lcc}
\hline
Freq. [GHz] & FWHM [arcmin] & \makecell{Noise level [$\mu$K-arcmin]} \\
\hline
40  & 70.5 & 26.5 \\
50  & 58.5 & 23.7 \\
60  & 51.1 & 15.1 \\
68  & 41.6 & 14.1 \\
78  & 36.9 & 11.0 \\
89  & 33.0 & 8.7 \\
100 & 30.2 & 7.3  \\
119 & 26.3 & 5.4  \\
140 & 23.7 & 5.1  \\
166 & 28.9 & 3.9  \\
195 & 28.0 & 5.0  \\
235 & 24.7 & 7.6  \\
280 & 22.5 & 9.8  \\
337 & 20.9 & 15.5 \\
402 & 17.9 & 33.6 \\
\hline
\end{tabular}
\end{table}

The noise for the SAT is homogeneous, generated through Gaussian random sampling on the map domain with noise levels specified in Table \ref{tab:exp_params_sat}.

Finally, the observed maps are derived by summing the lensed CMB maps, foreground maps, and noise maps.

\subsection{Needlet Internal Linear Combination}\label{sec: nilc}
Component separation using the ILC (Internal Linear Combination) method \cite{basak2012needlet} has been widely applied in CMB observations. In this work, we perform component separation using the NILC (Needlet Internal Linear Combination) method independently for each experiment. The expansion of sky maps in spherical needlets allows for the localization of statistical properties in both pixel space and harmonic space, thereby enhancing the ability to account for variations in foreground intensity across the sky and the dominance of instrumental noise at small scales.
Under the assumption that each component of the observed maps is independent, the observed field can be expressed as
\begin{equation}
    X_{\ell m}^{\mathrm{obs}, c} = b_\ell^c X_{\ell m}^{\mathrm{CMB}} + b_\ell^c X_{\ell m}^{\mathrm{FG}, c} + X_{\ell m}^{\mathrm{N}, c},
\end{equation}
where $X \in \{ \Theta, E, B \}$ represents the harmonic coefficient of a field, $b_\ell^c$ is the beam function, and $c$ denotes the frequency channel.
We then perform a filter $h_\ell^{j}$ in harmonic space to extract observation at the very scale $j$,
\begin{equation}
    X_{\ell m}^{c, j} = h_\ell^{j} X_{\ell m}^{c},
\end{equation}
this corresponds to expanding the observed harmonic coefficient with spherical needlets at scale $j$, pixel $k$:
\begin{equation}
    \beta_{jk}^{X, c} = \beta_{jk}^{\mathrm{CMB}} + \beta_{jk}^{\mathrm{FG}, c} + \beta_{jk}^{\mathrm{N}, c},
\end{equation}
with corresponding spherical needlets given by:
\begin{equation}
    \begin{aligned}
        \beta_{jk}^{\mathrm{CMB}} &= \sqrt{\lambda_j} \sum_{\ell m} h_\ell^j b_\ell X_{\ell m}^{\mathrm{CMB}} Y_{\ell m}(\xi_j), \\
        \beta_{jk}^{\mathrm{FG}, c} &= \sqrt{\lambda_j} \sum_{\ell m} h_\ell^j b_\ell^c X_{\ell m}^{\mathrm{FG}, c} Y_{\ell m}(\xi_j), \\
        \beta_{jk}^{\mathrm{N}, c} &= \sqrt{\lambda_j} \sum_{\ell m} h_\ell^j \frac{b_\ell}{b_\ell^c} X_{\ell m}^{\mathrm{N}, c} Y_{\ell m}(\xi_j).
    \end{aligned}
\end{equation}

Since all the observations actually measure signal $bX^{\mathrm{CMB}}$ with some error $bX^{\mathrm{FG}} + X^{\mathrm{N}}$, consists in averaging all these measurements giving a specific weight $w_i$ to each of them, we naturally questing for a solution of the form in needlet space:
\begin{equation}
    \beta_{jk}^{\mathrm{NILC}} = \sum_{c=1}^{n_c} w_{jk}^c \beta_{jk}^{X, c},
\end{equation}
the weight $w_{jk}^c$ is determined by minimizing the variance of the reconstructed CMB, subject to the unbiasedness constraint ($\sum_{c=1}^{n_c} w_{jk}^c=1$). Using the method of Lagrange multipliers to enforce the constraint, the solution is given by:
\begin{equation}
    \hat{\omega}_{jk}^c = 
    \frac{
        \sum_{c'} \left[\hat{\mathbf{R}}_{jk}^{-1}\right]^{cc'} a^{c'}
    }{
        \sum_c \sum_{c'} a^c \left[\hat{\mathbf{R}}_{jk}^{-1}\right]^{cc'} a^{c'}
    },
\end{equation}
where $\left[\hat{\mathbf{R}}_{jk}^{-1}\right]^{cc'}$ is the inverse of the cross frequency covariance matrix:
\begin{equation}
        \hat{\mathbf{R}}_{jk}^{cc'} = \frac{1}{n_k} \sum_{k'} \omega_j(k,k') \beta_{jk}^{c}\beta_{jk}^{c'},
\end{equation}
where $\omega_j(k,k')$ selects the domain of $n_k$ pixels around the $k$-th pixel over which we perform our averaging to estimate the covariance.
Finally, the cleaned CMB map is given by:
\begin{equation}
        \hat X_\text{NILC}(\hat{\mathbf{n}}) = \sum_{\ell m}\sum_{jk} \beta_{jk}^{\mathrm{NILC}} \sqrt{\frac{4\pi}{\lambda_j}} h_{\ell}^j Y_{\ell m}(\xi_j)Y_{\ell m}(\hat{\mathbf{n}}).
\end{equation}
The harmonic transformations mentioned therein can be executed using \texttt{healpy} / \texttt{HEALPix} package \footnote{\url{https://healpix.sourceforge.io/}} \cite{2005ApJ...622..759G,Zonca2019}. A well-tested Python package to implement the NILC pipeline, is available at \texttt{openilc}\footnote{\url{https://github.com/dreamthreebs/openilc}}.

For the two ground-based LATs, we use beam-deconvolved maps at six frequencies as inputs, as detailed in Table~\ref{tab:exp_params_lat}. For the satellite-based SAT, maps at fifteen frequencies are employed, as outlined in Table~\ref{tab:exp_params_sat}. Component separation is performed independently for the observed $\Theta$, $E$, and $B$ maps. Additionally, E-to-B leakage is corrected during the transformation from Stokes parameters $Q$ and $U$ to $E$ and $B$ using the "E-mode recycling" method introduced by \cite{Liu:2022beb}. The final outputs consist of cleaned $\Theta$, $E$, and $B$ maps for each experiment, denoted as $X^{\text{NILC,obs},i}$, where $i$ represents the respective experiment.  

Furthermore, we linearly combine the corresponding noise and foreground maps (i.e., nulling the signal in the observed maps) for each experiment using the weights obtained above. The resulting output maps, $X^{\text{NILC,null},i}$, are used exclusively for inverse-weight calculations in harmonic space for map combination and for estimating the effective white noise level (see Section \ref{sec: rec}).

\subsection{Map combination}\label{sec: map_coadd}
We combine the NILC-cleaned maps from each experiment using inverse-variance weighting in harmonic space. The inverse-variance combination is given by:
\begin{equation}\label{eq:inv_comb}
    \begin{aligned}
        &\chi_{\ell m} = \sum_i \omega_{i, \ell} \chi_{i, \ell m}, \\
        &\omega_{i, \ell} = \frac{N_{i, \ell}^{-1}}{\sum_i N_{i, \ell}^{-1}}, \\
    \end{aligned}
\end{equation}
where $\omega_{i, \ell}$ represents the weight of the $i$-th experiment, $N_{i, \ell}$ is the power spectrum of the noise and residual foregrounds from the $i$-th experiment, denoted as $X^{\text{NILC,null},i}$, $\chi_{i, \ell m}$ corresponds to the harmonic coefficients of the fields $\Theta, E, B$ for the $i$-th experiment after foreground cleaning, and $\chi_{\ell m}$ represents the combined harmonic coefficients.

\begin{figure}[htbp]
    \centering
    \subfigure[Observation regions of two example experiments.]{
        \includegraphics[width=0.45\linewidth]{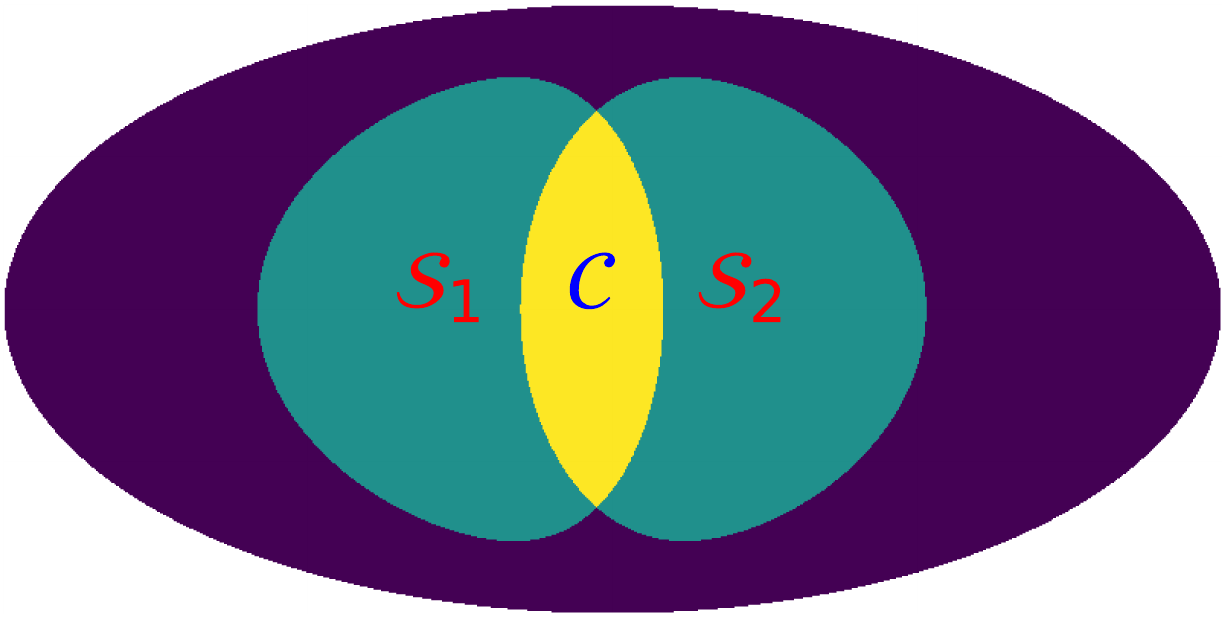}
    }
    \hspace{0.05\linewidth} 
    \subfigure[Multipole range coverage of two example experiments.]{
        \includegraphics[width=0.45\linewidth]{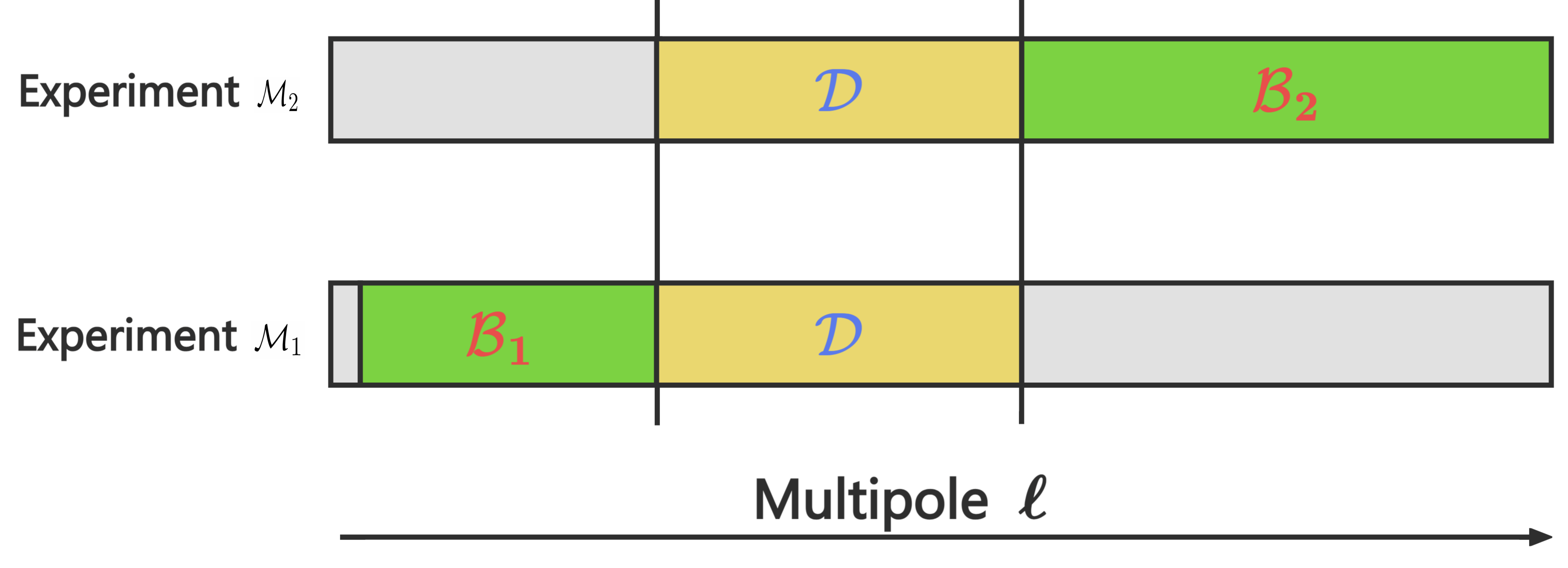}
    }
    \caption{Illustration of the sky coverage and multipole range for two example experiments used in the map combination. Panel (a) shows the sky regions observed by each experiment, while panel (b) represents the corresponding multipole range coverage.}
    \label{fig:example_exps}
\end{figure}

It is important to note that inverse-variance weighting can only be applied to fields that cover the same sky regions and share the same multipole ranges; otherwise, they will be directly mapped onto a null sky map.  
For instance, consider two observed maps from different experiments,  $\mathcal{M}_1$ and $\mathcal{M}_2$, as illustrated in Fig.~\ref{fig:example_exps}. Due to differences in experimental configurations—such as aperture size, location, atmospheric conditions, and scan strategy—their respective sky coverages differ. We denote the exclusive sky regions of these experiments as $\mathcal{S}_1$ and $\mathcal{S}_2$, while their overlapping region is denoted as $\mathcal{C}$.  
Similarly, their multipole coverage also differs. We define these ranges as $\mathcal{L}_1 = \mathcal{B}_1 + \mathcal{D}$ and $\mathcal{L}_2 = \mathcal{B}_2 + \mathcal{D}$, where $\mathcal{B}_1$ and $\mathcal{B}_2$ represent the experiment-specific multipole ranges, and $\mathcal{D}$ denotes the common multipole range.

The combination strategy follows these steps:  
\begin{enumerate}
    \item \textbf{Identify the sky regions}: First, determine the overlapping region ($\mathcal{C}$) and the exclusive regions for each experiment ($\mathcal{S}_1$ and $\mathcal{S}_2$) using the corresponding masks shown in Fig.~\ref{fig:mask_plot}. These masks have been apodized using the "Smooth" function (effectively a Gaussian window) implemented in \texttt{NaMaster}\cite{alonso2023namaster}, with an apodization scale of $6'$ to reduce artifacts caused by sharp mask edges.  
    
    \item \textbf{Identify the multipole ranges}: Next, determine the common multipole range ($\mathcal{D}$) and the experiment-specific multipole ranges ($\mathcal{B}_1$ and $\mathcal{B}_2$). 
    
    \item \textbf{Decompose information by region and scale}: The maps corresponding to each region and multipole range are extracted using a spherical harmonic transform, with the masks obtained in the previous step applied. This results in the following harmonic coefficients:  
    \[
    a_{\mathcal{B}_1+\mathcal{D}}^{\mathcal{S}_1}, \quad b_{\mathcal{B}_2+\mathcal{D}}^{\mathcal{S}_2}, \quad a_{\mathcal{D}}^{\mathcal{C}}, \quad b_{\mathcal{D}}^{\mathcal{C}}, \quad a_{\mathcal{B}_1}^{\mathcal{C}}, \quad b_{\mathcal{B}_2}^{\mathcal{C}}.
    \]
    Here, $m_{\mathcal{L}}^{\mathcal{S}}$ represents the harmonic coefficients of experiment $m$ ($a$ for $\mathcal{M}_1$, $b$ for $\mathcal{M}_2$) in region $\mathcal{S}$ and at multipole range $\mathcal{L}$.

    \item \textbf{Combine the maps in the pixel domain, region by region}:  
        \begin{enumerate}
            \item For region $\mathcal{S}_1$ and $\mathcal{S}_2$, the output is simply $a_{\mathcal{B}_1+\mathcal{D}}^{\mathcal{S}_1}$ and $b_{\mathcal{B}_2+\mathcal{D}}^{\mathcal{S}_2}$, respectively. It is worth noting that these cases require no spherical harmonic transformation in practice; a simple reprojection at the map level suffices.
 
            \item For the common region $\mathcal{C}$, perform a weighted combination as given by Eq.~(\ref{eq:inv_comb}) on $a_{\mathcal{D}}^{\mathcal{C}}$ and $b_{\mathcal{D}}^{\mathcal{C}}$. The final output for this region is:  
                \[
                \omega_{1} a_{\mathcal{D}}^{\mathcal{C}} + \omega_{2} b_{\mathcal{D}}^{\mathcal{C}} + a_{\mathcal{B}_1}^{\mathcal{C}} + b_{\mathcal{B}_2}^{\mathcal{C}}.
                \]
        \end{enumerate}
\end{enumerate}

As a result, the combined map is inverse-variance weighted only in the overlapping regions at the common scales, while preserving the remaining information. This approach is reasonable, as it fully utilizes the available information for weighted combination—i.e., no additional information is accessible for weighting in the exclusive regions or at the private scales.  

We emphasize that our analysis employs a diagonal inverse-variance weighting in harmonic space for map combination, together with the Wiener filtering applied in harmonic space in the LT construction (Section \ref{sec: delensing}). Taken together, these two steps yield a sub-optimal performance for cut-sky analyses compared to a more sophisticated methodology described in \cite{eriksen2004power}, which may be regarded as a more complete Wiener-filtered treatment of the observations.
This decision is based on two considerations. First, we have tested the alternative method and found it to be computationally demanding. Second, we do not expect a significant improvement in delensing efficiency from this adjustment. As described by \cite{manzotti2017cmb}, the key factor determining delensing efficiency is the accuracy of the lensing estimate, rather than contamination in the CMB maps undergoing delensing.

In this paper, we first combine the NILC-cleaned maps from the two LATs (hereafter denoted as \texttt{LATN+LATS}), both of which cover a multipole range of $(0, 6143)$, over the sky regions shown in Fig.~\ref{fig:mask_plot}. The outputs are the combined $\Theta$, $E$, and $B$ maps.  
These maps serve two purposes. First, they are used for internal lensing reconstruction. Second, they are further combined with the NILC-cleaned maps from the SAT (hereafter denoted as \texttt{LATN+LATS+SAT}), where the combined LAT covers the multipole range $(0, 6143)$, and the SAT covers $(0, 300)$. The final combined maps are then utilized for lensing B-mode template construction and parameter constraints. 
For comparative analysis, we additionally generate a \texttt{LATS+SAT} case by applying identical combination procedures to the NILC-cleaned maps from the LAT South (LATS) and SAT instruments.

\begin{figure}[htbp]
    \centering
    \makebox[\textwidth][c]{
        \subfigure{
            \includegraphics[width=0.5\linewidth]{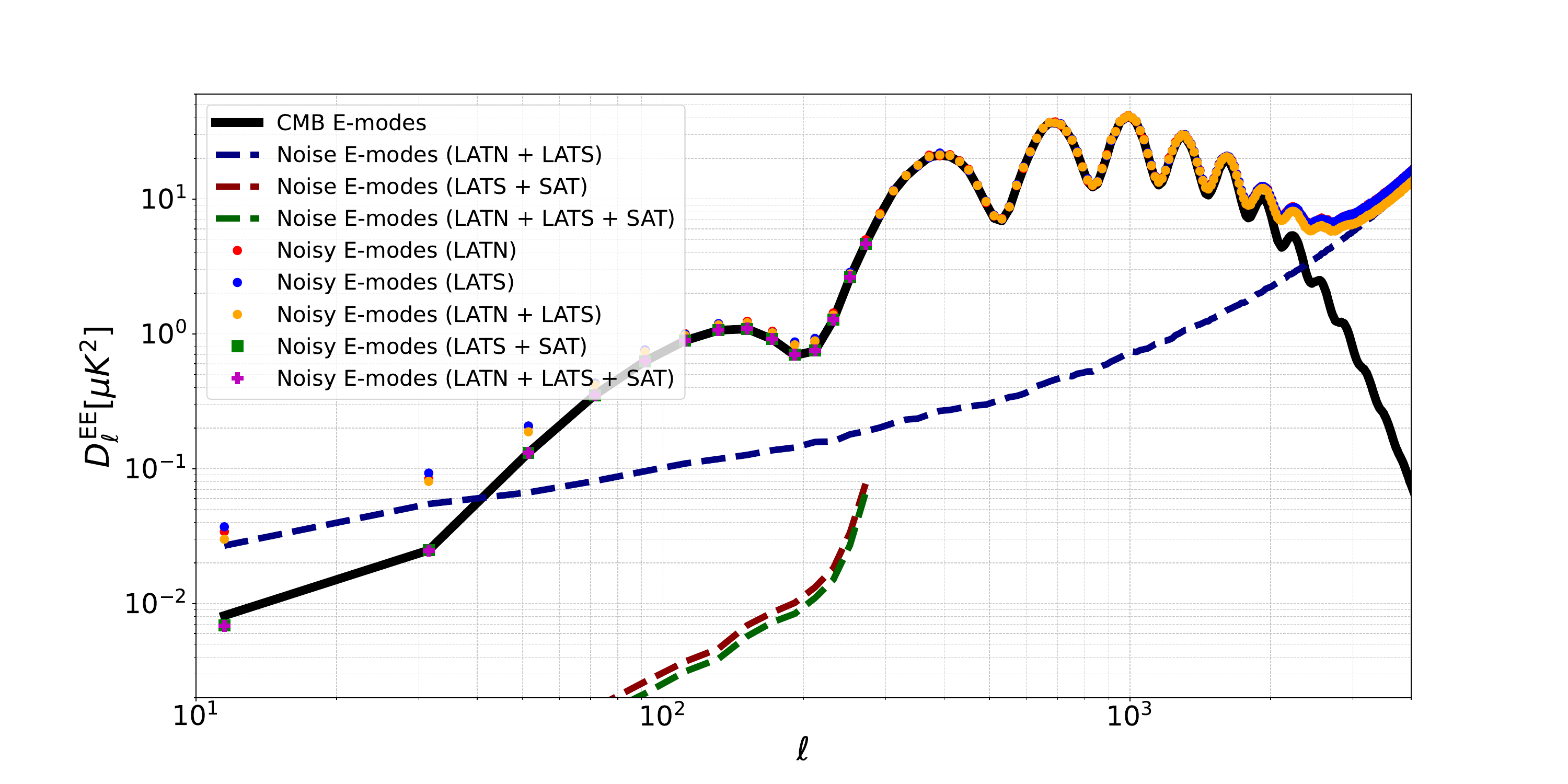}
        }
        \hfill
        \subfigure{
            \includegraphics[width=0.5\linewidth]{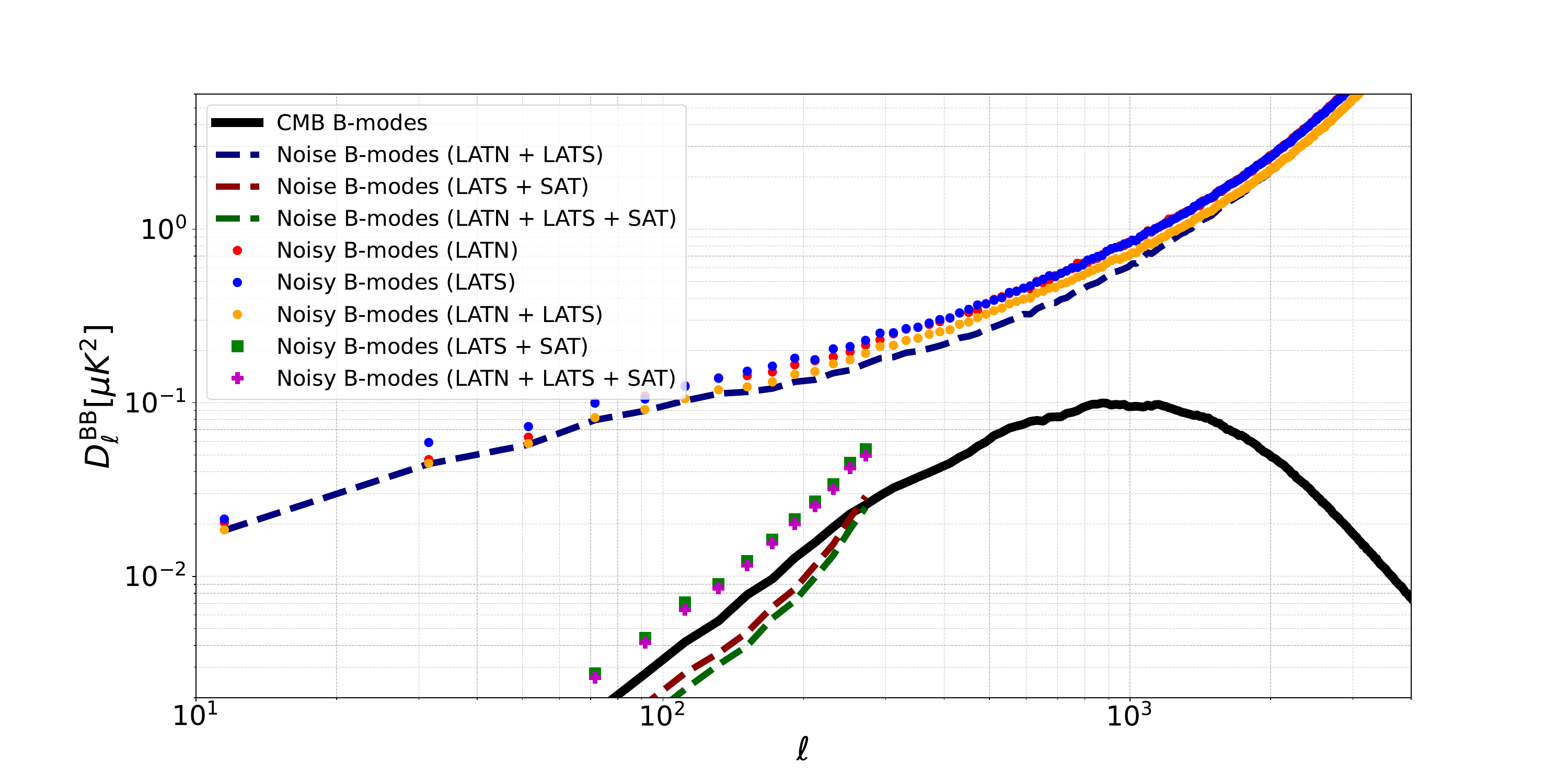}
        }
    }
    \caption{The power spectra of the selected polarization maps, both before and after map coaddition, are shown. The black solid lines represent the theoretical predictions. The 'Noise' labels indicate contributions from both instrumental noise residuals and foreground residuals after NILC cleaning, while the 'Noisy' labels refer to the NILC-cleaned maps. Labels with multiple experiments in parentheses correspond to the coadded maps, as described in Section~\ref{sec: map_coadd}. Note that for cases involving \texttt{SAT}, data are shown only for $\ell < 280$, where the \texttt{SAT} contribution dominates. Additionally, the \text{LATN} and \text{LATS} do not fully overlap due to differences in their sky coverage, which is relevant for the Galactic foregrounds.}
    \label{fig: power_coadd}
\end{figure}

The power spectra of selected polarization maps are shown in Fig.~\ref{fig: power_coadd}. It is evident that incorporating data from LATN (red dots) and LATS (blue dots) leads to a significant reduction in noise (see yellow dots), owing to the \textbf{20\%} sky overlap between the two experiments. Building on this, we further combine the data with SAT (see magenta plus signs). As expected, SAT contributes most of the signal at large scales ($\ell<300$), which is crucial for constraining $r$. 
It should be noted that for analyses incorporating SAT data (specifically the \texttt{LATS+SAT} and \texttt{LATN+LATS+SAT} combinations), we restrict the displayed data to multipoles $\ell < 280$. Beyond this threshold, as the SAT contribution becomes negligible, the power spectra are expected to align with those derived from the \texttt{LATS} and \texttt{LATN+LATS} datasets respectively.

\subsection{Lensing Reconstruction}\label{sec: rec}
\subsubsection{Internal Lensing Reconstruction}\label{sec: rec_in}
We use the combined maps from the two ground-based LAT observations (\texttt{LATN+LATS}) to perform lensing reconstruction using the Quadratic Estimator (QE) method, following the implementation described in \cite{carron2017maximum,aghanim2020planck}. To evaluate the improvement gained by incorporating \texttt{LATN} data, we also reconstruct the lensing potential using only the \texttt{LATS} maps.  

For polarization reconstruction, we use a multipole range of $(200, 5200)$, excluding multipoles below 200 to mitigate biases introduced in the delensing procedure. These biases arise from the overlap between the B-modes used for reconstructing $\phi$ and the B-modes intended for delensing (i.e., the observed B-modes), as noted by \cite{lizancos2021impact,teng2011cosmic}.  
For temperature reconstruction, we restrict the analysis to multipoles in the range $(200, 5200)$ to relieve contamination from large-scale atmospheric noise and residual small-scale extragalactic foreground contributions, as discussed in \cite{van2014cmb,baleato2022impact}.

We implement the reconstruction using the Python package \texttt{plancklens}\cite{aghanim2020planck}. The spin-1 real-space (unnormalized) lensing deflection QE is given by:
\begin{equation}
    {_{1}}\hat{d}(\hat{\mathbf{n}}) = - \sum_{s=0,\pm2} {_{-s}}\bar{X}(\hat{\mathbf{n}}) \big[\sharp_s X^{\text{WF}}\big](\hat{\mathbf{n}}).
\end{equation}
Here, $X = \{ \Theta,E,B \}$ from observation, $\sharp$ denotes the spin-raising operator, which is associated with the gradient of a field \cite{Okamoto:2003zw}, and the subscript $s$ represents the spin of the fields.  
The estimator takes as input the gradient of the Wiener-filtered maps, $_sX^{\text{WF}}(\hat{\mathbf{n}})$, and the real-space inverse-variance filtered maps, $_s\bar{X}(\hat{\mathbf{n}})$.

For filtering, we apply the optimal filtering method described in \cite{mirmelstein2019optimal}, implemented using the multi-grid preconditioned conjugate gradient algorithm. The iteration tolerance level $\epsilon$ is set to $10^{-5}$ for both temperature filtering and polarization filtering, which is considered sufficiently precise for accurate lensing reconstruction.
The root-mean-square (RMS) map required by this method is chosen to be an isotropic noise RMS map (we do not incorporate the hit-count variations in this work), with effective noise levels given in Table~\ref{tab:lat_nlev}. 
These noise levels are approximately determined from the power spectra of the null signal simulation, $X^{\text{NILC,null},i}$, for each experiment, thereby accounting for contributions from both instrumental noise and foreground residuals:
\begin{equation}\label{eq:nlev}
    \hat{\mathbf{n}}_\text{lev} = \sqrt{\sum_{\ell} \frac{2\ell+1}{4\pi}\hat{C}_{\ell}^\text{noise} \cdot \Omega_\text{pix} }\frac{60 \cdot 180}{\pi} [\mu \text{K-arcmin}],
\end{equation}  
where the summation range is identical to that used in the lensing reconstruction,
$\hat{C}_{\ell}^\mathrm{noise}$ denotes the power spectrum of the null-signal simulation $X^{\mathrm{NILC,null},i}$, and
$\Omega{\mathrm{pix}} = \frac{4\pi}{12 \ast \mathrm{NSIDE}^2}$ is the pixel area of a \texttt{HEALPix} map.
Additionally, we apply an apodized mask to reduce the creation of spurious gradients at the mask boundary \cite{namikawa2013bias} with an apodization scale of $6'$, which has been shown to significantly accelerate the convergence of the iterative filtering process, particularly for temperature filtering.

\begin{table}[htbp]
\centering
\caption{The effective white noise level used for lensing reconstruction filtering includes contributions from both instrumental noise and foreground residuals. As expected, in the overlapping regions, both the temperature and the polarization noise decreases by nearly $1/\sqrt{2}$ due to the multiple realizations of both instrumental noise and foreground.
}
\label{tab:lat_nlev}
\begin{tabular}{lccc}
\hline
nlev ($\mu$K-arcmin) & LATN exclusive regions & LATS exclusive regions & Overlapping regions\\
\hline
Temperature  & 20.0 & 18.6 & 13.6 \\
Polarization  & 8.0 & 8.0  & 5.7\\
\hline
\end{tabular}
\end{table}

Using the following relation, we derive a raw estimator for the lensing potential:
\begin{equation}
    _{\pm1}\hat d (\hat{\mathbf{n}}) \equiv \left(\sum_{LM} \frac{\hat g_{LM} \pm i\hat c_{LM}}{\sqrt{L(L+1)}}\right) {_{\pm1}Y_{LM}(\hat{\mathbf{n}})},
\end{equation}
where the gradient mode $\hat g_{LM}$ is the quantity of interest, while the curl mode $\hat c_{LM}$ is expected to be negligible.

In practice, non-ideal factors such as the survey boundary and inhomogeneous noise introduce biases in the reconstruction. In this paper, we estimate the mean-field biases by averaging over simulation realizations and subsequently subtracting the estimated bias from the raw QEs. 
The final estimator is given by:
\begin{equation}\label{eq:qe_norm}
    \hat \phi_{LM} = \frac{1}{\mathcal{R}_L^{\phi}} \left(\hat g_{LM} - \langle \hat g_{LM}\rangle_{MC} \right),
\end{equation}
where the gradient mode $\hat g_{LM}$ is obtained from the quadratic terms of the inverse-filtered and Wiener-filtered fields. The response function $\mathcal{R}_L^{\phi}$ serves as a normalization factor and must be computed accordingly. We apply an approximate normalization at the map level, calculated analytically following \cite{Okamoto:2003zw}. Although this semi-analytic normalization is only accurate within $1\%-2\%$, it is sufficient for the purpose of map-level delensing.

\subsubsection{External Lensing Reconstruction}\label{sec: rec_ex}
Under current CMB experimental conditions, internal lensing reconstruction typically suffers from a low signal-to-noise ratio (S/N) on small angular scales, primarily due to the dominance of reconstruction noise (notably the $N^0$ bias). To overcome this limitation, the incorporation of external large-scale structure (LSS) tracers—such as the Cosmic Infrared Background (CIB) and galaxy number density—has emerged as a promising strategy for enhancing the S/N of lensing reconstruction, especially at small scales. When properly filtered, these tracers can effectively complement internal delensing by providing additional information on the lensing potential at high multipoles \cite{smith2012delensing,sherwin2015delensing,simard2015prospects}.

Although the dominant contribution to large-scale lensed B-modes originates from lensing at intermediate angular scales ($L \sim \mathcal{O}(100)$; see Fig.2 of \cite{simard2015prospects}), numerous studies have demonstrated that the inclusion of external tracers can nonetheless lead to a significant improvement in delensing efficiency \cite{yu2017multitracer,namikawa2024litebird,manzotti2018future,baleato2022delensing,hertig2024simons}.

To quantify the improvement in delensing performance achieved by incorporating large-scale structure (LSS) tracer data, we combine the internal lensing reconstruction from \texttt{LATS} and \texttt{LATN+LATS} with simulated external tracers—specifically the GNILC (Generalized NILC) CIB map from \emph{Planck} PR2 \cite{remazeilles2011foreground,aghanim2016planck} and galaxy number density measurements from the \emph{Euclid} survey \cite{laureijs2011Euclid}. These combinations are hereafter referred to as \texttt{LATS+SAT+External} and \texttt{LATN+LATS+SAT+External}, respectively.

The simulation of large-scale structure (LSS) tracer maps is based on theoretical models for their power spectra, which primarily comprise two components: (1) the underlying matter distribution, incorporating appropriate tracer biases, and (2) a noise term that accounts for experimental uncertainties, mainly arising from shot noise.
We adopt the same configuration as used in our previous work \cite{chen2025delensing_lss}.
A detailed description of the LSS models and the simulation setup can be found therein. For further background and reviews, see also \cite{toffolatti2014planck,hall2010angular,namikawa2024litebird,namikawa2022simons,amendola2018cosmology,abell2009lsst,yu2017multitracer,laureijs2011Euclid}.

For the CIB, we use only the simulated maps at 353 GHz, where the GNILC CIB map has been shown to exhibit the highest correlation with CMB lensing \cite{yu2017multitracer}. Moreover, given the high signal-to-noise ratio (S/N) of the CIB auto-power spectrum over the relevant angular scales, the contribution of instrumental noise is negligible. Consequently, we omit instrumental noise in our theoretical modeling of the CIB. We assume the CIB fully overlaps with the sky coverage of the SAT's configuration in this work, and the large-scale CIB ($L<100$) are excluded  to avoid the contamination from Galactic foreground residuals \cite{namikawa2024litebird}.

For galaxy number density, we adopt the configuration of the \emph{Euclid} survey, which achieves an effective galaxy number density of approximately 30 arcmin$^{-2}$ \cite{laureijs2011Euclid}. To fully exploit the redshift information provided by \emph{Euclid}, we divide the galaxy sample into five tomographic redshift bins, each serving as a tracer of the underlying matter distribution.
It is worth noting that we assume the \emph{Euclid} survey fully overlaps with the sky coverage of the SAT's configuration. In practice, some deviation is expected due to the differing scan strategies and scientific objectives of the two surveys. However, thanks to the forthcoming fourth- and fifth-generation galaxy surveys---including DESI~\cite{abareshi2022overview}, LSST~\cite{ivezic2019lsst}, the Nancy Grace Roman Space Telescope~\cite{spergel2013wide}, Euclid~\cite{mellier2024euclid}, CSST~\cite{gong2025introduction}, MUST~\cite{zhao2024multiplexed}, the MegaMapper telescope~\cite{schlegel2022megamapper}, the 11\,m Maunakea Spectroscopic Explorer (MSE)~\cite{asai2024exploring}, the 12\,m Wide-field Spectroscopic Telescope (WST)~\cite{sheinis2023maunakea}, and the $\sim$12\,m Extremely Large Spectroscopic Survey Telescope (ESST)~\cite{su2024optical}---we expect that combining these surveys will provide a deep and wide galaxy sample suitable for realistic data processing. A more detailed investigation is left to future work.

The tracers are combined through a linear combination of $I_i$:
\begin{equation}
    I = \sum_i c^i I_i,
\end{equation}
where $I_i$ denotes the MV (minimum-variance) quadratic estimator (Eq.~(\ref{eq:qe_norm})), the CIB, and the galaxy number density from \emph{Euclid}. The coefficients $c^i$ are determined by maximizing the correlation coefficient between the combined tracer and the true lensing convergence $\kappa$; the explicit form of $c^i$ can be found in \cite{sherwin2015delensing}. It is important to note that the resulting combined tracer is no longer an unbiased estimator of the true convergence—instead, it effectively corresponds to a Wiener-filtered version. This approach is optimal for delensing, as it maximizes the overall delensing efficiency.\footnote{Here, "optimal" does not account for iterative lensing reconstruction \cite{carron2017maximum,belkner2024cmb}, and refers only to the combination of the quadratic estimator (QE) and LSS tracers.}

\subsection{Lensing B-mode Template Construction}\label{sec: delensing}
\subsubsection{Constructing LT on map-level}
In this paper, we construct the lensing B-mode template using two delensing methods, as discussed in Section \ref{sec: delensing_theory}. To assess the improvement gained by incorporating \texttt{LATN} data and LSS tracers, we construct the lensing $B$-mode template (LT) separately for four cases: \texttt{LATS+SAT}, \texttt{LATS+SAT+External}, \texttt{LATN+LATS+SAT}, and \texttt{LATN+LATS+SAT+External}. 
In the internal delensing cases (\texttt{LATS+SAT} and \texttt{LATN+LATS+SAT}), the LT is constructed only within the sky region overlapping between the ground-based \texttt{LATs} and the \texttt{SAT}, since the lensing field is not available outside this area. In the joint delensing cases (\texttt{LATS+SAT+External} and \texttt{LATN+LATS+SAT+External}), the LT can in principle be constructed over the full sky; however, for a consistent and fair baseline comparison, we restrict the analysis to the same overlapping sky region used in the internal delensing scenarios. Results that additionally include the LT outside the LAT--SAT overlapping area are presented in the Appendix \ref{sec: joint_delensing_app}.

The polarized maps for all cases are therefore masked with apodized masks to exclude sky regions beyond the LAT coverage and mitigate complications arising from sharp mask edges during the delensing procedure. 
This exclusion effectively prevents issues related to missing information at $\ell > 300$ in these regions, as discussed in Section~\ref{sec: map_coadd}.
The reconstructed lensing potential used for delensing is selected over the multipole range $20 < L < 5000$, and is Wiener-filtered as described in Section~\ref{sec: rec_results}.

Before performing the delensing operation, we first apply a Wiener filter to the combined observed QU maps to enhance the signal-to-noise ratio (S/N), given by:
\begin{equation}
	\begin{gathered}
		Q_{\ell m} \Rightarrow \frac{C_{\ell}^{EE}}{C_{\ell}^{EE} + N_{\ell}^{EE}} Q_{\ell m}, \\
		U_{\ell m} \Rightarrow \frac{C_{\ell}^{EE}}{C_{\ell}^{EE} + N_{\ell}^{EE}} U_{\ell m}, \\
	\end{gathered}
\end{equation}
where the combined E-mode noise power spectrum is given by the weighted combination of each experiments:
\begin{equation}
	N_{\ell}^{EE} = \sum_i \omega_{i, \ell}^2 N_{i, \ell}^{EE}.
\end{equation}

Regarding the Gradient-order lensing template method, we proceed by converting the spin-2 fields $Q + iU$ to spin-1 and spin-3 fields, and transform the spin-0 lensing potential field to spin-$-1$ and spin-1 fields using \texttt{CMBlensplus} \cite{namikawa2021CMBlensplus}. 
Following this conversion, we compute the gradient lensing template according to Section \ref{sec: delensing_theory}. The resulting QU represents the lensing effect on E-mode, from which the lensing B-mode can be separated using harmonic transformation and EB leakage  is corrected with an "E-mode Recycling" method described in \cite{Liu:2022beb}.

Regarding the Inverse-lensing method, we proceed by estimating the inverse deflection angle $ \mathbf{d^{inv}} $ from the filtered lensing potential map $\hat \phi^\text{MV}$ and from combined tracer. This is done by iteratively solving Eq.(\ref{hat_n}) using \texttt{CMBlensplus}. Subsequently, we remap the filtered observed QU maps to obtain the delensed QU maps. These are then subtracted from the observed QU maps to derive the noisy lensing template QU maps. The lensing template B map can be separated through harmonic transformation from QU fields, and EB leakage should also be corrected. 

\subsubsection{Bandpower calculation and bias estimation}\label{sec: cal_bp}
We compute the auto- and cross-bandpowers of lensing B-mode template (LT) and foreground-cleaned observed B-mode ($C_{b}^\text{LT}$ and $C_{b}^{\text{LT} \times \text{obs}}$) using the \texttt{NaMaster}\cite{alonso2023namaster} code. 
The multipole range for likelihood analysis (Section \ref{sec: likelihood}) spans from $\ell_{\text{min}} = 2$ to $\ell_{\text{max}} = 200$, with a binning width of $\Delta\ell = 40$.

As discussed in Sec.~\ref{sec: delensing}, the lensing template (LT) is defined only over the LAT–SAT overlap region, whereas the observed B-mode map covers the full survey footprint after applying the Galactic mask. For the auto-spectra, we apply apodized masks independently to each field (with a 0.5° apodization for the LT and 0.2° for the observed B modes, generated using the \texttt{NaMaster} \texttt{Smooth} function).
In contrast, when computing the cross-spectrum between the LT and the observed B modes, the mask must correspond to the intersection of their sky coverages. Only this common region contains valid LT information, and therefore only this region provides a physically meaningful contribution to the cross-spectrum. The resulting bandpowers are then included in the data vector (Eq.~\ref{eq: data_vec}) used for the likelihood analysis.

As for the corresponding bias terms analyzed in Section \ref{sec: bias}, we input the required CMB maps and the potential maps to the same pipeline as described above to simulate all the bias terms, their auto- and cross-bandpowers are then also calculated using the \texttt{NaMaster} code. 
An alternative method is to replace the reconstructed $\phi$ with the true $\phi$ in the baseline, which will cause some of the cross terms in the results to vanish. This allows us to isolate these terms by taking the difference between the baseline simulation results and these specific simulation results. We found that these two debiasing methods have similar effects, allowing us to choose either one flexibly. 
The average over 500 simulations of these terms serves as an estimate of the bias terms in the likelihood.

Note that the resulting lensing B-mode template (LT) is a filtered version of the lensed B-mode during to the suppression on signal from Wiener filter. So after debiasing, the auto- and cross-bandpowers are both filtered version and we then further compute the transfer function by signal-only simulation to compensate for the suppression on power-spectrum level. The transfer function for $C_{b}^\text{LT}$ is calculated as the ratio of the average power of the lensing B-mode template from signal-only simulation to the average lensed BB power. Besides, the transfer function for $C_{b}^{\text{LT} \times \text{obs}}$ is similarly calculated by replacing the numerator with the cross-spectrum between the constructed signal-only template and the signal lensed B-modes. We use these transfer functions to rescale $C_{b}^\text{LT}$ and $C_{b}^{\text{LT} \times \text{obs}}$ when fitting parameters.

\subsection{Likelihood analysis and \texorpdfstring{$r$}{r} constraint}\label{sec: likelihood}
We investigate the parameter space using \texttt{Cobaya} \cite{Torrado_2021,torrado2020cobaya}, which employs a Markov Chain Monte Carlo (MCMC) approach.  
To improve constraints on the tensor-to-scalar ratio \( r \), rather than subtracting the lensing B-mode template (LT) from the observation (map-level delensing), we treat it as an additional pseudo-channel and compute the auto- and cross-bandpowers to construct the theoretical model (cross-spectral method delensing). As noted by \cite{hertig2024simons}, both methods yield comparable uncertainties. Our map channels consist of \( B^\text{NILC,obs} \) and \( B^\text{LT} \).  
The dataset vector is then defined as  
\begin{equation}\label{eq: data_vec}
    \mathbf{\hat{X}}_{b} = \left[ C_{b}^{\text{NILC},\text{obs}}, C_{b}^{\text{LT} \times \text{obs}}, C_{b}^{\text{LT}} \right],
\end{equation}
which includes two auto-bandpowers and one cross-bandpower. We parameterize each element of \( \mathbf{\hat{X}}_{b} \) using a two-parameter model:
\begin{align}
    &C_{b}^{\text{NILC,obs}} = rC_{b}^\text{tens}|_{r=1} + A_LC_{b}^\text{lens} 
     + N_{b}^{BB,\text{NILC}}, \\
    &C_{b}^{\text{LT} \times \text{obs}} = {A_L}C_{b}^\text{lens}, \\
    &C_{b}^\text{LT} = A_LC_{b}^\text{lens} + N_{b}^\text{temp},
\end{align}
where the subscript $b$ denotes the bandpower bin, the parameter \( A_L \) scales the intensity of the lensing B-mode. \( N_{b}^{BB,\text{NILC}} \) represents the NILC residual power spectrum obtained from the null signal simulation \( B^\text{NILC,null} \) (see Section~\ref{sec: nilc}). And $N_{b}^\text{temp}$ is the bias term as we discussed in Section \ref{sec: bias}.

Considering that our analysis includes the reionization $B$-modes at low multipoles (where only a small number of modes are available), we adopt the Hamimeche--Lewis (HL) likelihood \cite{hamimeche2008likelihood} in the baseline likelihood analysis. The HL likelihood effectively captures the non-Gaussian distribution of bandpowers in this low-$\ell$ regime and thus provides a more accurate treatment than a simple Gaussian approximation:

\begin{equation}
	\begin{aligned}
        -2 \ln \mathcal{L}(C_{b} | \hat{C}_{b}) = \frac{2b + 1}{2} \mathrm{Tr} \left[ \left( \mathbf{C}_{f_{b}}^{-1/2} \mathbf{C}_{g_{b}} \mathbf{C}_{f_{b}}^{-1/2} \right)^2 \right]  
        = \mathbf{X}_{g_{b}}^T \mathbf{M_{b}}^{-1} \mathbf{X}_{g_{b}},
	\end{aligned}
\end{equation}
where $\mathbf{X}_{g_{b}}=\text{vecp}(C_{g_{b}})$ is the vector of distinct elements of the transformed power spectra $C_{gb}$. For a comprehensive review, we refer the reader to reference \cite{hamimeche2008likelihood}. A discussion about the alternative choice of the likelihood function is given in Appendix \ref{sec: likelihood_choice}.

The covariance matrix $\mathbf{M_b}$ used in the likelihood analysis is derived from a baseline of 500 simulations, accounting for the effects of masking, cosmic variance, noise, foreground residuals, and other factors. The details of the simulation, foreground cleaning, and map combination procedures for the mock maps are provided in Section~\ref{sec: pre_processing}.  
Delensing is applied to these simulations to generate the lensing B-mode templates. Subsequently, all power spectra contained in $\mathbf{\hat X}_{b}$ are computed for each simulation using the \texttt{NaMaster} code, as described in Section~\ref{sec: delensing}. The covariance matrix is then estimated from these 500 realizations of $\mathbf{\hat X}_{b}$, with its elements given by:
\begin{equation}
    \mathbf{M}_{b,ij} = \langle \mathbf{\hat{X}}_{b,i} \mathbf{\hat{X}}_{b,j} \rangle - \langle \mathbf{\hat{X}}_{b,i} \rangle \langle \mathbf{\hat{X}}_{b,j} \rangle,
\end{equation}
where the average is taken over simulations, and the subscripts $i,j$ denote the elements of $\mathbf{\hat X}_{b}$. 

Since the covariance is estimated from a finite number of simulations, appropriate corrections are required to ensure an accurate likelihood analysis, as implemented in many previous works \cite{bianchini2025cmb,collaborations2018bicep2,ade2014measurement,abazajian2022cmb,adachi2020measurement,balkenhol2022parameter,choi2020atacama,dutcher2021measurements}.
We first apply the Hartlap–Anderson factor \cite{hartlap2007your} to the inverse covariance matrix, $\mathbf{M}_{b}^{-1}$, to correct for the bias introduced by Monte-Carlo noise in the covariance estimate.
In addition, constraining noisy elements of the covariance matrix (i.e., by conditioning the covariance estimate) is essential to reduce errors arising from the finite number of realizations.
Our choice of covariance-conditioning scheme is described in Appendix \ref{sec: likelihood_condition}.

\section{Delensing Pipeline Performance}\label{sec:results}
We now report the principal results from our delensing framework, organized into three key aspects: first, the reconstruction fidelity of the lensing potential quantified through both internal and external estimators; second, the effectiveness of lensing template generation in capturing B-mode contamination; and third, the cosmological parameter constraints derived from the delensed spectra. These results collectively establish the pipeline's capability to reduce lensing-induced variance while preserving primordial signal, ultimately enabling tighter limits on $r$ compared to raw (non-delensed) measurements.

\subsection{Results of lensing reconstruction}\label{sec: rec_results}
\begin{figure}
	\includegraphics[width=\columnwidth]{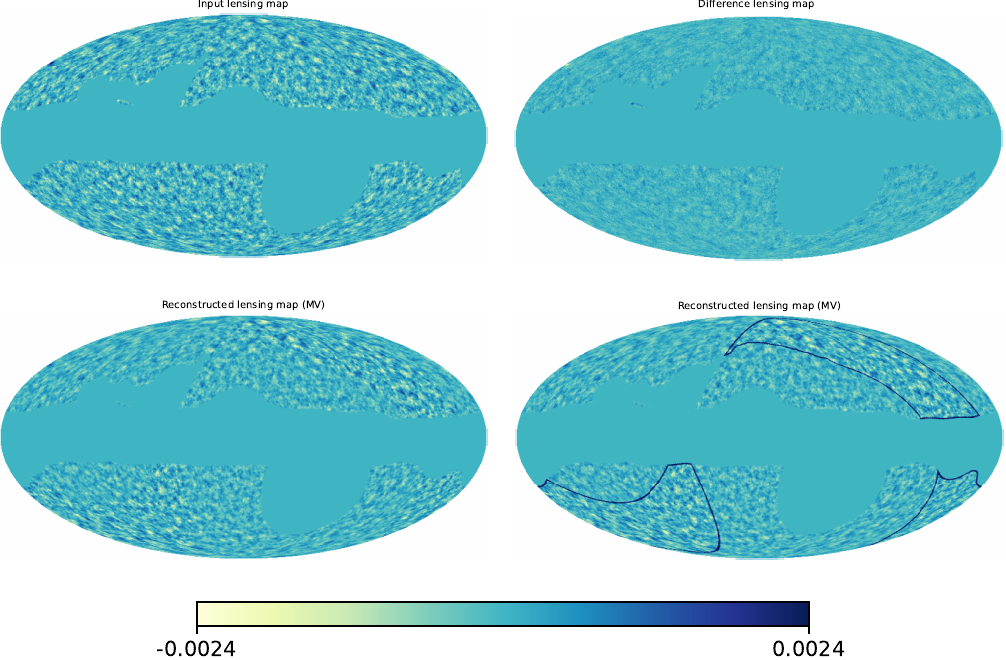}
	\caption{Reconstructed lensing map for the \texttt{LATN+LATS} case. 
The upper-left panel shows the input lensing map, and the lower-left panel shows the reconstructed map using the MV quadratic estimator (the overlap region between \texttt{LATN} and \texttt{LATS} is outlined in the lower-right panel). 
The upper-right panel shows the difference between the two maps. 
To better highlight the lensing structure, we plot the Wiener-filtered deflection-angle amplitude 
\( 
\hat{\alpha}^{\mathrm{WF}}_{LM}
=
\sqrt{L(L+1)}\, \hat{\phi}^{\mathrm{MV}}_{LM}
\,\frac{C^{\phi\phi,\mathrm{fid}}_L}{C^{\phi\phi,\mathrm{fid}}_L + N^{(0),\mathrm{ana}}_L}
\).
A clear improvement in signal-to-noise ratio is observed in the overlap region between \texttt{LATN} and \texttt{LATS}.
}
	\label{fig:phi_rec}
\end{figure}

The final internal reconstructed lensing potential, \(\hat{\phi}^\text{MV}_{LM}\), is shown in Figure \ref{fig:phi_rec}. To highlight the lensing structures, we plot the Wiener-filtered deflection angle amplitude, defined as 
\[
\hat{\alpha}^\text{WF}_{LM} = \sqrt{L(L+1)} \frac{ C^{\phi\phi,\text{fid}}_L}{C^{\phi\phi,\text{fid}}_L + N^{(0),\text{ana}}_L}\hat{\phi}^\text{MV}_{LM}.
\]
The left panel presents the input data, while the middle panel shows the reconstructed deflection angle. The right panel displays their difference. It is evident that the reconstructed deflection successfully captures most of the features of the input data, and the homogeneous distribution of brightness and darkness in the difference map confirms the effectiveness of the reconstruction.

\begin{figure}
	\includegraphics[width=\columnwidth]{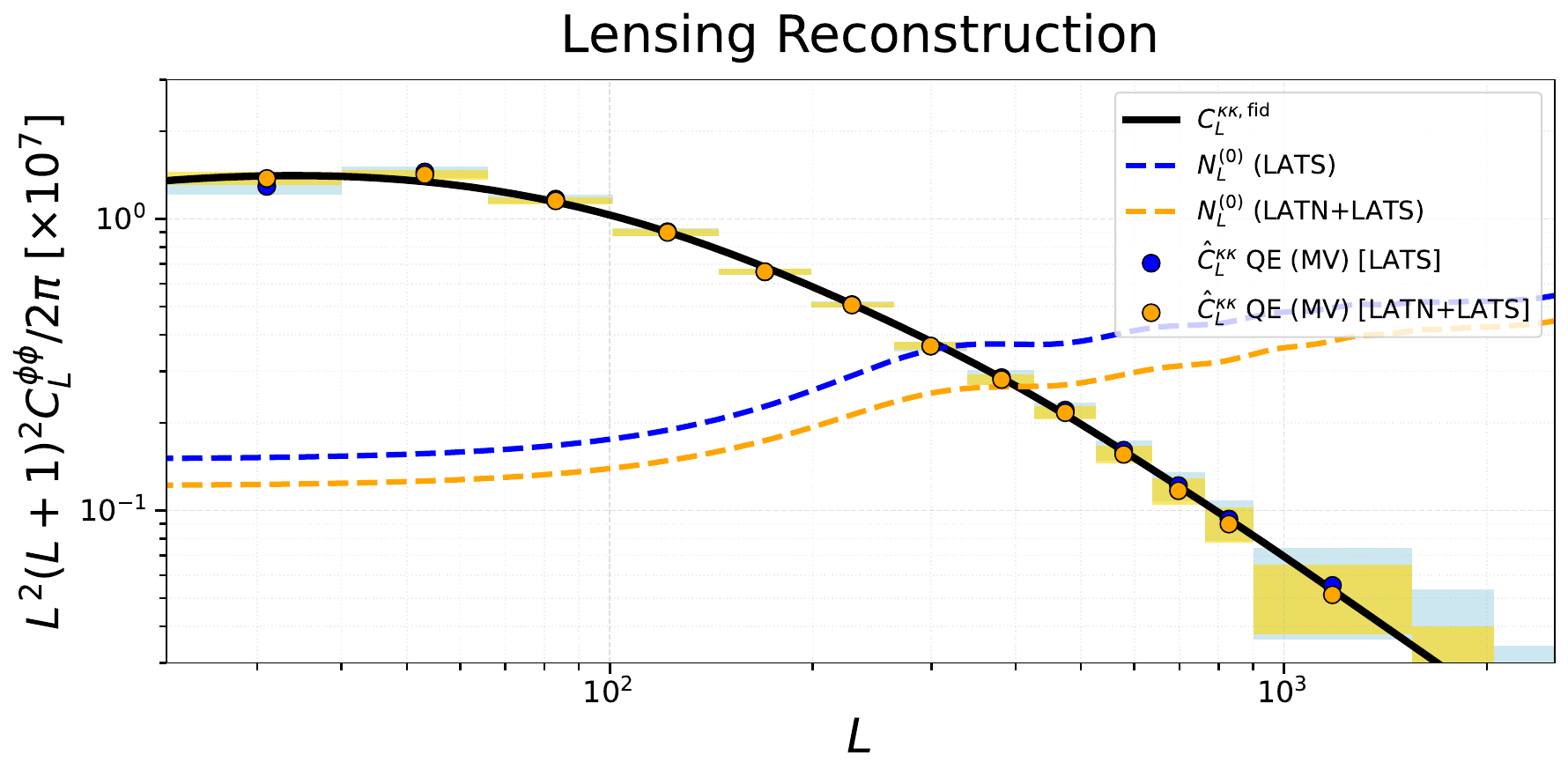}
	\caption{The lensing power spectra from the internal reconstruction (MV) of a single simulation for both the \texttt{LATS} and \texttt{LATN+LATS} cases are shown. The light-blue and gold shaded regions represent the $1\sigma$ standard deviation estimated from 400 simulations for the LATS and \texttt{LATN+LATS} cases, respectively. The terms "$N^{(0)}_L$" denote the semi-analytical leading-order reconstruction noise, which is used for Wiener filtering.}
	\label{fig:lensing_power}
\end{figure}

The power spectrum estimate of the reconstructed lensing convergence from a single simulation for both the \texttt{LATS} and \texttt{LATN+LATS} cases are shown in Fig.~\ref{fig:lensing_power}. These spectra have been binned and rescaled with a multiplicative correction following \cite{aghanim2020planck}.  
We observe a significant reduction in the standard deviation on large scales for the \texttt{LATN+LATS} case, primarily due to the increased sky coverage enabled by the addition of LATN. Furthermore, a notable reduction in reconstruction noise (dashed lines) is evident for the \texttt{LATN+LATS} case, attributed to the suppression of noise and foreground residuals resulting from the overlap between \texttt{LATN} and \texttt{LATS}.

\begin{figure}
	\includegraphics[width=\columnwidth]{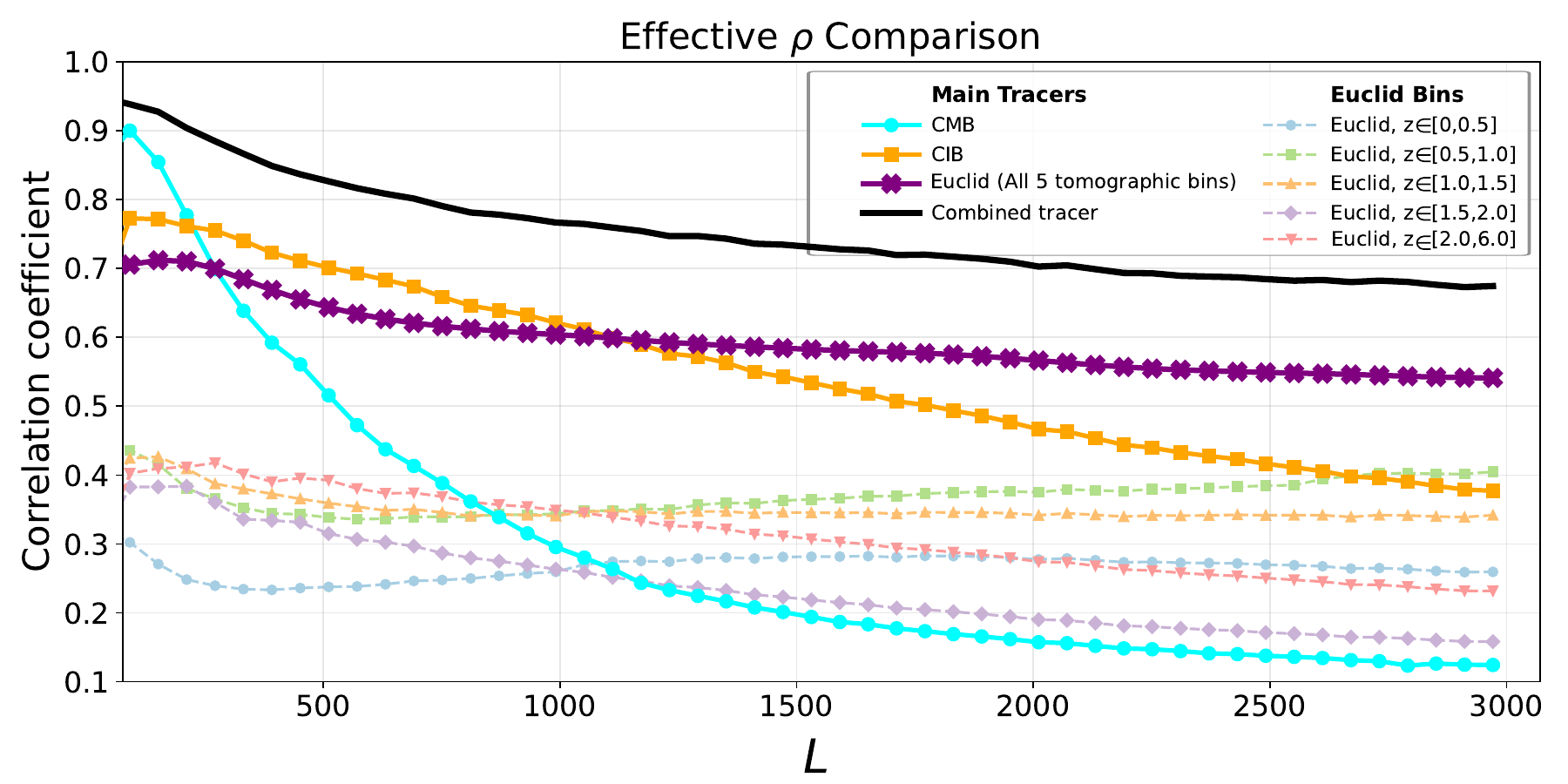}
	\caption{The correlation coefficient $\rho$ of each tracer, as well as the combined tracer (black), with the true convergence is shown. Here "CMB" represents the internal reconstruction for \texttt{LATN+LATS} case. It is worth noting that improvements in $\rho$ at different angular scales contribute unequally to the effectiveness of delensing. In particular, the inclusion of external tracers significantly enhances the correlation on smaller scales compared to internal lensing reconstruction alone, which can lead to a non-negligible improvement in delensing performance.}
	\label{fig:tracer_rhos}
\end{figure}

Figure~\ref{fig:tracer_rhos} shows the correlation coefficient $\rho = C_L^{\kappa \times I}/\sqrt{C_L^{\kappa \kappa}C_L^{I I}}$ between each tracer and the true convergence field. Our results indicate that combining external tracers such as CIB and \emph{Euclid} with internal CMB lensing reconstruction does not significantly enhance the correlation on large scales, where internal reconstruction already achieves a high signal-to-noise ratio (S/N). However, as expected, the external tracers substantially improve the correlation on smaller angular scales. This scale-dependent enhancement is particularly valuable, as it can contribute meaningfully to improving delensing performance under current experimental configurations.

\subsection{Results of LT construction}
\begin{figure}
\centering
\includegraphics[width=\textwidth]{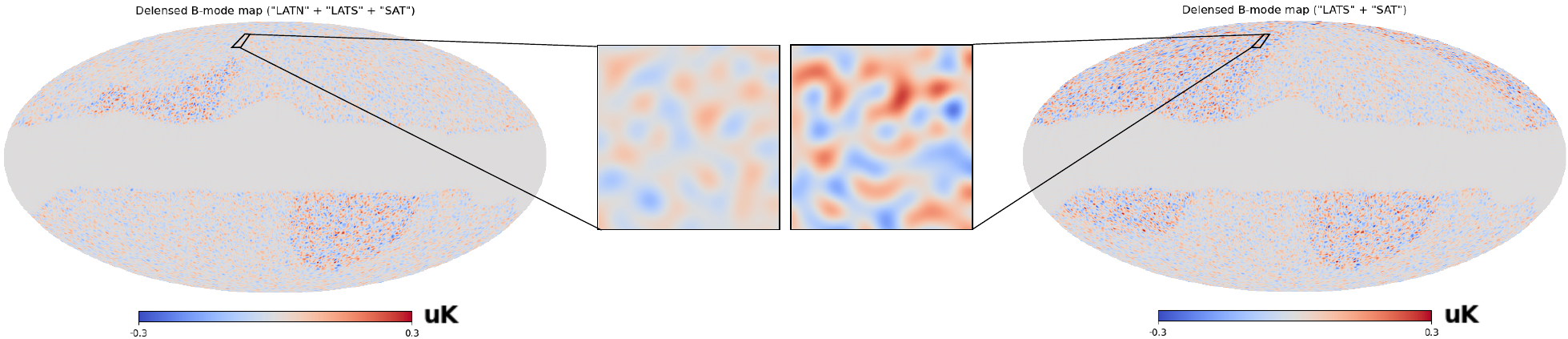}
\caption{
    The delensed $B$-mode maps obtained using the gradient-order method for both the \texttt{LATN+LATS+SAT} and \texttt{LATS+SAT} cases. 
    The two middle zoomed-in subplots depict a $10^\circ \times 10^\circ$ region centered at $(50^\circ, 70^\circ)$ in Galactic coordinates. 
    Shaded regions indicate the lensing $B$-mode reduction. Compared to \texttt{LATS+SAT} ($43\%$ sky coverage), the inclusion of LATN data expands coverage to $61\%$, improving the lensing template reconstruction.
}
\label{fig:delensed_b_map}  
\end{figure}

We plot one of the delensed B-mode maps from our simulation set for \( \ell < 200 \) in Figure~\ref{fig:delensed_b_map}. The delensed B-mode map is obtained by subtracting the gradient-order lensing B-mode template from the observation. After removing the noise component from the noisy lensing template at the map level, we divide the result by the square root of the transfer function, as calculated above, to compensate for signal suppression caused by filtering.
It is evident that both the \texttt{LATN+LATS+SAT} and \texttt{LATS+SAT} cases mitigate part of the lensed B-modes. However, the former allows a larger fraction of the sky to be delensed, thanks to the additional data from LATN.

\begin{figure}
\centering
\includegraphics[width=1\textwidth]{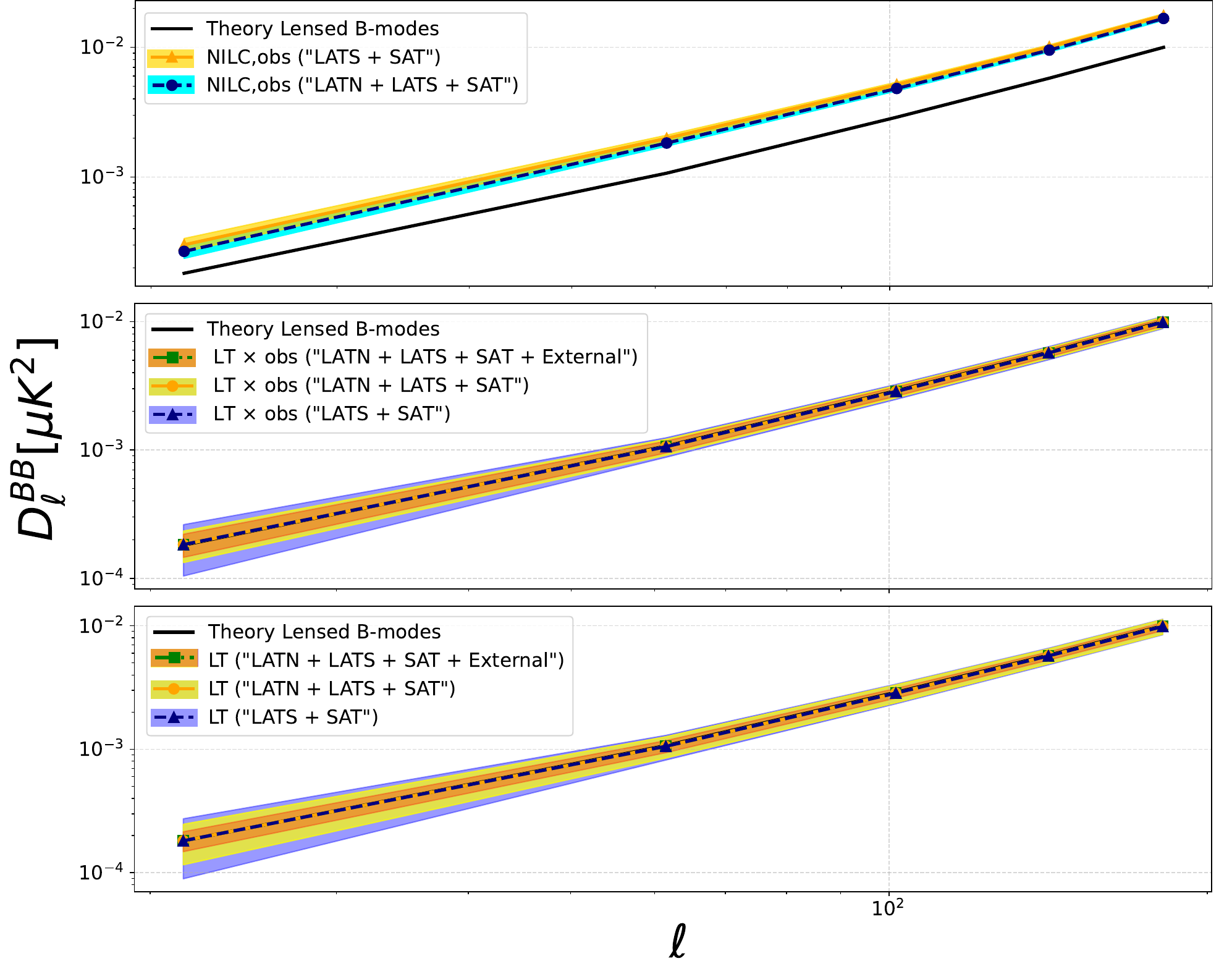}
\caption{The power spectra of the NILC-cleaned B-modes and the lensing B-mode template (LT) are shown for two cases, with all spectra properly debiased. The LT was generated using the gradient-order method, and the power spectra have been debiased and rescaled using the transfer functions. The black lines represent the theoretical predictions based on our models with best-fit parameters.  
In the top panel, the blue and green shaded areas indicate the $3\sigma$ standard deviation computed from 500 observation simulations for the \texttt{LATN+LATS+SAT}(same as \texttt{LATN+LATS+SAT+External}) and \texttt{LATS+SAT}(same as \texttt{LATS + SAT + External}) cases, respectively. 
In the lower two panels, the red, yellow and blue shaded areas represent the $2\sigma$ standard deviation of the LT for the \texttt{LATN+LATS+SAT+External}, \texttt{LATN+LATS+SAT} and \texttt{LATS+SAT} cases, respectively.  
As observed, all mean spectra in the lower two panels, after debiasing, align well with the theoretical predictions, which are precisely overlaid by the other curves. Moreover, a clear reduction in uncertainty is evident when comparing the LT derived from the \texttt{LATN+LATS+SAT} case to that from the \texttt{LATS+SAT} case.}
\label{fig: power_all}
\end{figure}

The power spectra of the NILC-cleaned B-modes and the lensing B-mode template (LT) are shown in Fig.~\ref{fig: power_all}. Compared to the \texttt{LATS+SAT} case, the power spectrum of the NILC-cleaned B-modes in the \texttt{LATN+LATS+SAT} case exhibits a slight reduction, attributed to its overlap with the other two experiments.  
Moreover, we observe a notable reduction in the standard deviation of the lensing B-mode template power spectrum in the \texttt{LATN+LATS+SAT} case. This improvement is primarily attributed to the increased sky coverage available for delensing, which enhances the statistical power of the reconstructed lensing signal. The reduction in standard deviation directly translates into a tighter constraint on the tensor-to-scalar ratio ($r$), as will be demonstrated in Section~\ref{sec: params}. 
Furthermore, the inclusion of LSS tracers (\texttt{LATN+LATS+SAT+External} case) leads to an additional reduction in the standard deviation by providing a more accurate estimate of the lensing potential. 


\subsection{Results on the \texorpdfstring{$r$}{r} constraint}\label{sec: params}

\begin{figure}[htbp]
    \centering
    \subfigure[Covariance Matrix for \texttt{LATS+SAT} ]{
        \includegraphics[width=0.45\linewidth]{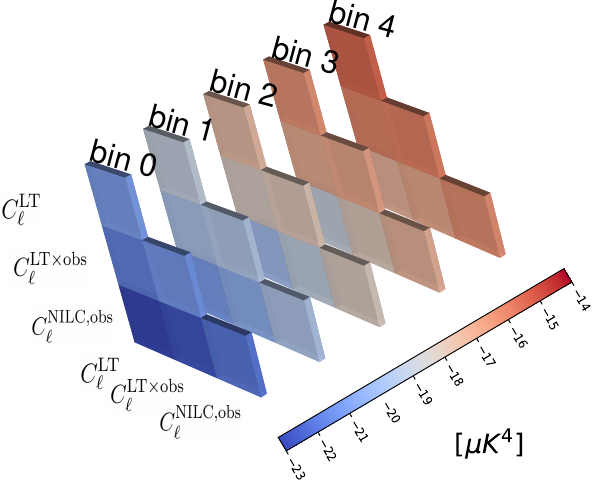}
    }
    \hspace{0.01\linewidth} 
    \subfigure[Covariance Matrix for \texttt{LATN+LATS+SAT}]{
        \includegraphics[width=0.45\linewidth]{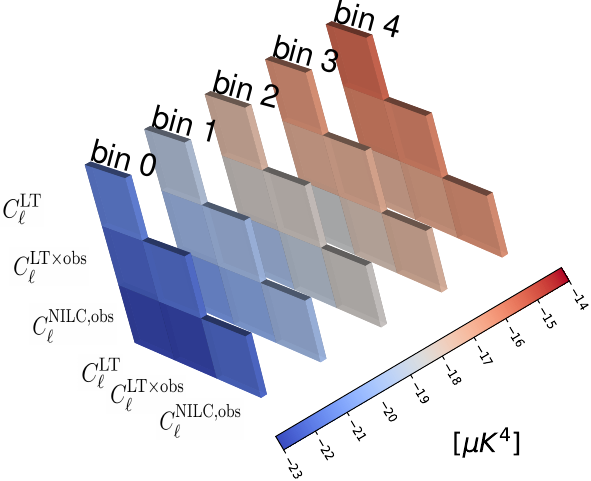}
    }
    \caption{The covariance matrix of the data vector $\mathbf{\hat{X}}_{b}$ for the \texttt{LATS+SAT} and \texttt{LATN+LATS+SAT} cases. Notice: Only the diagonal blocks ($\text{Cov}(C_{\text{bin} \_ i^{th}}^A,C_{\text{bin} \_ i^{th}}^B)$) are shown, and the covariance values are shown in logarithmic scale for clarity. The five slices represent the covariance matrices from bin 0 to bin 4, corresponding to the multipole range from $\ell = 2$ to $\ell = 200$, which are used for parameter constraints. We find an overall reduction in the matrix value for the \texttt{LATN+LATS+SAT} case (right panel).}
    \label{fig:covmat}
\end{figure}

The covariance matrices of the data vector $\mathbf{\hat{X}}_{b}$ for the \texttt{LATS+SAT} and \texttt{LATN+LATS+SAT} cases are shown in Fig.~\ref{fig:covmat}. A notable reduction in covariance is observed when comparing the \texttt{LATN+LATS+SAT} case to the \texttt{LATS+SAT} case, with the former exhibiting bluer overall. This reduction is primarily attributed to the increased sky coverage available for delensing.

\begin{table}
	\centering
	\caption{Input value used for B-mode simulation and prior imposed on each parameters for MCMC sampling. $\mathcal{U}(a,b)$ denotes uniform distribution between $[a,b]$.}
	\label{tab:prior}
	\begin{tabular}{lcccr} 
		\hline
		Parameter & Input value & Prior\\
		\hline
		$r$ & 0 &$\mathcal{U}(-0.2,0.2)$ \\
		$A_L$ & 1.000 &$\mathcal{U}(0,1.2)$ \\
		\hline
	\end{tabular}
\end{table}

\begin{figure}[htbp]
    \centering
    \includegraphics[width=1\textwidth]{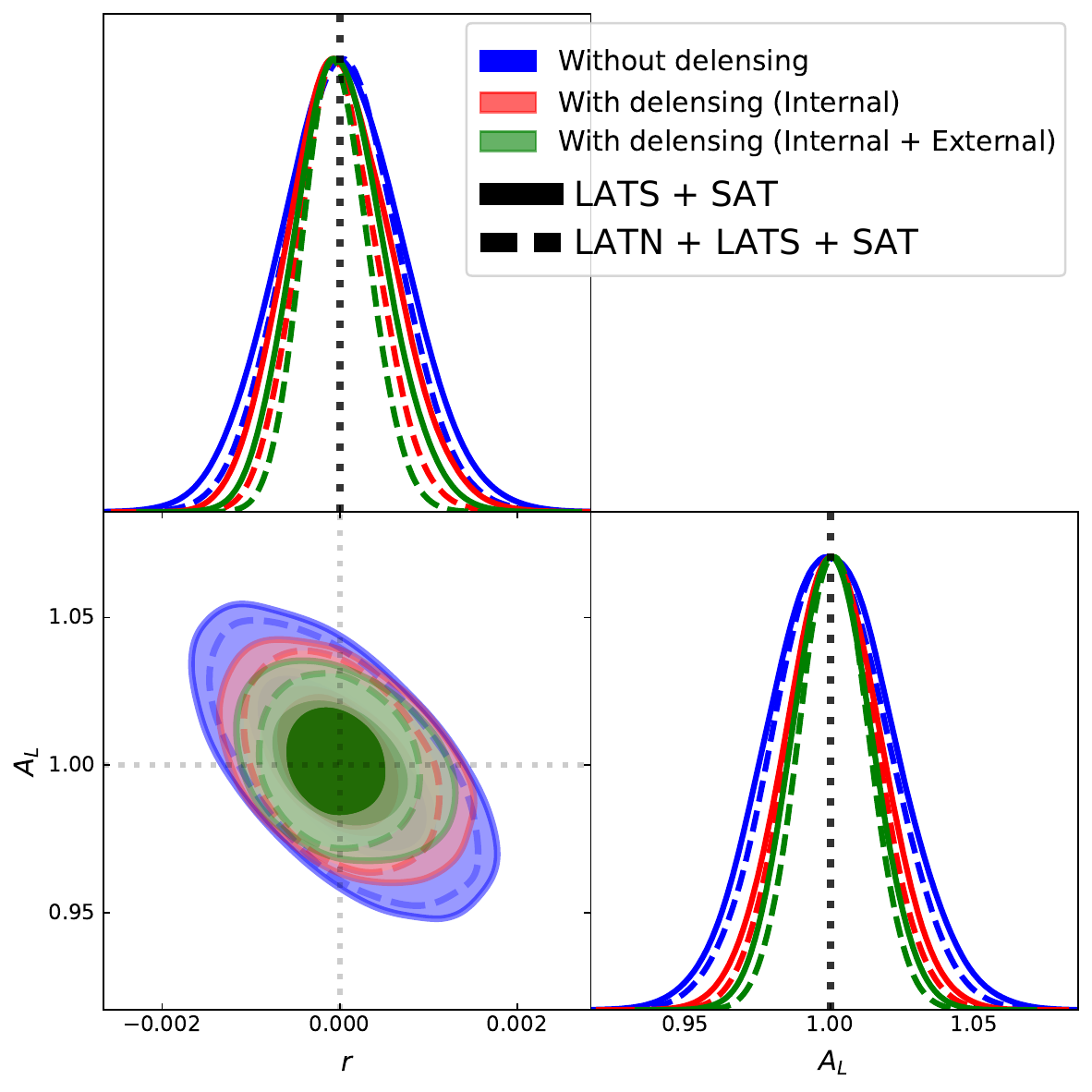}
    \caption{Posterior distributions of the baseline model parameters are shown for the \texttt{LATS+SAT} and \texttt{LATN+LATS+SAT} cases, with delensing and without delensing. The delensing is performed with the gradient-order method. The Gaussian approximation was used when performing the likelihood analysis. The lensing template (LT) was constructed using the gradient-order method. It is evident that the uncertainty in $r$ is reduced by including the \texttt{LATN} data in the likelihood analysis, for both "With delensing" and "Without delensing" cases.}
    \label{fig: posterior}
\end{figure}

We present the posterior distributions for the parameters using the two-parameter model with simulated data in Fig.~\ref{fig: posterior}. 
The degeneracy between $r$ and $A_L$ is progressively alleviated with the inclusion of the lensing B-mode template, as shown in the contour plots. Additionally, we observe a significant reduction in the uncertainty of $r$ with the inclusion of LATN data, regardless of whether the delensing procedure is performed.  

The summarized results from the Gradient-order method and Inverse-lensing method are presented in Table~\ref{tab:posterior_temp} and Table~\ref{tab:posterior_remap}, respectively. For the \texttt{LATS+SAT} configuration, the uncertainty in the tensor-to-scalar ratio \(r\) decreases from approximately \(0.70 \times 10^{-3}\) to \(0.58 \times 10^{-3}\) when the lensing template (LT) is incorporated as an independent observation channel in the likelihood, corresponding to an improvement of about 17\%.
We further find that the inclusion of \texttt{LATN} data leads to an additional reduction in the uncertainty of \(r\). Even in the absence of delensing, adding \texttt{LATN} improves the constraint by nearly 11\%. More remarkably, when the lensing template is constructed with \texttt{LATN} data included, the uncertainty in \(r\) is reduced by a further 18\%, from \(0.58 \times 10^{-3}\) to \(0.45 \times 10^{-3}\). This level of improvement is comparable to that achieved by incorporating external large-scale structure (LSS) tracers, which yields an approximately 13\% reduction (\(0.58 \times 10^{-3}\) to \(0.50 \times 10^{-3}\)).
Once \texttt{LATN} data are included, the marginal benefit of adding external LSS information is correspondingly reduced, resulting in an additional improvement of about 10\% in the uncertainty of \(r\), from \(0.45 \times 10^{-3}\) to \(0.38 \times 10^{-3}\).
We also find a strong consistency between the two delensing methods for most cases, further reinforcing the robustness of our results.

\begin{table*} 
	\centering 
	\caption{The mean and $1\sigma$ standard deviation of each parameter using Gradient-order method, with the \texttt{LATS+SAT} case, \texttt{LATS+SAT+External} case, \texttt{LATN+LATS+SAT} case and \texttt{LATN+LATS+SAT+External} case. HL likelihood and "diag full" covariance conditioning were used in likelihood analysis. Notice that in the main text, the terms "Without delensing" and "With delensing" are used interchangeably with "Before adding LT" and "After adding LT" shown here, respectively.}
	\label{tab:posterior_temp}
	\begin{adjustbox}{max width=\textwidth} 
	\begin{tabular}{lccccccc} 
		\hline
		& & & & \multicolumn{4}{c}{Gradient-order Method} \\
		\cmidrule(lr){5-8}  
		Parameter & Input value & \makecell{Before adding LT \\ (\texttt{LATS+SAT})} & \makecell{Before adding LT \\ (\texttt{LATN+LATS+SAT})} & \makecell{After adding LT \\ (\texttt{LATS+SAT})} & \makecell{After adding LT \\ (\texttt{LATS+SAT+External})} & \makecell{After adding LT \\ (\texttt{LATN+LATS+SAT})} & \makecell{After adding LT \\ (\texttt{LATN+LATS+SAT+External})} \\
		\hline
		$r (\times 10^3)$ & 0 & $0.021 \pm 0.699$ & $0.025 \pm 0.618$ & $0.014 \pm 0.582$ & $-0.025 \pm 0.496$ & $-0.016 \pm 0.450$ & $-0.031 \pm 0.380$\\
		$A_L$ & 1 & $1.000 \pm 0.022$ & $1.000 \pm 0.020$ & $1.001 \pm 0.017$ & $1.001 \pm 0.014$ & $1.001 \pm 0.015$ & $1.001 \pm 0.012$ \\
		\hline
	\end{tabular}
	\end{adjustbox}
\end{table*}

\begin{table*} 
	\centering 
	\caption{The mean and $1\sigma$ standard deviation of each parameter using Inverse-lensing method, with the \texttt{LATS+SAT} case and \texttt{LATN+LATS+SAT} case. HL likelihood and "diag full" covariance conditioning were used in likelihood analysis. Notice that in the main text, the terms "Without delensing" and "With delensing" are used interchangeably with "Before adding LT" and "After adding LT" shown here, respectively. We do not present cases with \texttt{+External} since there is no reason to expect any evident difference in the Gradient-order method.}
	\label{tab:posterior_remap}
	\begin{adjustbox}{max width=\textwidth} 
	\begin{tabular}{lccccc} 
		\hline
		& & & & \multicolumn{2}{c}{Inverse-lensing Method} \\
		\cmidrule(lr){5-6}  
		Parameter & Input value & \makecell{Before adding LT \\ (\texttt{LATS+SAT})} & \makecell{Before adding LT \\ (\texttt{LATN+LATS+SAT})} & \makecell{After adding LT \\ (\texttt{LATS+SAT})} & \makecell{After adding LT \\ (\texttt{LATN+LATS+SAT})} \\
		\hline
		$r (\times 10^3)$ & 0 & $0.021 \pm 0.699$ & $0.025 \pm 0.618$ & $-0.006 \pm 0.586$ & $0.003 \pm 0.460$ \\
		$A_L$ & 1 & $1.000 \pm 0.022$ & $1.000 \pm 0.020$ & $1.001 \pm 0.016$ & $1.001 \pm 0.015$  \\
		\hline
	\end{tabular}
	\end{adjustbox}
\end{table*}

\section{Conclusions}\label{sec: conclusion}

In this paper, we present a pipeline designed to improve full-sky constraints on \(r\) through large-scale B-mode delensing for future ground-based and satellite CMB observations. 
As baseline, we consider a scenario (denoted as \texttt{LATS+SAT}), where a ground-based large-aperture telescope in the Southern Hemisphere (\texttt{LATS}) enables the detection of the lensing potential, while a satellite-based small-aperture telescope (SAT) facilitates the detection of large-scale polarization with a significantly low noise level. 
Furthermore, we find that incorporating an additional ground-based LAT in the Northern Hemisphere (denoted as \texttt{LATN+LATS+SAT}) provides two key advantages. On the one hand, the overlap between experiments helps reduce the noise level. On the other hand, the expanded sky coverage increases the area over which the lensing B-mode template can be constructed, thereby enhancing the overall effectiveness of B-mode delensing. These two factors together significantly improve the constraints on \(r\), further reducing its uncertainty compared to the \texttt{LATS+SAT} case. We also incorporate the external LSS tracers to the delensing procedure as comparison.

We apply the NILC method to remove foreground components in the observations from both LATs and SAT, utilizing six frequency channels for LATs and fifteen for SAT. We observe a significant reduction in the power of the cleaned maps, owing to the method’s objective of minimizing the total variance.  
For polarization, we expect noise and foreground residuals to have minimal impact on the delensing procedure, as discussed in our prior work \cite{chen2025delensing}. For temperature maps, it is well established that the ILC-based cleaning is less effective at suppressing compact foregrounds. Due to their non-Gaussianity and correlations with the underlying matter distribution, residual extragalactic foregrounds can bias the internally-reconstructed lensing signal at the several-percent level \cite{schaan2019foreground,osborne2014extragalactic,sailer2020lower,darwish2023optimizing}. Such bias can then propagate into the delensing procedure and subsequently affect the likelihood analysis \cite{baleato2022impact}. A similar bias may also arise from the non-Gaussianity of Galactic foregrounds \cite{abril2025impact,beck2020impact}.
However, \cite{abril2025impact} demonstrates that, for an SO-like experiment employing MV HILC, the bias induced by Galactic foregrounds remains safely below the reconstruction uncertainty. Consistent conclusions are reported in \cite{beck2020impact}, where no significant additional bias is found after foreground cleaning, enabling an unbiased measurement of the tensor-to-scalar ratio after delensing. In addition, our own analysis incorporating explicit foreground-mitigation strategies (Appendix~\ref{sec: bh_app}) yields constraints on $r$ that agree well with our baseline results.
Based on these findings, we conclude that the baseline analysis remains robust against residual foreground contamination. Nonetheless, a more systematic optimization and evaluation of foreground-induced reconstruction biases will be pursued in future work.

We perform internal lensing reconstruction using simulated data from ground-based LATs, incorporating both temperature and polarization information. External LSS tracers such as CIB and galaxy number density are further included in the pipeline.

The lensing B-mode template (LT) is constructed using simulated data from the combined observations of the satellite mission (SAT) and ground-based LATs, employing two delensing approaches.
For parameter estimation, we extend the likelihood function to incorporate all auto- and cross-bandpowers between the LT and the observed B-modes.
Our results show that internal delensing using southern-hemisphere data alone (\texttt{LATS+SAT}) reduces the uncertainty in the tensor-to-scalar ratio 
$r$ by approximately 17\% compared to the no-delensing case.
Furthermore, the inclusion of northern LAT (\texttt{LATN}) enables full-sky internal delensing and yields an additional $\sim$18\% reduction in the uncertainty of 
$r$, comparable to the improvement from including external LSS tracers ($\sim$13\%).
Once \texttt{LATN} is incorporated, the marginal gain from LSS tracers diminishes to around 10\%.

We therefore conclude that for future full-sky missions targeting primordial gravitational waves, achieving full-sky internal delensing critically depends on the sky coverage of ground-based LATs. For LATs with SO-like sensitivity, the inclusion of a northern LAT provides an improvement in constraints on $r$ comparable to that achieved by incorporating external LSS data. This contribution is expected to become even more significant for future, ultra-sensitive LATs when used for full-sky delensing in support of space-based full-sky PGW detection.

\appendix
\section{Quantification of the Delensing bias terms}\label{sec: delens_bias_ana}
As discussed in Section~\ref{sec: bias}, the delensing procedure introduces several bias terms in both the auto- and cross-power spectra of the lensing B-mode template and the observed B-modes. These terms must be properly accounted for in the likelihood analysis to avoid introducing significant biases in the inferred value of~$r$. In this section, we examine the magnitude and physical origin of these contributions for both the Gradient-Order Template Method and the Inverse-Lensing Method. We identify the effects that are common to both approaches, as well as those that arise specifically from the distinct characteristics of each method. For clarity of discussion, we reproduce below several key expressions previously introduced in Section~\ref{sec: bias}.

For the Gradient-Order Template Method, the lensing B-mode template can be written as:
\begin{equation}\label{EQ:template_app}
	\begin{aligned}
    		B^\text{temp} &= \mathcal{B}^{(1)}[\mathcal{W}^E(E^\text{lens}+E^\text{res,NILC}) \ast \mathcal{W}^{\phi}(\phi + \phi^\text{noise})] \\
		  &= \begin{aligned}[t] &\mathcal{B}^{(1)}[\mathcal{W}^E E^\text{lens} \ast \mathcal{W}^{\phi}\phi] 
		  	+ \left\{ \mathcal{B}^{(1)}[\mathcal{W}^E E^\text{res,NILC} \ast \mathcal{W}^{\phi}(\phi + \phi^\text{noise})] 
		  	+ \mathcal{B}^{(1)}[\mathcal{W}^E E^\text{lens} \ast \mathcal{W}^{\phi}\phi^\text{noise}] \right\}
		  \end{aligned} \\
		  &= B^\text{temp}_S + B^\text{temp}_N,
	\end{aligned}
\end{equation}
where $\mathcal{B}^{(1)}[E\ast\phi]$ represents the operation of constructing the gradient-order template using E-mode filed and $\phi$. There are two sources of template noise: one arising from residual foregrounds and noise in the NILC maps, and the other originating from the spurious lensing effect induced by the lensing reconstruction noise.

\begin{figure}
	\includegraphics[width=\columnwidth]{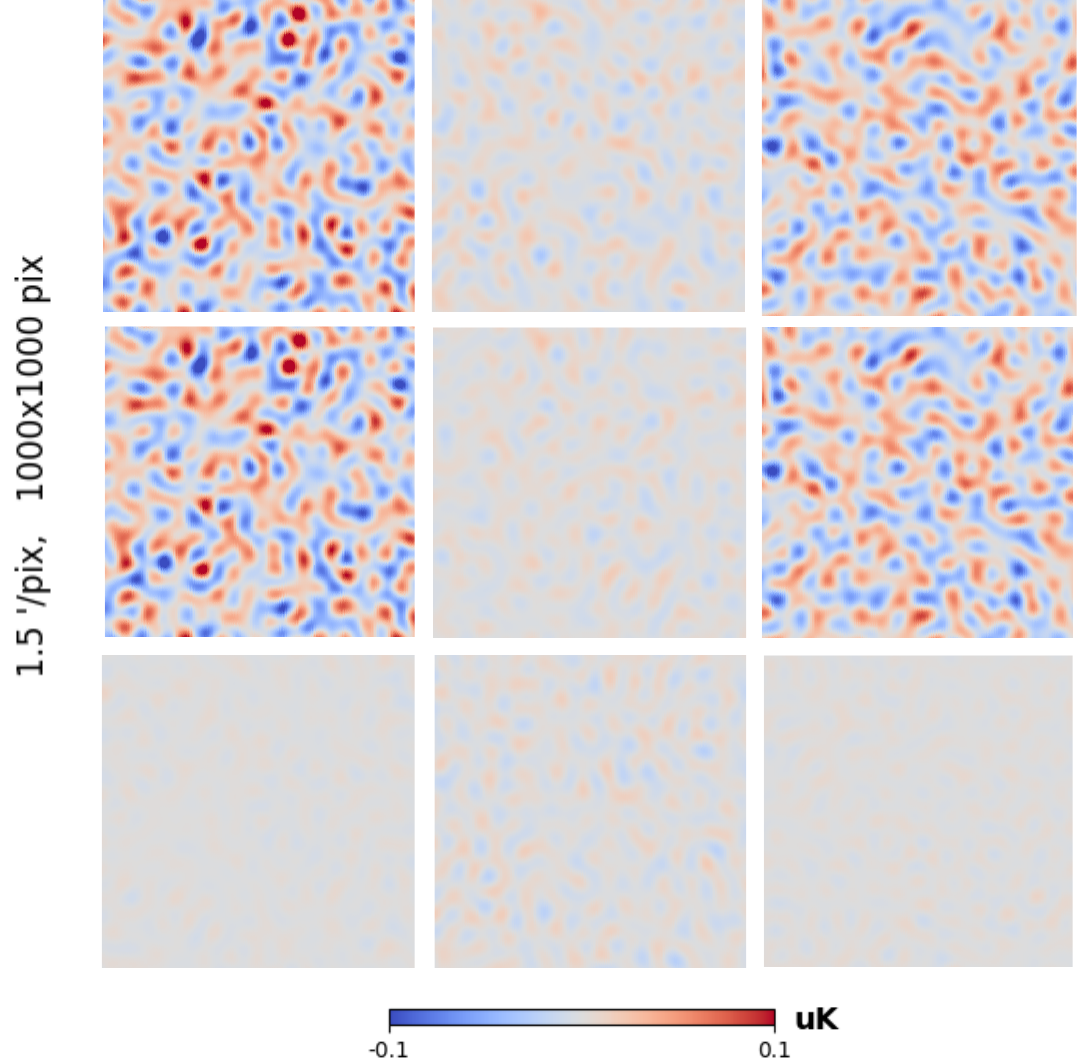}
    \caption{A comparison of the terms shown in Eq.~\ref{EQ:remap_template_app} (upper three panels), 
    Eq.~\ref{EQ:template_app} (middle three panels), and their differences (lower three panels). 
    The multipole range is restricted to $\ell < 200$ to match that used in the likelihood analysis. 
    The first column shows the signal component of the LT template, $B^{\text{temp}}_{S}$. 
    The second column shows the bias introduced by lensing of the noise:
    $\mathcal{B}^{(1)}[\mathcal{W}^{E} E^{\text{res,NILC}} \ast \mathcal{W}^{\phi}(\phi + \phi^{\text{noise}})]$ 
    for the Gradient-order template method, and 
    $\mathcal{W}^{E} B^{\text{res,NILC}} - 
    \mathcal{B}(\mathcal{W}^{E} B^{\text{res,NILC}} \star \mathbf{d}^{\mathrm{inv}})$ 
    for the Inverse-lensing method.
    The third column shows the bias due to the spurious lensing effect sourced by lensing reconstruction noise:
    $\mathcal{B}^{(1)}[\mathcal{W}^{E} E^{\text{lens}} \ast \mathcal{W}^{\phi} \phi^{\text{noise}}]$ 
    for the Gradient-order method, and 
    $[\mathcal{W}^{E} B^{\text{lens}} - 
    \mathcal{B}(\mathcal{W}^{E} B^{\text{lens}} \star \mathbf{d}^{\mathrm{inv}}_{\text{noise}})]
    - B^{\text{de,hi}}$ 
    for the Inverse-lensing method. 
    Overall, we find that the two methods exhibit similar behavior on large angular scales across all three terms.
    }
	\label{fig:delens_terms_map}
\end{figure}

For the Inverse-Lensing Method, the remapped B-modes is given by:
\begin{equation}\label{EQ:remap_app}
	\begin{aligned}
    		B^\text{del} &= \mathcal{B} [\mathcal{W}^E (B^\text{lens} + B^\text{res,NILC}) \star \mathbf{d^{inv}} ] \\
				& \approx  \mathcal{B} [\mathcal{W}^E (B^\text{lens} + B^\text{res,NILC}) \star (\mathbf{d^{inv}_\phi} + \mathbf{d^{inv}_\text{noise}}) ] \\
				& \approx \left\{ \mathcal{B} [\mathcal{W}^E B^\text{lens} \star \mathbf{d^{inv}_\phi} ] \right\} 
				+ \left\{ [ \mathcal{B} [\mathcal{W}^E B^\text{lens} \star \mathbf{d^{inv}_\text{noise}} ] - \mathcal{W}^E B^\text{lens}] + \mathcal{B} [\mathcal{W}^E  B^\text{res,NILC} \star \mathbf{d^{inv}}  ]  + B^\text{de,hi} \right\} \\
				&= B^\text{del}_S + B^\text{del}_{N}, 
	\end{aligned}
\end{equation}
where $\mathcal{B} [B\star \mathbf{d^{inv}}]$ represents the inverse
remapping operation of lensed B mode with $\mathbf{d^{inv}}$ to obtain
the delensed B-modes. Notice that we have applied a linear approximation when moving from the second row to the third, and all higher-order terms neglected in this step are absorbed into $B^{\mathrm{de,hi}}$. This should be distinguished from the higher-order lensing contributions contained in the term $\mathcal{B}[\, B \star \mathbf{d}^{\mathrm{inv}} ]$.
The lensing B-mode template:
\begin{equation}\label{EQ:remap_template_app}
\begin{aligned}
B^\text{temp} &= \mathcal{W}^E B^{obs} - B^\text{del} \\
&= \mathcal{W}^E[B^\text{lens} + B^\text{res,NILC}] - [B^\text{del}_S + B^\text{del}_{N}] \\
&\approx [\mathcal{W}^E B^\text{lens} - B^\text{del}_S] + [\mathcal{W}^E B^\text{res,NILC} - B^\text{del}_N] \\
&= [\mathcal{W}^E B^\text{lens} - B^\text{del}_S] 
    + \left\{
      \begin{aligned}
        &\mathcal{W}^E B^\text{res,NILC}
          -\mathcal{B}(\mathcal{W}^E B^\text{res,NILC}\star\mathbf{d^{inv}}) \\
        &\quad + [\mathcal{W}^E B^\text{lens}
          -\mathcal{B}(\mathcal{W}^E B^\text{lens}\star\mathbf{d^{inv}_\text{noise} })] -B^\text{de,hi}
      \end{aligned}
      \right\} \\
&= B^\text{temp}_S + B_N^\text{temp},
\end{aligned}
\end{equation}
where $B_N^{\mathrm{temp}}$ corresponds to $B^{\mathrm{temp}}_N$ in Eq.~\ref{EQ:template_app}. We find that the two noise terms share a similar physical origin, although the former includes higher-order contributions through $B^{\mathrm{de,hi}}$ beyond the gradient approximation. A comparison of these terms is presented in Figure~\ref{fig:delens_terms_map}. Overall, we find that the two methods yield similar behavior on large angular scales.

Therefore, the auto-power spectrum of the lensing template, given by \( B^\text{temp} = B^\text{temp}_S + B_N^\text{temp} \), is:
\begin{equation}\label{EQ:auto_temp_app}
	\begin{aligned}
        C^\text{temp} &= \langle B^\text{temp}_S B^\text{temp}_S \rangle + \langle B_N^\text{temp} B_N^\text{temp} \rangle + 2\langle B^\text{temp}_S B_N^\text{temp} \rangle.
	\end{aligned}
\end{equation}
For the cross-power spectrum with the observed B-modes, \( B^\text{obs} = B^\text{lens} + B^\text{res,NILC} \), we have:
\begin{equation}\label{EQ:cross_temp_app}
	\begin{aligned}
        C^{\text{cross}} &= \langle B^\text{lens} B_S^\text{temp} \rangle + \langle B^\text{lens} B_N^\text{temp} \rangle + \langle B^\text{res,NILC} B^\text{temp} \rangle.
	\end{aligned}
\end{equation}

\begin{figure}[htbp]
    \centering
    \subfigure[Some terms in the auto-power spectra. Shown are the auto-power of the $B_N^\text{temp}$ and its cross-power with $B_S^\text{temp}$. ]{
        \includegraphics[width=0.47\linewidth]{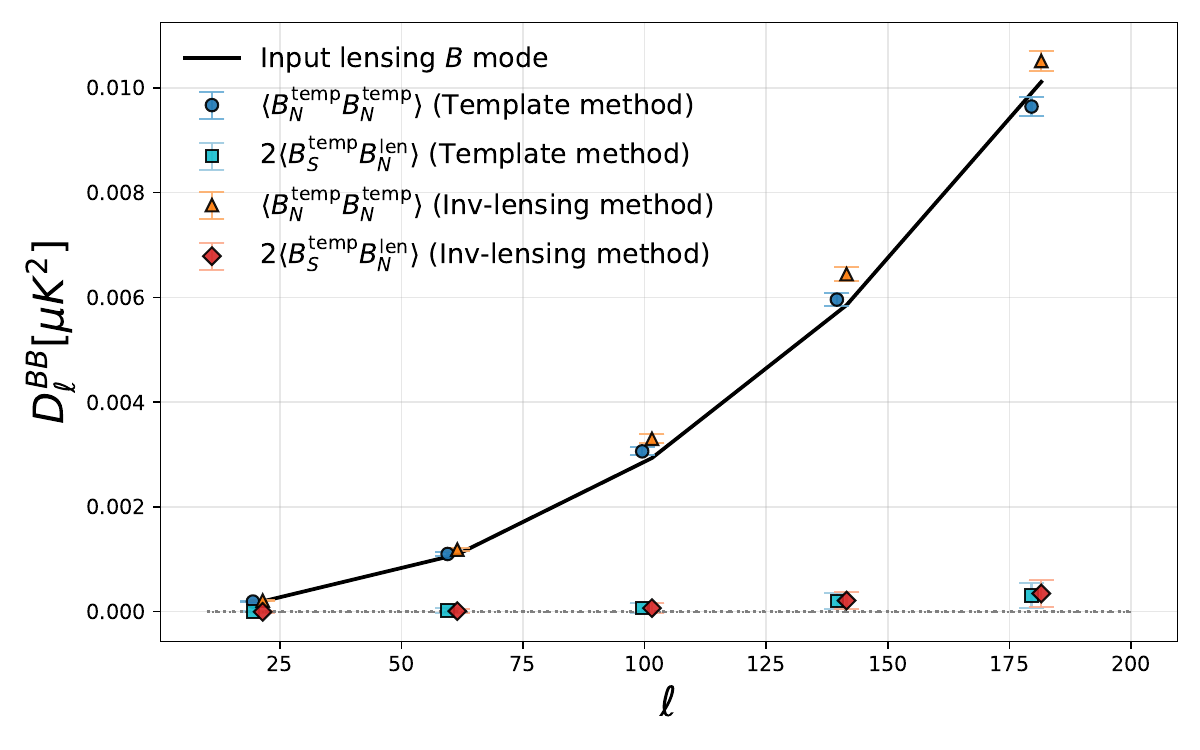}
    }
    \hspace{0.01\linewidth} 
    \subfigure[Some terms in the cross-power spectra. Shown are the noisy cross-power spectra between $B^\text{obs}$ and $B^\text{temp}$ and the sum of both cross bias.]{
        \includegraphics[width=0.47\linewidth]{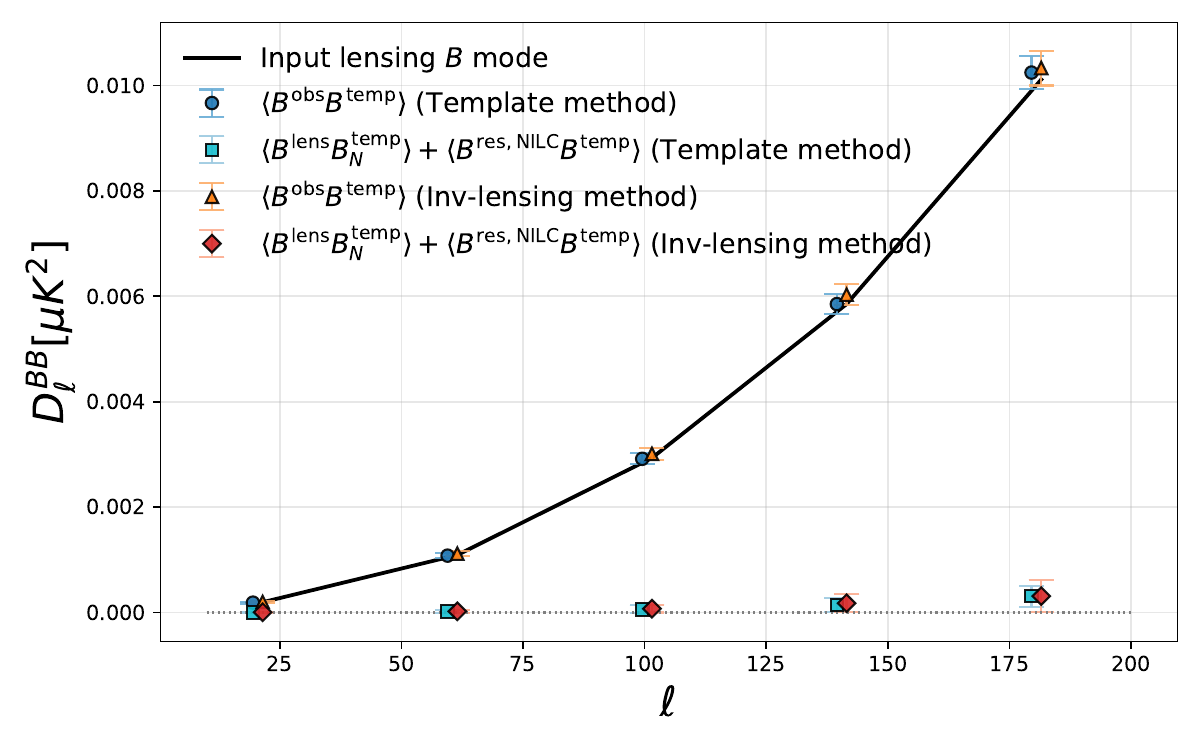}
    }
    \caption{Selected auto- and cross-power spectrum terms of LT for both delensing methods are shown.
    All spectra have been rescaled following the procedure in Section~\ref{sec: cal_bp}.
    For visual clarity, certain data points are slightly offset along the multipole axis.
    Note: The apparent agreement between the LT-noise power spectrum (left panel) and the input lensing 
    $B$-mode spectrum is purely coincidental. The LT-noise spectrum is shaped by residuals in the observed maps 
    as well as lensing reconstruction noise, and this visual similarity should not be overinterpreted.}
    \label{fig:delens_terms}
\end{figure}

A comparison between these terms is shown in Figure~\ref{fig:delens_terms} for the \texttt{LATN+LATS+SAT} case. For the other cases considered in this work, they are likewise expected to be of the same order as the \texttt{LATN+LATS+SAT} case. As illustrated in the left panel, the cross-bias term $\langle B_S^{\mathrm{temp}} B_N^{\mathrm{temp}} \rangle$ is consistent with zero for both delensing methods. For the Gradient-Order Template Method, as indicated by Eq.~\ref{EQ:template_app}, the cross-bias term can be schematically written as
\begin{equation}\label{EQ:cross_bias}
    \begin{aligned}
        \langle B_S^{\mathrm{temp}} B_N^{\mathrm{temp}} \rangle 
        \propto\ 
        \langle E^{\mathrm{lens}} E^{\mathrm{res,NILC}}\, \phi\, \hat{\phi} \rangle 
        + 
        \langle E^{\mathrm{lens}} E^{\mathrm{lens}}\, \phi\, \phi^{\mathrm{noise}} \rangle ,
    \end{aligned}
\end{equation}
where non-trivial six-point correlation terms are expected if the internal lensing reconstruction uses the $EB$ quadratic estimator (noting that $\hat{\phi}^{EB} \propto \mathcal{Q}[E^{\mathrm{LAT}},B^{\mathrm{LAT}}]$). Fortunately, as discussed in \cite{lizancos2021impact}, this bias is highly local in multipole space. Hence, removing large-scale polarization modes from the lensing reconstruction---e.g., excluding $\ell < 200$ in this work, which is the same range where the lensing template is constructed and the likelihood analysis is performed---provides an effective means to suppress the bias. We have adopted exactly this mitigation strategy in Section~\ref{sec: rec_in}.
For the Inverse-Lensing Method, we likewise find the cross-bias term to be consistent with zero. This behavior is expected, since the structure of $B_N^{\mathrm{temp}}$ is analogous for the two methods, even though the Inverse-Lensing Method includes additional contributions arising from higher-order lensing effects. These higher-order contributions do not generate a detectable cross bias within the multipole range considered here.

The auto-power spectrum of $B_N^{\mathrm{temp}}$ is found to be of the same order as $C^{\mathrm{lens}}$. A further decomposition of this term shows that, for the Gradient-Order Template Method, most of its power originates from the auto-spectrum of 
$\mathcal{B}^{(1)}[\mathcal{W}^E E^{\mathrm{lens}} \ast \mathcal{W}^{\phi}\phi^{\mathrm{noise}}]$, 
while for the Inverse-Lensing Method it is dominated by 
$[\mathcal{W}^E B^{\mathrm{lens}} - \mathcal{B}(\mathcal{W}^E B^{\mathrm{lens}} \star \mathbf{d}^{\mathrm{inv}}_{\mathrm{noise}} )] - B^{\mathrm{de,hi}}$.
In both cases, the power is sourced by the spurious lensing of the lensed CMB fields induced by the reconstruction noise.
The overall contributions from the two methods are comparable. However, a subtle difference arises because the Inverse-Lensing Method contains additional noise contributions from higher-order lensing effects (see the middle column of Figure~\ref{fig:delens_terms_map}). This effect becomes slightly more pronounced at higher multipoles, as reflected in the behavior of $\langle B_N^{\mathrm{temp}} B_N^{\mathrm{temp}} \rangle$ shown in Figure~\ref{fig:delens_terms}.

For the cross-power spectrum $C^{\mathrm{cross}}$ (right panel of Figure~\ref{fig:delens_terms}), the cross-bias terms are found to be consistent with zero for both delensing methods. 
A similar four-point correlation bias could in principle arise from the use of $EB$ pairs in the internal lensing reconstruction. However, this effect is strongly suppressed by removing large-scale polarization modes from the reconstruction, as discussed above. 
The agreement between the input lensing $B$-mode power and $\langle B^{\mathrm{obs}} B^{\mathrm{temp}} \rangle$ further confirms the minimal cross-bias.

\section{Alternative choice in the likelihood analysis}\label{sec: likelihood_app}
\subsection{Covariance conditioning Schemes}\label{sec: likelihood_condition}
The specific implementations of covariance conditioning vary substantially across analyses \cite{collaborations2018bicep2,ade2014measurement,abazajian2022cmb,adachi2020measurement,balkenhol2022parameter,choi2020atacama,dutcher2021measurements}, depending on the scientific goals and data characteristics. For instance, the BICEP/Keck Collaboration \cite{collaborations2018bicep2} retain correlations in the first off-diagonal bins but suppress noise in the covariance estimate by discarding elements whose expectation values are identically zero. In contrast, the Polarbear Collaboration \cite{ade2014measurement} omit all off-diagonal elements within each block, effectively ignoring mode-coupling.

\begin{figure}
	\includegraphics[width=\columnwidth]{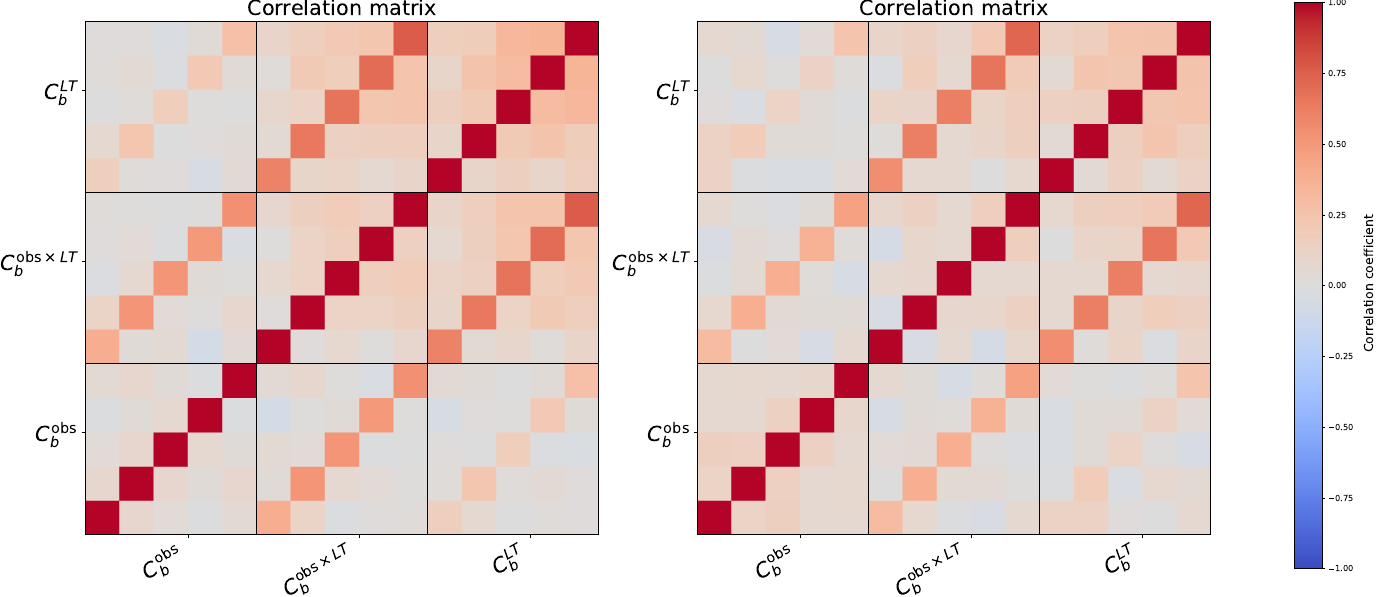}
	\caption{Correlation matrix of the data vector $[C_b^{\mathrm{obs}},\, C_b^{\mathrm{obs} \times \mathrm{LT}},\, C_b^{\mathrm{LT}}]$, computed using five $\ell$ bins ($\Delta\ell=40$), for the ``With delensing (Internal) (LATN+LATS+SAT)'' case (left panel) and the ``With delensing (Internal) (LATS+SAT)'' case (right panel). All correlations are normalized by the main diagonal variances.}
	\label{fig:response_covmat}
\end{figure}

In this work, we consider three conditioning schemes along the $\ell$-bin dimension: “full”, which retains all covariance elements; “nearest”, which keeps only the first off-diagonal elements; and “diagonal”, which drops all off-diagonal elements within each block.
As illustrated in Figure~\ref{fig:response_covmat}, the covariance between $C_b^{\mathrm{obs}}$ and $C_b^{\mathrm{obs}}$ is particularly susceptible to Monte Carlo noise arising from the finite number of simulations used in the covariance estimation. We therefore apply an analogous set of three conditioning schemes to the covariance between different power-spectrum components, i.e., $[C_b^{\mathrm{obs}}, C_b^{\mathrm{obs}\times \mathrm{LT}}, C_b^{\mathrm{LT}}]$.
Combining the conditioning choices for $\ell$ bins and for the component blocks yields a total of nine possible conditioning schemes considered in our analysis.

\begin{figure}[htbp]
    \centering
    \subfigure[The effective $\chi^2$ with "diag full" covariance conditioning.]{
        \includegraphics[width=0.45\linewidth]{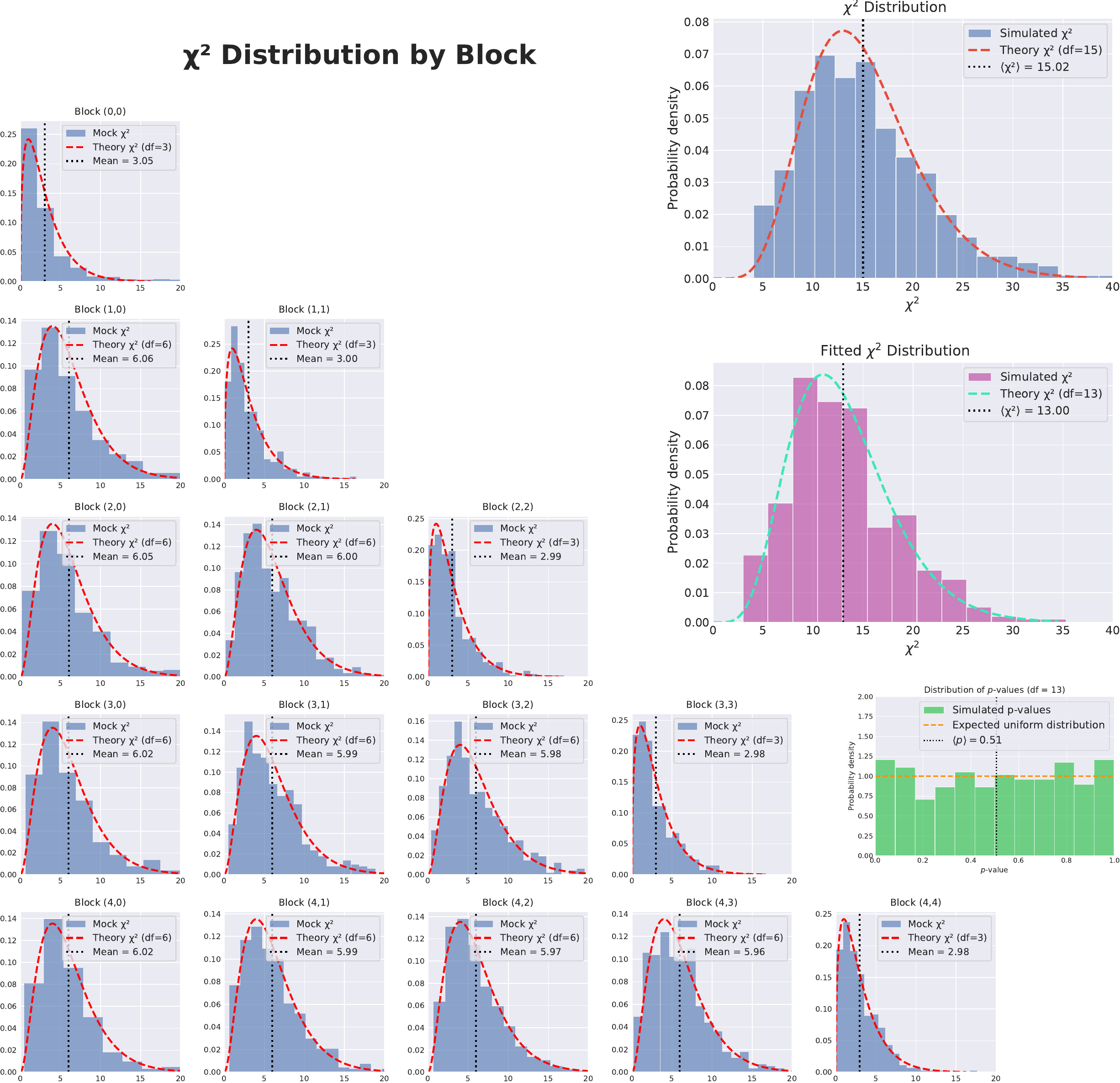}
    }
    \hspace{0.01\linewidth} 
    \subfigure[The effective $\chi^2$ with "nearest full" covariance conditioning.]{
        \includegraphics[width=0.45\linewidth]{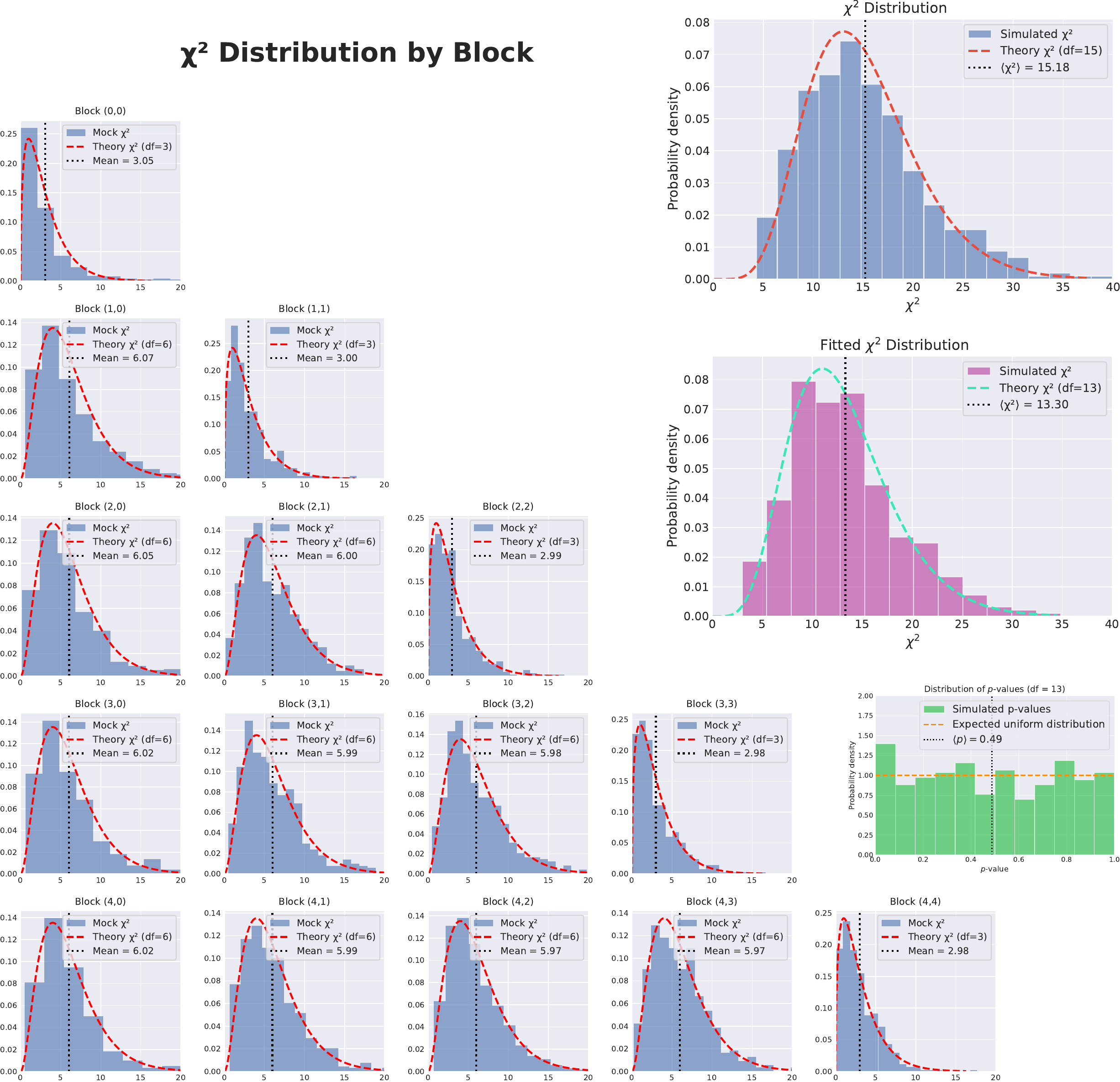}
    }
    \caption{Effective $\chi^2$ distributions (blue histograms) derived from 500 mock data vectors, compared to the theoretical expectation, for the ``With delensing (Internal) (LATN+LATS+SAT)'' case. The lower triangle shows the $\chi^2$ distributions for each individual $\ell$-bin block, while the upper-right panel corresponds to the full set of blocks. Overall, the mock $\chi^2$ distributions are consistent with the theoretical values across all $\ell$-bin blocks, serving as a goodness-of-fit criterion \cite{adachi2020measurement,hamimeche2008likelihood}. Additionally, we present the fitted $\chi^2$ distributions (pink histograms), where the degrees of freedom (df) are reduced to 13 due to the two free fit parameters. The corresponding p-values exhibit a uniform distribution on (0,1), as expected for a correctly specified model, demonstrating that the fit residuals are statistically consistent with the assumed covariance and that no significant systematic mismatch is detected.}
    \label{fig:chi2_distribution}
\end{figure}

Following \cite{adachi2020measurement,hamimeche2008likelihood}, we define a goodness-of-fit statistic
\begin{equation}
\chi^2_{\mathrm{eff}}
\;\equiv\;
-2\ln \mathcal{L}_\text{HL}\!\left(\hat{\mathbf{C}}_b\,\big|\,\mathbf{C}_b^{\mathrm{th}}(r=0,A_L=1)\right),
\end{equation}
which, when evaluated at the fiducial theory with $r=0$ and $A_L=1$, is approximately distributed as a chi-squared variable with
$\mathrm{dof}=N_{\mathrm{data}}$, where $N_{\mathrm{data}}$ is the total length of the data vector (i.e., the total number of bins across all unique spectra).
Table~\ref{tab:response_conditioning} summarizes the fitting results and the mean $\chi^2$ values for the \texttt{LATN+LATS+SAT} ``With delensing (Internal)'' case under different conditioning schemes.
The corresponding $\chi^2$ distributions for the ``diag--full'' and ``nearest--full'' schemes, derived from 500 mock data vectors, are shown in Figure~\ref{fig:chi2_distribution};
we also show the fitted chi-squared distributions obtained from the minimum values, $\chi^2_{\min}$, of the same 500 mock realizations.
Conceptually, the fixed-parameter statistic $\chi^2_{\mathrm{eff}}$ primarily probes the fidelity of the assumed noise model---namely, the overall calibration and the correlation structure encoded in the covariance matrix $\mathbf{M}$, as well as the likelihood's tail behavior---since no freedom remains for the model to absorb mis-specification (expected $\mathrm{dof}=N_{\mathrm{data}}$).
In contrast, the distribution of $\chi^2_{\min}$ (expected $\mathrm{dof}=N_{\mathrm{data}}-k$, with $k$ the number of fitted parameters) mainly assesses the adequacy of the model family after parameter optimization; its reduced value,
\begin{equation}
\chi^2_{\nu}\;\equiv\;\frac{\chi^2_{\min}}{N_{\mathrm{data}}-k},
\end{equation}
diagnoses the overall error calibration but is less sensitive to fine-grained correlation mis-specification.

We first exclude the “full–full’’ scheme, as Monte Carlo noise is known to strongly contaminate the off-diagonal elements. Schemes marked with slashes are removed because their resulting $\chi^2$ values are inconsistent with the expected degrees of freedom of the data vector (3 components and 5 $\ell$ bins) \cite{dutcher2021measurements}. The “full–diag”, “diag–diag”, and “nearest–diag’’ schemes are also excluded, as they are inferior to the remaining options—either discarding too much information or retaining excessively noisy covariance elements.

Among the remaining options, the “diag–full’’ and “nearest–full’’ schemes yield consistent $\chi^2$ values (see Figure~\ref{fig:chi2_distribution}) and comparable constraints on $r$. Similar behavior is observed for other cases and likelihood combinations, which we therefore omit for brevity. We adopt the “diag–full’’ conditioning scheme as the baseline for this work, and also report results obtained with the “nearest–full’’ scheme in Table~\ref{tab:posterior_temp_conditioning} for completeness.

\begin{table}[htbp]
\centering
\caption{HL likelihood results for the ``With delensing (Internal) (LATN+LATS+SAT)'' case using different covariance conditioning schemes. The two words in each conditioning choice indicate the scheme applied to the $\ell$ bins and the components, respectively. Conditioning choices marked with slashes are excluded due to $\chi^2$ inconsistency, while ``full full'' is excluded because Monte Carlo noise significantly affects the off-diagonal elements. The preferred choices, ``\textbf{\textit{diag full}}'' and ``\textbf{\textit{nearest full}}'', are highlighted in tilt.
 The remaining schemes are not considered, as they either discard too much information or retain excessively noisy elements.}
\label{tab:response_conditioning}
\begin{tabular}{lcccc}
\toprule
\textbf{Conditioning Choice} & $r(\times10^3)$ & $A_L$ & $\langle \chi^2_{\mathrm{eff}} \rangle$ (df=15) & $\langle \chi^2_{\mathrm{min}} \rangle$ (df=13)\\
\midrule
\cancel{\textbf{full full}} & $-0.012 \pm 0.462$ & $1.001 \pm 0.017$ & 15.05 & 13.32 \\
\textbf{full diag} & $0.048 \pm 0.565$ & $1.001 \pm 0.013$ & 15.11 & 12.70\\
\cancel{\textbf{full nearest}} & $-0.034 \pm 0.398$ & $1.001 \pm 0.015$ & 19.26 & 16.50\\
\textbf{\textit{diag full}} & $-0.016 \pm 0.450$ & $1.001 \pm 0.015$ & 15.02 & 13.00\\
\cancel{\textbf{diag diag}} & $0.035 \pm 0.536$ & $1.001 \pm 0.012$ & 15.09 & 11.77\\
\cancel{\textbf{diag nearest}} & {$-0.026 \pm 0.409$} & {$1.001 \pm 0.014$} & {18.69} & 15.91\\
\textbf{\textit{nearest full}} & $-0.018 \pm 0.455$ & $1.001 \pm 0.016$ & 15.18 & 13.30\\
\textbf{nearest diag} & $0.052 \pm 0.566$ & $1.001 \pm 0.012$ & 15.23 & 12.45\\
\cancel{\textbf{nearest nearest}} & {$-0.039 \pm 0.399$} & {$1.001 \pm 0.015$} & {19.07} & 16.16\\
\bottomrule
\end{tabular}
\end{table}

\begin{table*} 
	\centering 
	\caption{Mean values and $1\sigma$ uncertainties of each parameter obtained using the Gradient-Order method for the \texttt{LATS+SAT}, \texttt{LATS+SAT+External}, \texttt{LATN+LATS+SAT}, and \texttt{LATN+LATS+SAT+External} cases. The HL likelihood and ``nearest full'' covariance conditioning were used in the analysis. These results are comparable to those presented in the main text using the ``diag full'' conditioning scheme (Table~\ref{tab:posterior_temp}).}
	\label{tab:posterior_temp_conditioning}
	\begin{adjustbox}{max width=\textwidth} 
	\begin{tabular}{lccccccc} 
		\hline
		& & & & \multicolumn{4}{c}{Gradient-order Method} \\
		\cmidrule(lr){5-8}  
		Parameter & Input value & \makecell{Before adding LT \\ (\texttt{LATS+SAT})} & \makecell{Before adding LT \\ (\texttt{LATN+LATS+SAT})} & \makecell{After adding LT \\ (\texttt{LATS+SAT})} & \makecell{After adding LT \\ (\texttt{LATS+SAT+External})} & \makecell{After adding LT \\ (\texttt{LATN+LATS+SAT})} & \makecell{After adding LT \\ (\texttt{LATN+LATS+SAT+External})} \\
		\hline
		$r (\times 10^3)$ & 0 & $0.021 \pm 0.699$ & $0.025 \pm 0.618$ & $0.020 \pm 0.600$ & $0.004 \pm 0.522$ & $-0.018 \pm 0.455$ & $-0.034 \pm 0.385$\\
		$A_L$ & 1 & $1.000 \pm 0.022$ & $1.001 \pm 0.020$ & $1.001 \pm 0.018$ & $1.001 \pm 0.015$ & $1.001 \pm 0.016$ & $1.001 \pm 0.012$ \\
		\hline
	\end{tabular}
	\end{adjustbox}
\end{table*}

\subsection{Alternative likelihood function}\label{sec: likelihood_choice}
We use the Hamimeche–Lewis (HL) likelihood \cite{hamimeche2008likelihood} in the baseline analysis, as it effectively captures the non-Gaussian distribution of bandpowers.

When a sufficiently large number of modes is available (i.e., at high $\ell$), the likelihood can be well approximated by a Gaussian form through the central limit theorem:
\begin{equation}
        -2 \ln \mathcal{L}(C_{b} \mid \hat{C}_{b})
            = (\mathbf{C}_{b}-\mathbf{\hat C}_{b})^{T}
            \mathbf{M}_{b}^{-1}
            (\mathbf{C}_{b}-\mathbf{\hat C}_{b})
            + \ln |\mathbf{M}_{b}|,
\end{equation}
where the vector $\mathbf{\hat C}_{b}$ contains the bandpower estimates derived from the observed cut-sky maps—computed using the \texttt{NaMaster} code—and $\mathbf{M}_{b}$ is the corresponding covariance matrix.

The Gaussian likelihood is computationally efficient, but using an incorrect sampling distribution can bias parameter estimates and mischaracterize uncertainties. It is therefore crucial to assess the validity of the Gaussian approximation for the bandpower distribution, especially on large angular scales and for surveys with limited sky coverage.

Since the covariance matrix in this work is estimated from a finite number of simulations, its inverse, $\mathbf{\hat M}_{b}^{-1}$, is not an unbiased estimator of the true inverse covariance, even if $\mathbf{\hat M}_{b}$ itself is unbiased. This bias arises because $\mathbf{\hat M}_{b}^{-1}$ follows an inverse-Wishart distribution, whose expectation value is scaled by a factor $\alpha = \frac{N - p - 2}{N - 1}$. Hartlap and Anderson \cite{hartlap2007your} proposed a simple correction factor to rescale $\mathbf{\hat M}_{b}^{-1}$, which can be directly applied when using a Gaussian likelihood.

However, \cite{sellentin2015parameter} pointed out that this correction does not fully capture the statistical scatter of the estimator $\mathbf{\hat M}_{b}^{-1}$. They instead treat the covariance matrix as a sample and marginalize the likelihood over the inverse-Wishart distribution of the true covariance matrix conditioned on the observed sample. This procedure naturally leads to an adapted multivariate $t$-distribution, known as the Sellentin–Heavens (SH) likelihood, which is given by:
\begin{equation}
	\begin{gathered}
        -2 \ \text{ln} \ \mathcal{L}(C_{b}|\hat{C_{b}}) = N_\text{sim}\ \text{ln} (1 + \frac{(\mathbf{C}_{b}-\mathbf{\hat C}_{b})^T \mathbf{M_{b}^{-1}} (\mathbf{C}_{b}-\mathbf{\hat C}_{b})}{N_\text{sim}-1}) - 2 \ \text{ln} \frac{c}{\sqrt{2 \pi \mathbf{M_{b}}}},
	\end{gathered}
\end{equation}
where $c$ is a normalization constant\cite{sellentin2015parameter}.

The parameter fitting results obtained using these alternative likelihoods are presented in Table~\ref{tab:posterior_temp_GS} (Gaussian likelihood) and Table~\ref{tab:posterior_temp_SH} (SH likelihood), corresponding to the baseline results obtained with the HL likelihood (Table~\ref{tab:posterior_temp}). Overall, we do not observe significant differences in the uncertainties on $r$. However, the bias in $r$ can be noticeably larger when using the GS or SH likelihood compared to the HL likelihood.

\begin{figure}[htbp]
    \centering
    \includegraphics[width=1\textwidth]{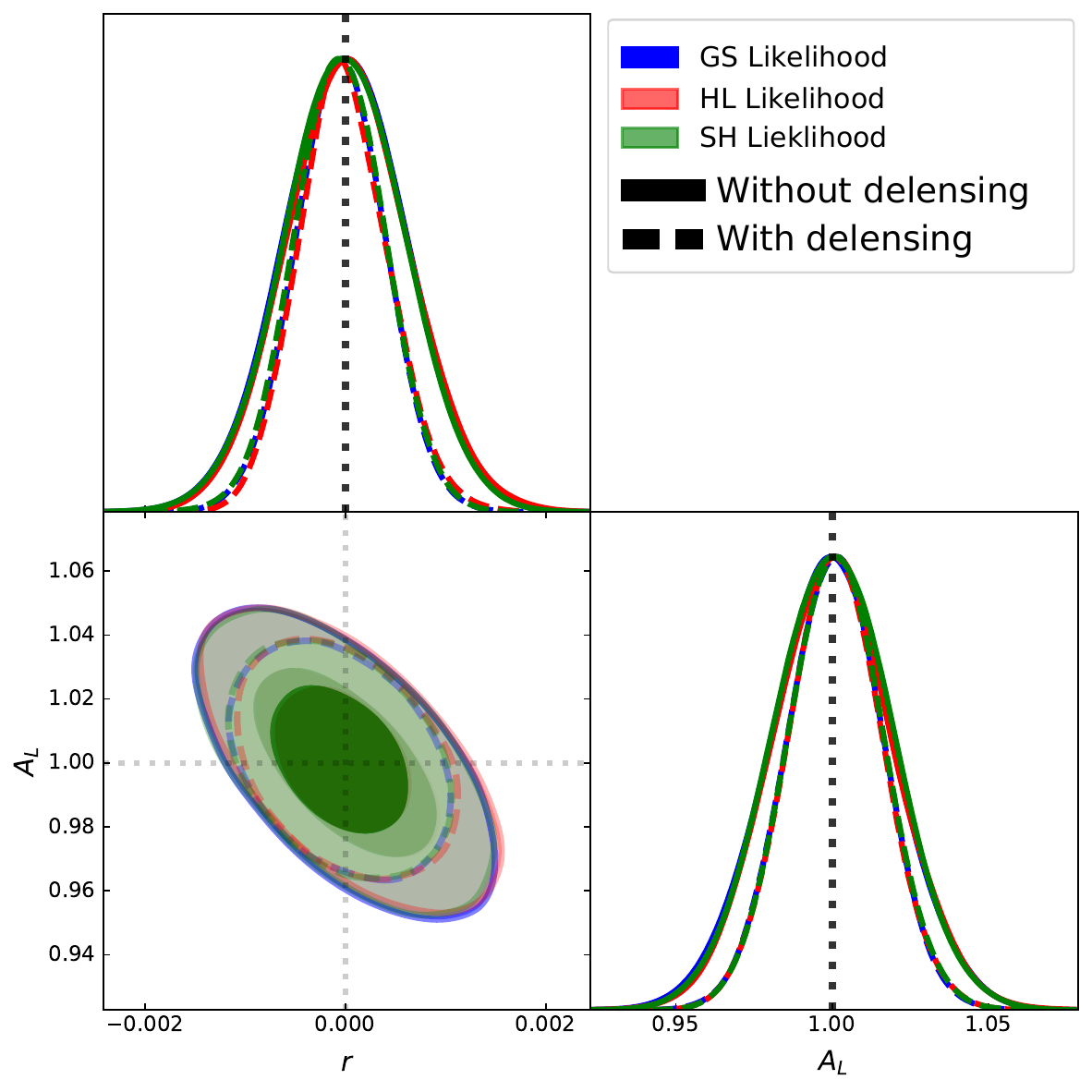}
    \caption{Posterior distributions of the baseline model parameters for the \texttt{LATN+LATS+SAT} cases, shown both with and without delensing. Results obtained using three different likelihoods are presented for comparison.}
    \label{fig: posterior_likelihood}
\end{figure}

Figure~\ref{fig: posterior_likelihood} shows the posterior distributions of the parameters for the \texttt{LATN+LATS+SAT} cases using all three likelihoods. The results demonstrate overall consistency across the different likelihood choices.

\begin{table*} 
	\centering 
	\caption{The mean and $1\sigma$ standard deviation of each parameter using Gradient-order method, with the \texttt{LATS+SAT} case, \texttt{LATS+SAT+External} case, \texttt{LATN+LATS+SAT} case and \texttt{LATN+LATS+SAT+External} case. GS likelihood and "diag full" covariance conditioning were used in likelihood analysis. The results are comparable to those shown in the main text with HL likelihood (Table.\ref{tab:posterior_temp}).}
	\label{tab:posterior_temp_GS}
	\begin{adjustbox}{max width=\textwidth} 
	\begin{tabular}{lccccccc} 
		\hline
		& & & & \multicolumn{4}{c}{Gradient-order Method} \\
		\cmidrule(lr){5-8}  
		Parameter & Input value & \makecell{Before adding LT \\ (\texttt{LATS+SAT})} & \makecell{Before adding LT \\ (\texttt{LATN+LATS+SAT})} & \makecell{After adding LT \\ (\texttt{LATS+SAT})} & \makecell{After adding LT \\ (\texttt{LATS+SAT+External})} & \makecell{After adding LT \\ (\texttt{LATN+LATS+SAT})} & \makecell{After adding LT \\ (\texttt{LATN+LATS+SAT+External})} \\
		\hline
		$r (\times 10^3)$ & 0 & $0.012 \pm 0.712$ & $0.010 \pm 0.609$ & $-0.050 \pm 0.574$ & $-0.083 \pm 0.505$ & $-0.050 \pm 0.447$ & $-0.067 \pm 0.376$\\
		$A_L$ & 1 & $1.000 \pm 0.022$ & $1.000 \pm 0.020$ & $1.001 \pm 0.017$ & $1.001 \pm 0.014$ & $1.001 \pm 0.014$ & $1.001 \pm 0.012$ \\
		\hline
	\end{tabular}
	\end{adjustbox}
\end{table*}

\begin{table*} 
	\centering 
	\caption{The mean and $1\sigma$ standard deviation of each parameter using Gradient-order method, with the \texttt{LATS+SAT} case, \texttt{LATS+SAT+External} case, \texttt{LATN+LATS+SAT} case and \texttt{LATN+LATS+SAT+External} case. SH likelihood and "diag full" covariance conditioning were used in likelihood analysis. The results are comparable to those shown in the main text with HL likelihood (Table.\ref{tab:posterior_temp}).}
	\label{tab:posterior_temp_SH}
	\begin{adjustbox}{max width=\textwidth} 
	\begin{tabular}{lccccccc} 
		\hline
		& & & & \multicolumn{4}{c}{Gradient-order Method} \\
		\cmidrule(lr){5-8}  
		Parameter & Input value & \makecell{Before adding LT \\ (\texttt{LATS+SAT})} & \makecell{Before adding LT \\ (\texttt{LATN+LATS+SAT})} & \makecell{After adding LT \\ (\texttt{LATS+SAT})} & \makecell{After adding LT \\ (\texttt{LATS+SAT+External})} & \makecell{After adding LT \\ (\texttt{LATN+LATS+SAT})} & \makecell{After adding LT \\ (\texttt{LATN+LATS+SAT+External})} \\
		\hline
		$r (\times 10^3)$ & 0 & $0.009 \pm 0.705$ & $-0.007 \pm 0.603$ & $-0.045 \pm 0.578$ & $-0.082 \pm 0.505$ & $-0.055 \pm 0.449$ & $-0.069 \pm 0.378$\\
		$A_L$ & 1 & $1.000 \pm 0.022$ & $1.000 \pm 0.019$ & $1.001 \pm 0.017$ & $1.001 \pm 0.014$ & $1.001 \pm 0.014$ & $1.001 \pm 0.012$ \\
		\hline
	\end{tabular}
	\end{adjustbox}
\end{table*}

\section{Joint delensing on full sky}\label{sec: joint_delensing_app}
To construct a lensing template (LT), one requires both the CMB $E$-modes and the lensing potential at the \textit{same sky positions} and over a \textit{broad multipole range}. Therefore, outside the overlapping area between the ground-based LATs and the SAT, the primary limitation is the \textit{absence of a lensing reconstruction} at those positions, which prevents the construction of an LT.

\begin{figure}[htbp]
    \centering
    \includegraphics[width=1\textwidth]{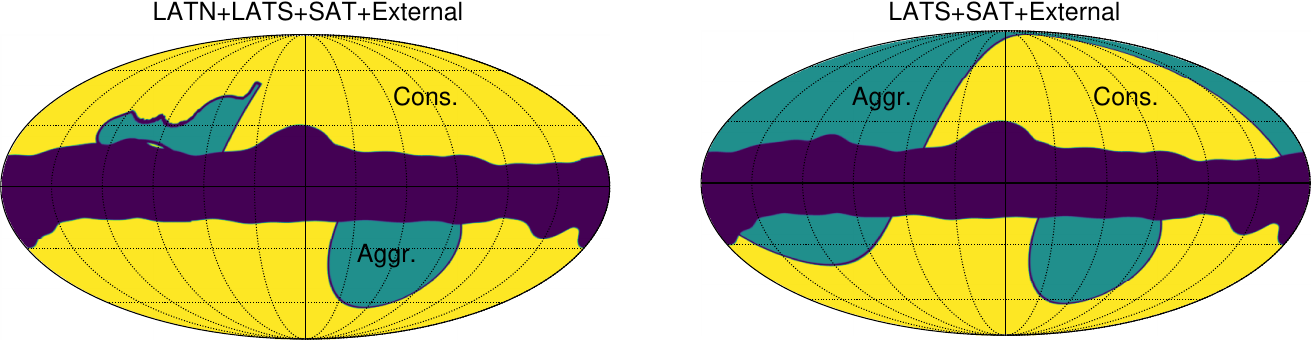}
    \caption{Sky areas used in the joint-delensing cases (\texttt{LATS+SAT+External} and \texttt{LATN+LATS+SAT+External}).
    The LAT–SAT overlapping region is labeled “Cons.”, while the full-sky region includes the LAT–SAT non-overlapping area is labeled “Aggr.”. }
    \label{fig: mask_joint}
\end{figure}

For the joint-delensing cases (\texttt{LATS+SAT+External} and \texttt{LATN+LATS+SAT+External}), one might expect that external large-scale structure (LSS) tracers could provide lensing information outside the LAT–SAT overlap region, enabling the construction of the LT in those areas (the green region in Figure~\ref{fig: mask_joint}). In these cases, within the LAT-covered region, the LT is constructed using $E$-modes from the combined LAT and SAT maps and a lensing proxy that combines internal and external reconstructions. In contrast, in the region not covered by the LATs, the LT must rely on SAT-only $E$-modes together with external reconstruction as the lensing proxy. However, in this region, SAT provides $E$-modes only up to $\ell < 300$, and the external lensing proxy has relatively low signal-to-noise on large angular scales ($\ell < 200$; see Figure~\ref{fig:tracer_rhos} for illustration). These two limitations together lead to a significantly reduced signal-to-noise ratio for the LT.

\begin{figure}[htbp]
    \centering
    \includegraphics[width=1\textwidth]{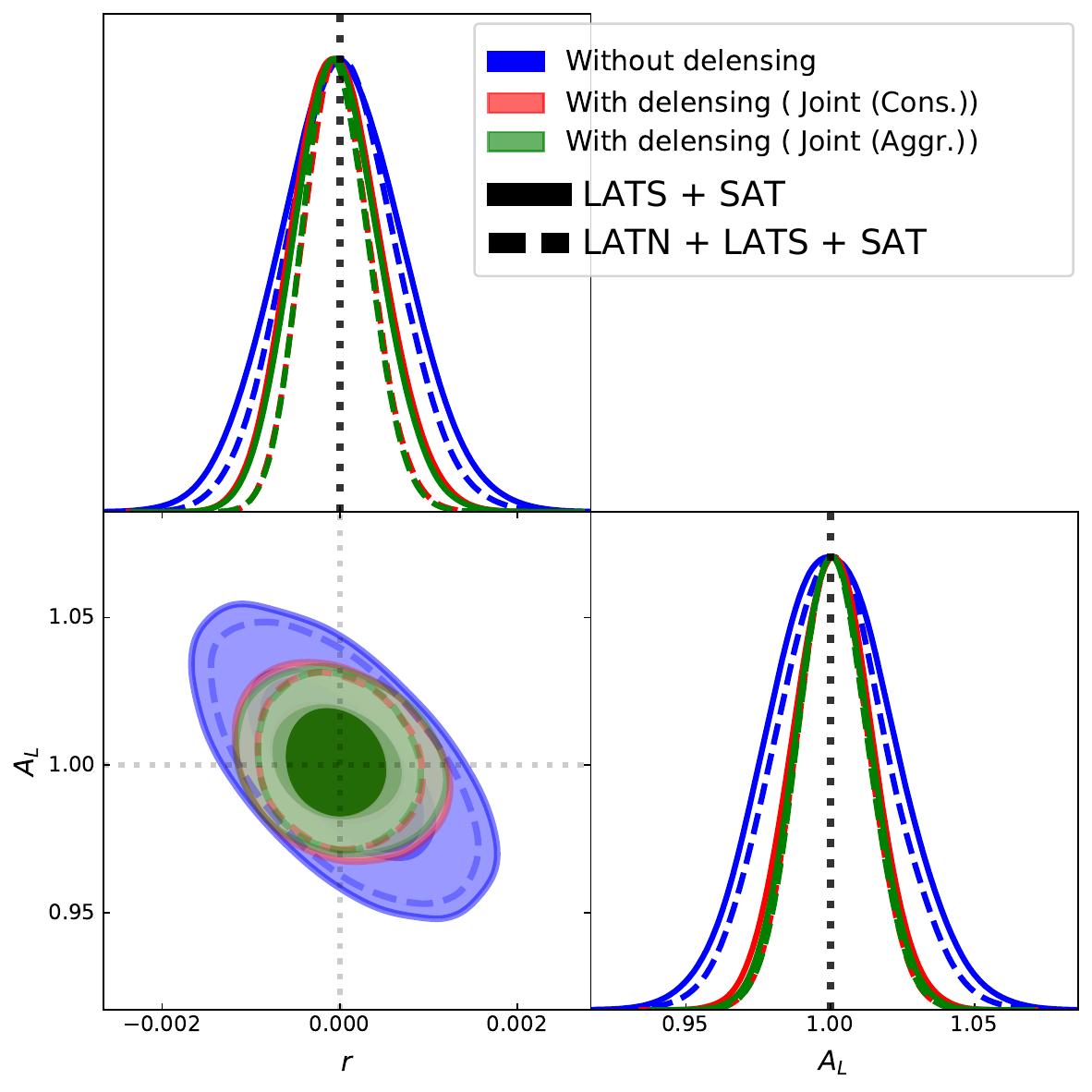}
    \caption{Posterior distributions of the baseline model parameters are shown for the joint-delensing cases (\texttt{LATS+SAT+External} and \texttt{LATN+LATS+SAT+External}), both with and without delensing. We compare two treatments of the LT: using only the LAT--SAT overlapping region (labeled ``Cons.'') and using the full-sky LT (labeled ``Aggr.''). As expected, the contribution from the LT outside the LAT--SAT overlap region is negligible due to its low signal-to-noise ratio.}
    \label{fig: posterior_joint}
\end{figure}

The results with/without the LT outside the LAT--SAT overlap region for the joint delensing cases are shown in Figure.\ref{fig: posterior_joint} and summarized in Table.\ref{tab:posterior_temp_joint}. As expected, we find only a little improvement on reduction on $\sigma(r)$ by including the LT outside the LAT--SAT overlap region.

Since the primary goal of this work is to highlight the impact of LATN on improving constraints on $r$, and given that the LT contribution within the LAT--SAT overlap region is negligible, we therefore restrict our baseline analysis to constructing the LT only within the LAT--SAT overlap region.

\begin{table*} 
	\centering 
	\caption{The mean and $1\sigma$ standard deviation of each parameter using Gradient-order method, for the joint-delensing cases (\texttt{LATS+SAT+External} and \texttt{LATN+LATS+SAT+External}). We compare two treatments of the LT: using only the LAT--SAT overlapping region (labeled ``Cons.'') and using the full-sky LT (labeled ``Aggr.'').}
	\label{tab:posterior_temp_joint}
	\begin{adjustbox}{max width=\textwidth} 
	\begin{tabular}{lccccccc} 
		\hline
		& & & & \multicolumn{4}{c}{Gradient-order Method} \\
		\cmidrule(lr){5-8}  
		Parameter & Input value & \makecell{Before adding LT \\ (\texttt{LATS+SAT})} & \makecell{Before adding LT \\ (\texttt{LATN+LATS+SAT})} & \makecell{After adding LT \\ (\texttt{LATS+SAT+External} (cons.))} & \makecell{After adding LT \\ (\texttt{LATS+SAT+External}(aggr.))} & \makecell{After adding LT \\ (\texttt\texttt{LATN+LATS+SAT+External}(cons.))} & \makecell{After adding LT \\ (\texttt{LATN+LATS+SAT+External}(aggr.))} \\
		\hline
		$r (\times 10^3)$ & 0 & $0.009 \pm 0.705$ & $-0.007 \pm 0.603$ & $-0.025 \pm 0.496$ & $-0.028 \pm 0.483$ & $-0.032 \pm 0.380$ & $-0.043 \pm 0.373$\\
		$A_L$ & 1 & $1.000 \pm 0.022$ & $1.000 \pm 0.019$ & $1.001 \pm 0.014$ & $1.001 \pm 0.013$ & $1.001 \pm 0.012$ & $1.001 \pm 0.012$ \\
		\hline
	\end{tabular}
	\end{adjustbox}
\end{table*}

\section{Alternative choice to mitigate extragalactic bias}\label{sec: bh_app}
As mentioned in Section~\ref{sec: sim}, non-Gaussian extragalactic foregrounds introduce biases in CMB lensing reconstruction and delensing. A number of works have proposed strategies to mitigate these effects~\cite{darwish2023optimizing,osborne2014extragalactic,maccrann2024atacama,qu2024atacama,sailer2023foreground,schaan2019foreground,sailer2020lower,maccrann2024atacama,lizancos2025halo}, based on multi-frequency information and modified lensing estimators.

We use NILC for map cleaning and the standard Quadratic Estimator (hereafter HO) method~\cite{okamoto2003cosmic,carron2017maximum,aghanim2020planck} in our baseline analysis. This combination is known to introduce biases in lensing reconstruction and in the subsequent delensing procedure. In our baseline results, however, these biases are naturally propagated into the likelihood analysis through simulations, and therefore do not lead to significant biases in the inferred parameters.

However, for real observations, the modelling of these biases cannot be perfect due to inevitable limitations in simulations. Consequently, extragalactic foreground--induced biases must be suppressed prior to likelihood analysis. The mainstream mitigation strategies fall into two categories.
First, bright point sources and cluster tSZ can be handled at the map level, for example by subtracting templates constructed using matched filtering~\cite{melin2006catalog,vargas2023atacama,hasselfield2013atacama}, or by masking followed by inpainting~\cite{benoit2013full,raghunathan2019inpainting,bucher2012filling}. One may also fully exploit multi-frequency information and deproject specific foreground components during ILC procedure, at the cost of increased map variance~\cite{kusiak2023enhancing}.
Second, modified lensing estimators—such as bias-hardened estimators~\cite{namikawa2014bias,osborne2014extragalactic,sailer2020lower}, shear estimators~\cite{qu2023cmb,schaan2019foreground}, and symmetric multi-frequency–cleaned estimators~\cite{darwish2023optimizing,madhavacheril2018mitigating,darwish2021atacama}—can further suppress extragalactic-foreground biases, though typically at the expense of some signal-to-noise.

In this section, to complement the baseline setup regarding the treatment of extragalactic foregrounds, we replace the HO temperature estimator with a bias-hardened estimator (hereafter BH), while keeping the polarization estimators unchanged, following \cite{qu2024atacama}:
\begin{equation}
\hat \phi_{LM} = (\mathcal{R}_L^{\mathrm{MV,HO}})^{-1}
\left[
\frac{\hat \phi_{LM}^{\mathrm{MV,HO}}}{(\mathcal{R}_L^{\mathrm{MV,HO}})^{-1}}
- \frac{\hat \phi_{LM}^{\mathrm{TT,HO}}}{(\mathcal{R}_L^{\mathrm{TT,HO}})^{-1}}
+ \frac{\hat \phi_{LM}^{\mathrm{TT,BH}}}{(\mathcal{R}_L^{\mathrm{TT,BH}})^{-1}}
\right].
\end{equation}
Here, $\mathcal{R}$ denotes the lensing response. The resulting estimator is approximately equivalent to the normalized MV estimator, given that we do not consider polarized extragalactic foregrounds in this work.
This modified estimator is subsequently applied in the delensing pipeline and all following analysis steps, in the same manner as implemented in the baseline cases.

\begin{figure}[htbp]
    \centering
    \includegraphics[width=1\textwidth]{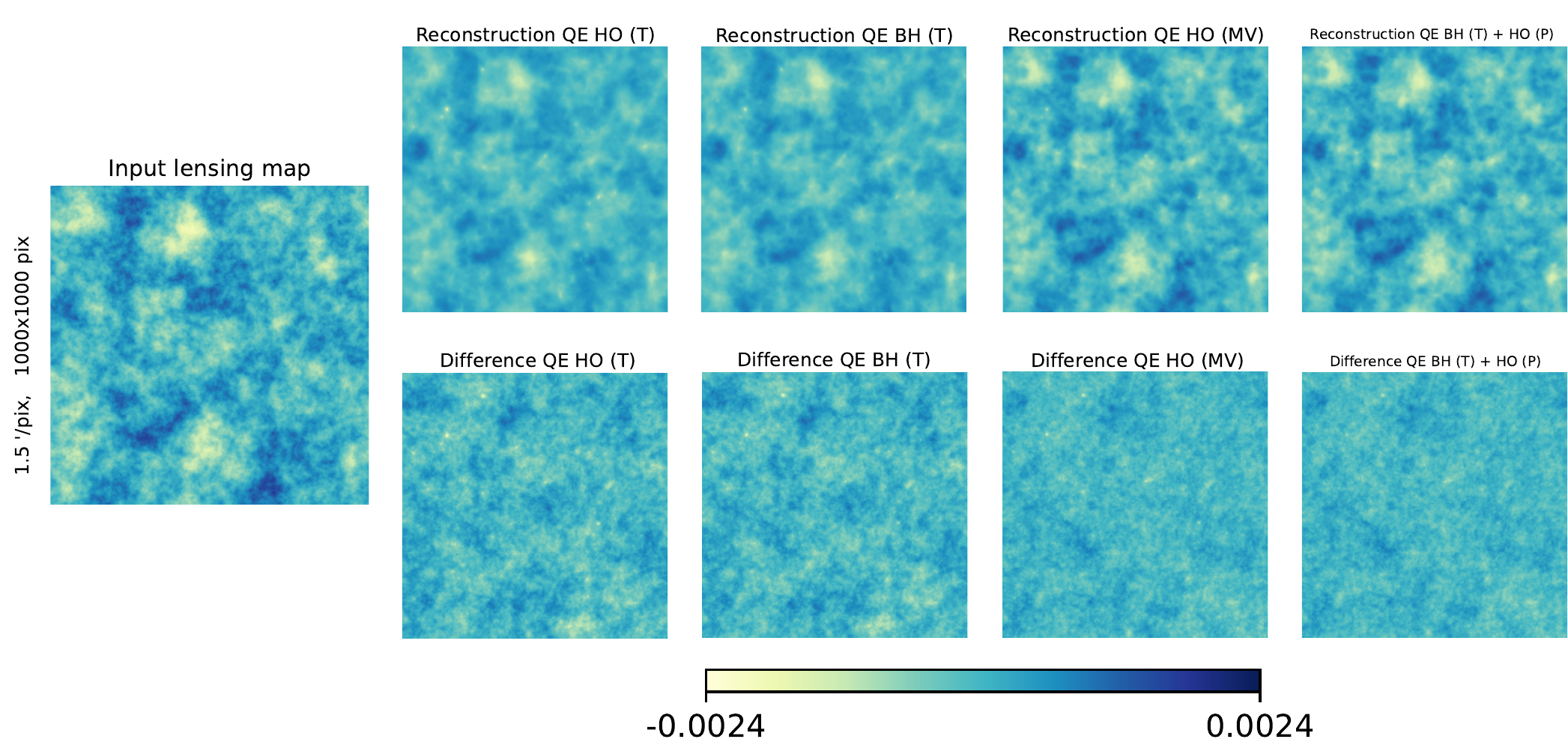}
    \caption{A patch of the reconstructed lensing map is shown. The upper panels display the reconstructed maps, while the lower panels show their differences with respect to the input.
To better highlight the lensing structure, we plot the Wiener-filtered deflection-angle amplitude 
\( 
\hat{\alpha}^{\mathrm{WF}}_{LM}
=
\sqrt{L(L+1)}\, \hat{\phi}^{\mathrm{MV}}_{LM}
\,\frac{C^{\phi\phi,\mathrm{fid}}_L}{C^{\phi\phi,\mathrm{fid}}_L + N^{(0),\mathrm{ana}}_L}
\). It is evident that the HO TT estimator is highly sensitive to point-like foreground-induced biases, whereas these biases are largely suppressed when using the BH TT estimator. The difference between the QE HO (MV) reconstruction and the QE BH(T)+HO(P) reconstruction is less pronounced, since the contribution from the polarization-based estimators dilutes the impact of temperature-related biases.}
    \label{fig: rec_map_patch}
\end{figure}

We show a patch of the lensing reconstruction maps in Figure \ref{fig: rec_map_patch}. It is evident that the temperature HO estimator (QE HO (T)) is sensitive to point-like biases, whereas these biases are largely suppressed when using the bias-hardened estimator (QE BH (T)).

\begin{figure}[htbp]
    \centering
    \includegraphics[width=1\textwidth]{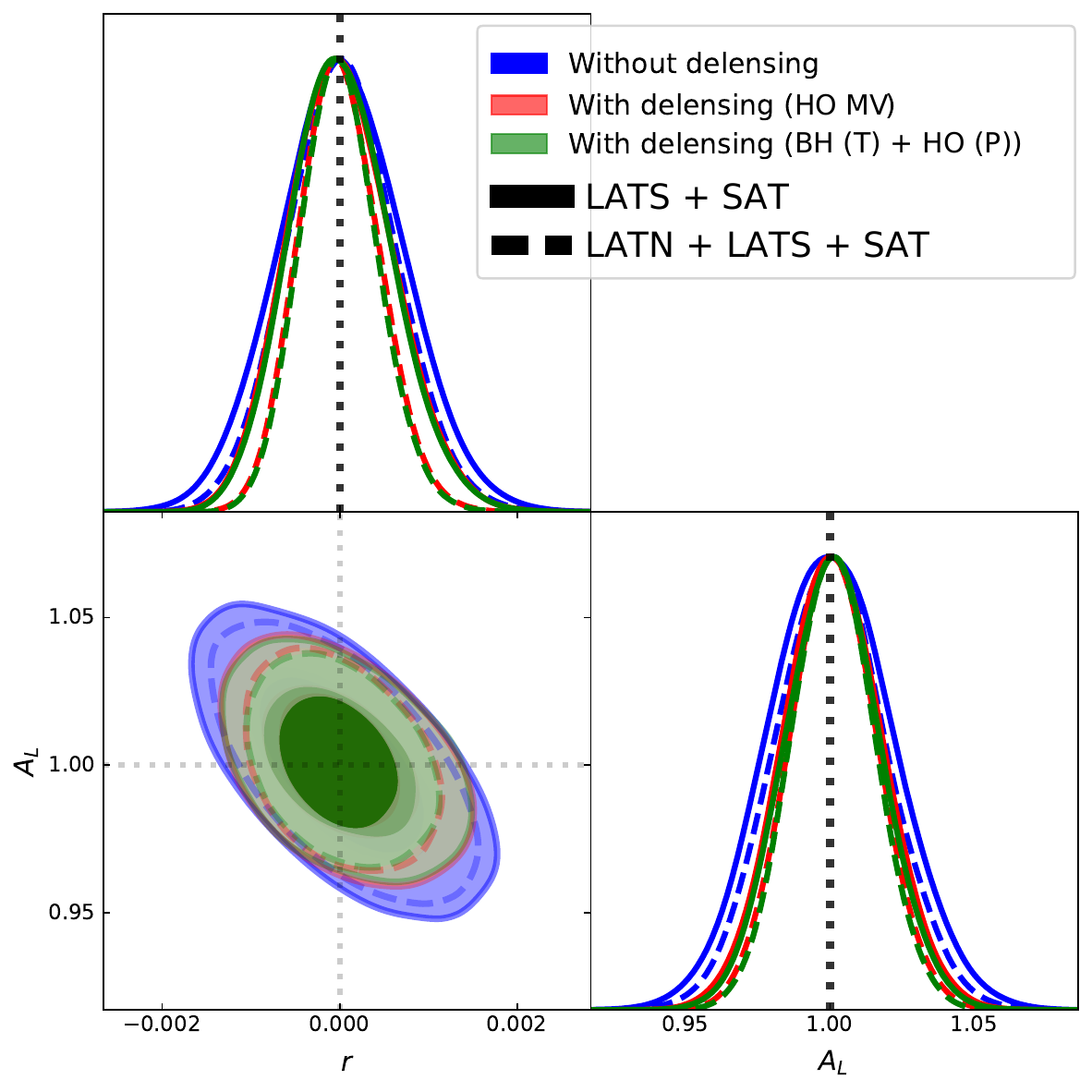}
    \caption{Posterior distributions of the baseline model parameters are shown for the internal–delensing cases (\texttt{LATS+SAT} and \texttt{LATN+LATS+SAT}), where the lensing template (LT) is constructed using either the HO MV estimator or the BH(T)+HO(P) estimator. We find that the LT constructed with the BH(T)+HO(P) estimator leads to slightly tighter constraints.}
    \label{fig: posterior_bh}
\end{figure}

The resulting parameter constraints for the internal delensing cases with different lensing estimators are shown in Figure~\ref{fig: posterior_bh} and summarized in Table~\ref{tab:posterior_temp_bh}. We find that the LT constructed with the BH(T)+HO(P) estimator yields slightly tighter constraints. This improvement arises because the BH temperature estimator reduces extragalactic-foreground–induced biases at the map level, leading to a cleaner LT map. Consequently, some of the bias-related contributions (see e.g.\cite{baleato2022impact} for detailed formula) to the LT power spectra are significantly suppressed, resulting in reduced uncertainties in the covariance matrix and, ultimately, tighter parameter constraints.

All in all, we conclude that although the baseline results do not explicitly correct for extragalactic-foreground–induced biases, the resulting parameter uncertainties remain reasonable—even when compared to those obtained with certain foreground-mitigation strategies applied. A more systematic optimization of these mitigation strategies will be explored in future work.

\begin{table*} 
	\centering 
	\caption{The mean and $1\sigma$ standard deviation of each parameter using Gradient-order method, with the \texttt{LATS+SAT} case and \texttt{LATN+LATS+SAT} case. Lensing estimate can be either HO MV or BH(T)+HO(P) are used for LT construction.}
	\label{tab:posterior_temp_bh}
	\begin{adjustbox}{max width=\textwidth} 
	\begin{tabular}{lccccccc} 
		\hline
		& & & & \multicolumn{4}{c}{Gradient-order Method} \\
		\cmidrule(lr){5-8}  
		Parameter & Input value & \makecell{Before adding LT \\ (\texttt{LATS+SAT})} & \makecell{Before adding LT \\ (\texttt{LATN+LATS+SAT})} & \makecell{After adding LT(HO MV) \\ (\texttt{LATS+SAT})} & \makecell{After adding LT(BH(T)+HO(P)) \\ (\texttt{LATS+SAT})} & \makecell{After adding LT(HO MV) \\ (\texttt{LATN+LATS+SAT})} & \makecell{After adding LT(BH(T)+HO(P)) \\ (\texttt{LATN+LATS+SAT})} \\
		\hline
		$r (\times 10^3)$ & 0 & $0.021 \pm 0.699$ & $0.025 \pm 0.618$ & $0.014 \pm 0.582$ & $0.019 \pm 0.575$ & $-0.016 \pm 0.450$ & $-0.005 \pm 0.440$\\
		$A_L$ & 1 & $1.000 \pm 0.022$ & $1.000 \pm 0.020$ & $1.001 \pm 0.017$ & $1.001 \pm 0.017$ & $1.001 \pm 0.015$ & $1.001 \pm 0.015$ \\
		\hline
	\end{tabular}
	\end{adjustbox}
\end{table*}

\acknowledgments

We thank Siyu Li, Yongping Li, Yepeng Yan, Zi-Xuan Zhang, Qian Chen for useful discussion. This study is supported by the National Natural Science Foundation of China No.12403005, the National Key R\&D Program of China No.2020YFC2201601.
We acknowledge the use of many python packages: \texttt{CAMB} \cite{lewis2011camb}, \texttt{healpy} and \texttt{HEALPix}\footnote{\url{https://healpix.sourceforge.io/}} \cite{2005ApJ...622..759G,Zonca2019}, \texttt{Lenspyx} \cite{carron2020lenspyx,reinecke2023improved},  \texttt{CMBlensplus} \cite{namikawa2021CMBlensplus}, \texttt{Plancklens}\cite{aghanim2020planck}, \texttt{PySM3} \cite{zonca2021python,thorne2017python}
,  \texttt{NaMaster} \cite{alonso2023namaster},  \texttt{Cobaya} \cite{Torrado_2021,torrado2020cobaya} and \texttt{pyccl} \cite{chisari2019core}.


\bibliographystyle{JHEP}
\bibliography{biblio.bib}


\end{document}